\documentclass[preprint,prd,amsmath,amssymb,amsthm,nofootinbib]{revtex4}
\usepackage{xcolor}

\usepackage[mathscr]{euscript}
\usepackage{bm}
\usepackage{graphicx}
\usepackage{subfigure}
\usepackage[colorlinks=true,linkcolor=red]{hyperref}
\newcommand{\bea}{\begin{eqnarray}}
\newcommand{\eea}{\end{eqnarray}}
\newcommand{\beq}{\begin{equation}}
\newcommand{\eeq}{\end{equation}}

\def\/{\over}

\begin{document}
\title{Entanglement dynamics for uniformly accelerated two-level atoms in the presence of a reflecting boundary}

\author{Shijing Cheng,
Hongwei Yu\footnote{Corresponding author: hwyu@hunnu.edu.cn } and
Jiawei Hu\footnote{Corresponding author: jwhu@hunnu.edu.cn}}

\affiliation{
Department of Physics and Synergetic Innovation Center for Quantum Effects and Applications, Hunan Normal University, Changsha, Hunan 410081, China}

\begin{abstract}

We study the entanglement dynamics for two uniformly accelerated two-level atoms in interaction with a bath of fluctuating electromagnetic fields in vacuum in the presence of a reflecting boundary. We consider two different alignments of atoms, i.e. parallel and vertical alignments with respect to the boundary.
In particular, we focus on the effects of the boundary, and acceleration on the entanglement dynamics, which are closely related to the orientations of  polarization. 
For the parallel case, the initial entanglement of two transversely polarizable atoms very close to the boundary can be preserved as if it were a closed system,  while for two vertically polarizable atoms, the concurrence evolves two times as fast as that in the free space.
In the presence of a boundary, entanglement revival is possible for two atoms initially in the symmetric state depending on the orientations of the atomic polarizations, which is in sharp contrast to the fact that the concurrence always decays monotonically in the free space.
Interestingly, two initially separable atoms, for which  entanglement generation can never happen in the free space with any given acceleration and separation, can get entangled in the presence of a boundary if they are aligned parallel to the boundary. 
The birth time of entanglement can be noticeably advanced or postponed for the parallel two-atom system placed close to the boundary, while the maximal concurrence during evolution can be significantly enhanced when the atoms are vertically aligned.
Moreover, two inertial atoms with different polarizations remain separable all the time,  while as the acceleration increases, the delayed birth of entanglement happens, and the nonzero concurrence can be enhanced.



\end{abstract}

\maketitle

\section{Introduction}

Quantum entanglement is one of the central concepts in quantum physics, and it plays an important role in many novel technologies such as  quantum communication \cite{C. H. Bennett1}, quantum teleportation \cite{C. H. Bennett2}, and so on. Unfortunately, quantum systems are inevitably influenced by the environment they are coupled to, and the system-environment interactions generally lead to decoherence \cite{decoherence}. Further studies show that two initially entangled atoms may get completely disentangled within a finite time,  which is known as entanglement sudden death \cite{T. Yu3,  T. Yu}, although the decoherence of a single atom occurs asymptotically.  However, decoherence is not the only consequence that environment entails. A common bath can also provide indirect interactions between otherwise independent atoms, which may lead to entanglement generation \cite{Braun,Kim,Schneider,Basharov,Jakobczyk,Reznik,Piani,Z. Ficek, R. Tanas,Floreanini,tanas2010,Tanas08}. For two atoms immersed in a common thermal bath, it has been shown that entanglement generation only happens in certain circumstances, while entanglement sudden death is a general feature \cite{tanas2010}. The destroyed entanglement can also be recreated, which is known as entanglement revival \cite{Ficek2}.

A uniformly accelerated observer perceives the Minkowski vacuum as a thermal bath with a temperature proportional to its proper acceleration, which is known as the Unruh effect~\cite{W. G. Unruh}.
So it is of interest to investigate the generation and evolution of entanglement for two uniformly accelerated atoms, and compare the results with those in a thermal bath at the corresponding Unruh temperature.
The entanglement generation for two uniformly accelerated atoms coupled with a bath of fluctuating scalar fields in the Minkowski vacuum with a vanishing separation has been studied, and the asymptotic entanglement of the atoms is shown to be exactly the same as that in a thermal bath at the Unruh temperature~\cite{Benatti}.
The time evolution of entanglement for a two-qubit system coupled with a bath of fluctuating scalar fields has been investigated \cite{Landulfo, Ostapchuk2, Hu}, and a comparison between the entanglement dynamics of accelerated atoms and that of static ones in a thermal bath shows that they are the same only in the limit of small acceleration. 
The entanglement dynamics of a quantum system composed of a stationary detector and a uniformly accelerating detector coupled to a common quantum scalar field has been studied in Ref. {\cite{Lin1}}, in which it has been shown that the relation between the disentanglement  time and acceleration is different when observed in the inertial and noninertial frame.
Furthermore, the entanglement creation process of two uniformly accelerated detectors coupled with a fluctuating massless scalar field in the Minkowski vacuum moving in opposite directions has been studied in Ref. \cite{Lin2}, and the result suggests that once quantum entanglement is created, it can last for a lifetime much longer than the natural period of the detectors in certain cases.


The studies above model the environment the atoms coupled to as quantum scalar fields in vacuum, while 
a more realistic  the environment would be a bath of fluctuating vacuum electromagnetic fields as opposed to that of scalar fields. In contrast to the scalar field case, it has been demonstrated that the spontaneous emission rate \cite{Takagi,Zhu06,radiation2,Zhou} and the Lamb shift \cite{Passante, radiation4} of an accelerated atom are not equivalent to those in a thermal bath at the Unruh temperature.
In the case of two atoms coupled with a common bath of electromagnetic vacuum fluctuations, it has been shown that the entanglement dynamics, including entanglement degradation, generation, revival and enhancement, is crucially dependent on the polarization directions of the two atoms~\cite{Yang}, which are irrelevant in the scalar field case.


In the above studies, the behaviors of entanglement dynamics are crucially dependent on the environment of vacuum fluctuations of quantum fields the two-qubit system are coupled to, which is characterized by the field correlation functions. 
It is well-known that the vacuum field modes are modified in the presence of a boundary.
The boundary effects on entanglement dynamics for a detector-field system has been explored in Ref. \cite{Behunin}.
Using linear entropy as a measurement of entanglement, the authors have discussed the early-time entanglement dynamics and late-time stationary limit of entanglement, and found that the late-time entanglement between the detector and fields decreases as the detector gets closer to the mirror when the system is in a stationary state.
Recently, entanglement generation for two atoms in interaction with fluctuating vacuum scalar fields near a reflecting boundary has been investigated, which shows that the presence of a boundary may offer more freedom in controlling entanglement generation \cite{Zhanga, Zhangb}.


Therefore it is interesting to study entanglement dynamics for two uniformly accelerated atoms in interaction with a bath of electromagnetic fields in the Minkowski vacuum in the presence of a reflecting boundary.
We expect new features to arise because of  the polarization of atoms as compared with the scalar field case \cite{Zhanga, Zhangb,Hu},  and of the presence of the boundary as compared with the free space case \cite{Yang}.   In particular, two different alignments of atoms in the presence of a boundary, i.e.  parallel and vertical alignments with respect to the boundary,  will be considered. 
For simplicity, we have neglected the effects of several factors in the present paper. First, we ignore the dipole-dipole interaction between the atoms, which is important when the two atoms are placed close to each other, so we do not discuss the case of vanishing interatomic separation, and assume that the separation of the two atoms is comparable to the characteristic wavelength of the atoms. For more details on the distance effects on entanglement dynamics, see, e.g. Refs. \cite {Lin5020,Hsiang}. In particular, the effects of  direct coupling between two detectors have been taken into account in Ref. \cite {Hsiang}. 
Second, we work in the Born-Markov approximation and neglect the memory effects. In fact, some of the results may change when the memory effects are considered. For example, the investigation of the non-Markovian dynamics of two static qubits in interaction with a common electromagnetic field reveals that, entanglement sudden death and the subsequent revivals are absent except when the qubits are sufficiently far apart \cite{Anastopoulos07}. 
Third, we do not take into account the decoherence (dynamics of the internal degrees of freedom) due to the  quantized center of mass motion (external degrees of freedom), as investigated in Ref. \cite{BLH68}. We hope to turn to these issues in a future work.

\section{the basic formalism}

In this section, we study the dynamics of an open quantum system composed of two decoupled, uniformly accelerated two-level atoms, which are weakly coupled with a bath of fluctuating quantum electromagnetic fields in  vacuum in the presence of a reflecting boundary. The Hamiltonian of the whole system takes the following form
\begin{equation}
H=H_{S}+H_{F}+H_{I}.
\end{equation}
Here  $H_{S}$ denotes the Hamiltonian of the two-atom system, which can be expressed as
\begin{equation}
H_{S}=\frac{\omega}{2}\sigma_{3}^{(1)}+\frac{\omega}{2}\sigma_{3}^{(2)},
\end{equation}
where $\sigma_{i}^{(1)}=\sigma_{i}\otimes\sigma_{0}$, $\sigma_{i}^{(2)}=\sigma_{0}\otimes\sigma_{i}$, with $\sigma_{i}\ (i=1,2,3)$ being the Pauli matrices, $\sigma_{0}$ the $2\times2$ unit matrix, and $\omega$ is the energy level spacing of the atoms. $H_{F}$ is the Hamiltonian of the external electromagnetic fields, the details of which are not necessary. $H_{I}$ represents the dipole interactions  between the two atoms and the fluctuating electromagnetic fields, which takes the form
\begin{equation}
H_{I}=-\mathbf{D}^{(1)}(\tau)\cdot \mathbf{E}[x^{(1)}(\tau)]-\mathbf{D}^{(2)}(\tau)\cdot \mathbf{E}[x^{(2)}(\tau)],
\end{equation}
where $\mathbf{D}^{(\alpha)}(\tau)$ $(\alpha=1,2)$ is the electric-dipole moment of the $\alpha$th atom,  and $\mathbf{E}[x^{(\alpha)}(\tau)]$ is the electric-field strength.
To obtain the master equation describing the evolution of the reduced density matrix of the two-atom system, we define two Lindblad operators as
\begin{eqnarray}
A^{(\alpha)}(\omega)\equiv A^{(\alpha)}=\mathbf{d}^{(\alpha)}\sigma_{-}, \ \ \ \ \
A^{(\alpha)}(-\omega)\equiv A^{(\alpha) \dag}=\mathbf{d}^{(\alpha)*}\sigma_{+},
\end{eqnarray}
where $\mathbf{d}^{(\alpha)}$ is the transition matrix element of the dipole operator of the $\alpha$th atom, written as $\mathbf{d}^{(\alpha)}=\langle 0 |\mathbf{D}^{(\alpha)}|1\rangle$.
Then the dipole operator of the $\alpha$th atom can be obtained in the interaction picture and takes the following form
\begin{equation}
\mathbf{D}^{(\alpha)}(\tau)=\mathbf{d}^{(\alpha)}\sigma_{-}e^{-i \omega \tau}+\mathbf{d}^{(\alpha)*}\sigma_{+}e^{i \omega \tau}.
\end{equation}
For simplicity, we assume that initially the two atoms are separable and are decoupled from the quantum electromagnetic fields, i.e. $\rho_{\rm tot}(0)=\rho(0)\otimes|0\rangle\langle0|$,
where $\rho(0)$  denotes the initial state of the two atoms, and $|0\rangle$ the vacuum state of the electromagnetic fields. The density matrix of the total system  satisfies the Liouville equation
\begin{equation}
\frac{\partial\rho_{\rm tot}(\tau)}{\partial \tau} = -i[H,\rho_{\rm tot}(\tau)].
\end{equation}
Under the Born-Markov approximation, the reduced density matrix of the two-atom system $\rho(\tau)=\mathrm{Tr}_{F}[\rho_{\rm tot}(\tau)]$ satisfies the Kossakowskl-Lindblad master equation \cite{Kossakowski, Lindblad},
\begin{equation}\label{m}
\frac{\partial\rho(\tau)}{\partial \tau} = -i[H_{\rm eff},\rho(\tau)] +\mathcal{D}[\rho(\tau)],
\end{equation}
where
\begin{equation}
H_{\rm eff}=H_{S}-\frac{i}{2}\sum\limits_{\alpha,\beta=1}^{2}\sum\limits_{i,j=1}^{3}  H_{ij}^{(\alpha\beta)}\sigma_{i}^{(\alpha)}\sigma_{j}^{(\beta)},
\end{equation}
and
\begin{equation}
\mathcal{D}[\rho(\tau)]=\frac{1}{2}\sum\limits_{\alpha,\beta=1}^{2}\sum\limits_{i,j=1}^{3}  C_{ij}^{(\alpha\beta)}[2\sigma_{j}^{(\beta)}\rho\sigma_{i}^{(\alpha)}-\sigma_{i}^{(\alpha)}\sigma_{j}^{(\beta)}\rho-\rho\sigma_{i}^{(\alpha)}\sigma_{j}^{(\beta)}].
\end{equation}
Introducing the Fourier transform of the electromagnetic field correlation function $\langle{E_{m}(x(\tau))E_{n}(x(\tau'))}\rangle$,
\begin{equation}\label{F}
\mathcal{G}_{mn}^{(\alpha\beta)}(\omega)=\int^{\infty}_{-\infty}d\Delta \tau e^{i\omega\Delta \tau}\langle{E_{m}(x(\tau))E_{n}(x(\tau'))}\rangle,
\end{equation}
where $\Delta \tau=\tau-\tau'$, the coefficient matrix $C_{ij}^{(\alpha\beta)}$  can then be expressed as
\begin{equation}\label{c}
C_{ij}^{(\alpha\beta)}=A^{(\alpha\beta)}\delta_{ij}-iB^{(\alpha\beta)}\epsilon_{ijk}\delta_{3k}-A^{(\alpha\beta)}\delta_{3i}\delta_{3j},
\end{equation}
where
\begin{equation}\label{b1}
A^{(\alpha\beta)}=\frac{1}{16}[\mathcal{G}^{(\alpha\beta)}(\omega)+\mathcal{G}^{(\alpha\beta)}(-\omega)],\ \
B^{(\alpha\beta)}=\frac{1}{16}[\mathcal{G}^{(\alpha\beta)}(\omega)-\mathcal{G}^{(\alpha\beta)}(-\omega)],
\end{equation}
with
\begin{equation}\label{b2}
\mathcal{G}^{(\alpha\beta)}(\omega)=\sum\limits_{m,n=1}^{3}d_{m}^{(\alpha)*}d_{n}^{(\beta)}\mathcal{G}_{mn}^{(\alpha\beta)}(\omega).
\end{equation}
Similarly, $H_{ij}^{(\alpha)}(\omega)$ in the above equations can be derived by replacing $\mathcal{G}_{mn}^{(\alpha\beta)}(\omega)$ with $\mathcal{K}_{mn}^{(\alpha\beta)}(\omega)$,  which is defined as
\begin{equation}
\mathcal{K}_{mn}^{(\alpha\beta)}(\omega)=\frac{P}{\pi i}\int_{-\infty}^{\infty}d\lambda\frac{\mathcal{G}_{mn}^{(\alpha\beta)}(\lambda)}{\lambda-\omega},
\end{equation}
with $P$ representing the principal value.

To investigate the dynamics of the two-atom system, first we shall solve the master equation in an appropriate basis. For convenience, we work in the coupled basis $\{|G \rangle=|00 \rangle,|A \rangle=\frac{1}{\sqrt{2}}(|10 \rangle-|01 \rangle),|S \rangle=\frac{1}{\sqrt{2}}(|10 \rangle+|01 \rangle),|E \rangle=|11 \rangle\}$.
Reexpressing Eq.~(\ref{c}) as
\bea\label{c1}
&&C_{ij}^{(11)}=A_{1}\delta_{ij}-iB_{1}\epsilon_{ijk}\delta_{3k}-A_{1}\delta_{3i}\delta_{3j},\nonumber\\
&&C_{ij}^{(22)}=A_{2}\delta_{ij}-iB_{2}\epsilon_{ijk}\delta_{3k}-A_{2}\delta_{3i}\delta_{3j},\nonumber\\
&&C_{ij}^{(12)}=C_{ij}^{(21)}=A_{3}\delta_{ij}-iB_{3}\epsilon_{ijk}\delta_{3k}-A_{3}\delta_{3i}\delta_{3j},
\eea
then a set of equations describing the evolution of the two-atom system, which are decoupled from other matrix elements, can be expressed, in the coupled basis,  as \cite{Z. Ficek2}
\begin{eqnarray}\label{evolution}
\dot{\rho}_{GG}&=&-2(A_{1}-B_{1}+A_{2}-B_{2})\rho_{GG}+(A_{1}+B_{1}+A_{2}+B_{2}-2A_{3}-2B_{3})\rho_{AA}\nonumber\\
&&+(A_{1}+B_{1}+A_{2}+B_{2}+2A_{3}+2B_{3})\rho_{SS}+(A_{1}+B_{1}-A_{2}-B_{2})(\rho_{AS}+\rho_{SA}),\nonumber\\
\dot{\rho}_{EE}&=&-2(A_{1}+B_{1}+A_{2}+B_{2})\rho_{EE}+(A_{1}-B_{1}+A_{2}-B_{2}-2A_{3}+2B_{3})\rho_{AA}\nonumber\\
&&+(A_{1}-B_{1}+A_{2}-B_{2}+2A_{3}-2B_{3})\rho_{SS}+(-A_{1}+B_{1}+A_{2}-B_{2})(\rho_{AS}+\rho_{SA}),\nonumber\\
\dot{\rho}_{AA}&=&-2(A_{1}+A_{2}-2A_{3})\rho_{AA}+(A_{1}-B_{1}+A_{2}-B_{2}-2A_{3}+2B_{3})\rho_{GG}\nonumber\\
&&+(A_{1}+B_{1}+A_{2}+B_{2}-2A_{3}-2B_{3})\rho_{EE}+(-B_{1}+B_{2})(\rho_{AS}+\rho_{SA}),\nonumber \\
\dot{\rho}_{SS}&=&-2(A_{1}+A_{2}+2A_{3})\rho_{SS}+(A_{1}-B_{1}+A_{2}-B_{2}+2A_{3}-2B_{3})\rho_{GG}\nonumber\\
&&+(A_{1}+B_{1}+A_{2}+B_{2}+2A_{3}+2B_{3})\rho_{EE}+(-B_{1}+B_{2})(\rho_{AS}+\rho_{SA}),\nonumber
\end{eqnarray}
\begin{eqnarray}
\dot{\rho}_{AS}&=&-2(A_{1}+A_{2})\rho_{AS}+(A_{1}-B_{1}-A_{2}+B_{2})\rho_{GG}+(-A_{1}-B_{1}+A_{2}+B_{2})\rho_{EE}\nonumber\\
&&+(-B_{1}+B_{2})(\rho_{SS}+\rho_{AA}),\nonumber\\
\dot{\rho}_{SA}&=&-2(A_{1}+A_{2})\rho_{SA}+(A_{1}-B_{1}-A_{2}+B_{2})\rho_{GG}+(-A_{1}-B_{1}+A_{2}+B_{2})\rho_{EE}\nonumber\\
&&+(-B_{1}+B_{2})(\rho_{SS}+\rho_{AA}),\nonumber\\
\dot{\rho}_{GE}&=&-2(A_{1}+A_{2})\rho_{GE},
\qquad\qquad\qquad\qquad
\dot{\rho}_{EG}=-2(A_{1}+A_{2})\rho_{EG},
\end{eqnarray}
where $\rho_{IJ}=\langle I|\rho|J\rangle$, $I,J\in\{G,E,A,S\}$.
Here if the initial density matrix is chosen as of the X form, i.e. the only nonzero elements are those along the diagonal and antidiagonal of the density matrix in the decoupled basis $\{|00\rangle,|01\rangle,|10\rangle,|11\rangle\}$, the X form will be  maintained during the whole evolution.

We characterize the degree of entanglement by concurrence \cite{W. K. Wootters}, which ranges from 0 for separable states, to 1 for maximally entangled states. For the X states, the concurrence  takes the form \cite{Z. Ficek3}
\begin{eqnarray}\label{K}
C[\rho(\tau)]=\mathrm{max}\{0,K_{1}(\tau),K_{2}(\tau)\},
\end{eqnarray}
where
\begin{eqnarray}
&&K_{1}(\tau)=\sqrt{[\rho_{AA}(\tau)-\rho_{SS}(\tau)]^{2}-[\rho_{AS}(\tau)-\rho_{SA}(\tau)]^{2}}-2\sqrt{\rho_{GG}(\tau)\rho_{EE}(\tau)},\nonumber\\
&&K_{2}(\tau)=2|\rho_{GE}(\tau)|-\sqrt{[\rho_{AA}(\tau)+\rho_{SS}(\tau)]^{2}-[\rho_{AS}(\tau)+\rho_{SA}(\tau)]^{2}}.
\end{eqnarray}

In the following, we explore the entanglement dynamics of a uniformly accelerated two-atom system in the presence of a reflecting boundary. We focus on the effects of the boundary on the entanglement dynamics of the atoms.  In particular, two different configurations are considered, i.e.
both parallel and vertical alignments with respect to the boundary.

\begin{figure}[!htbp]
\centering
\includegraphics[width=0.4\textwidth]{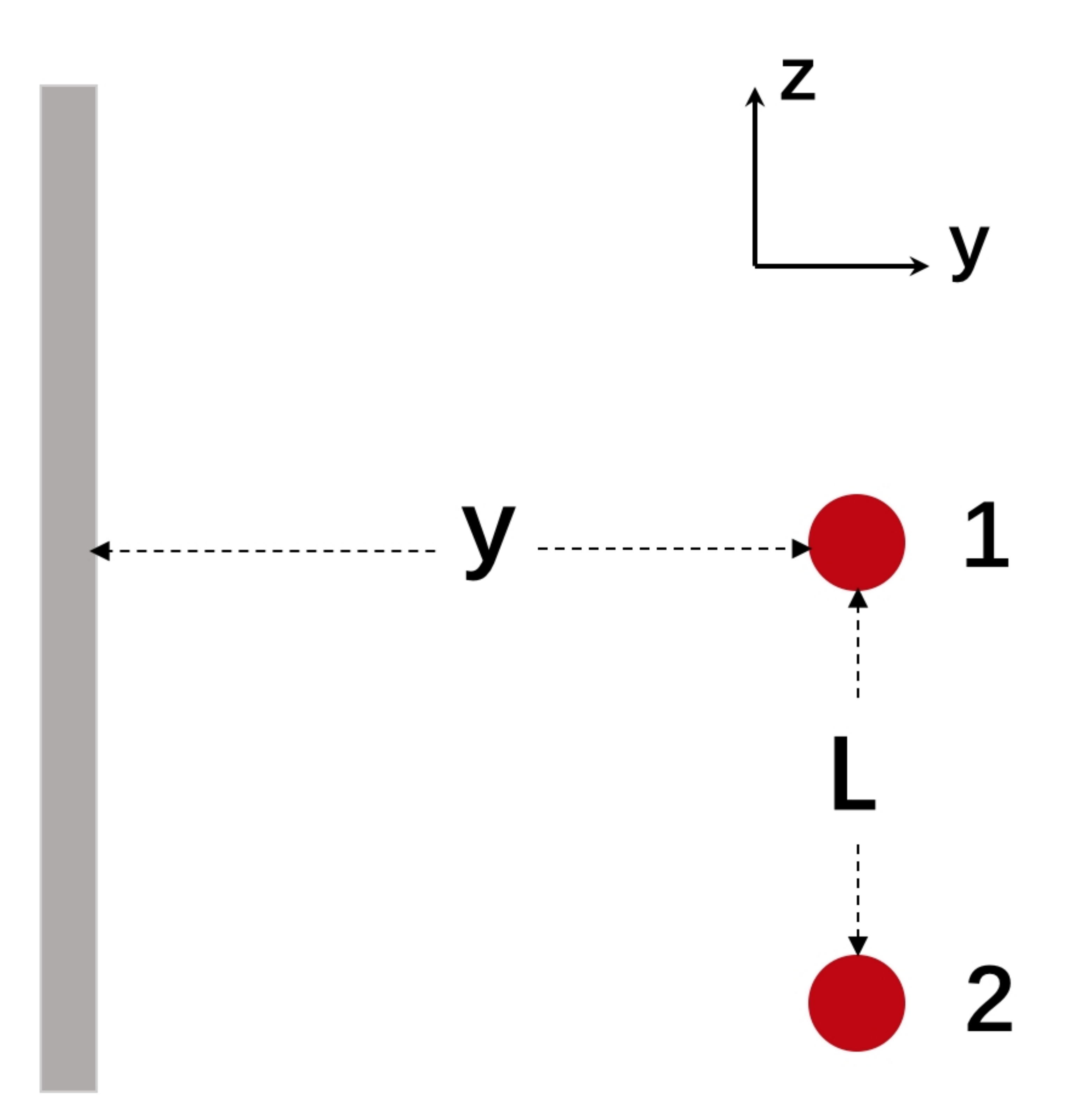}\qquad\qquad
\includegraphics[width=0.4\textwidth]{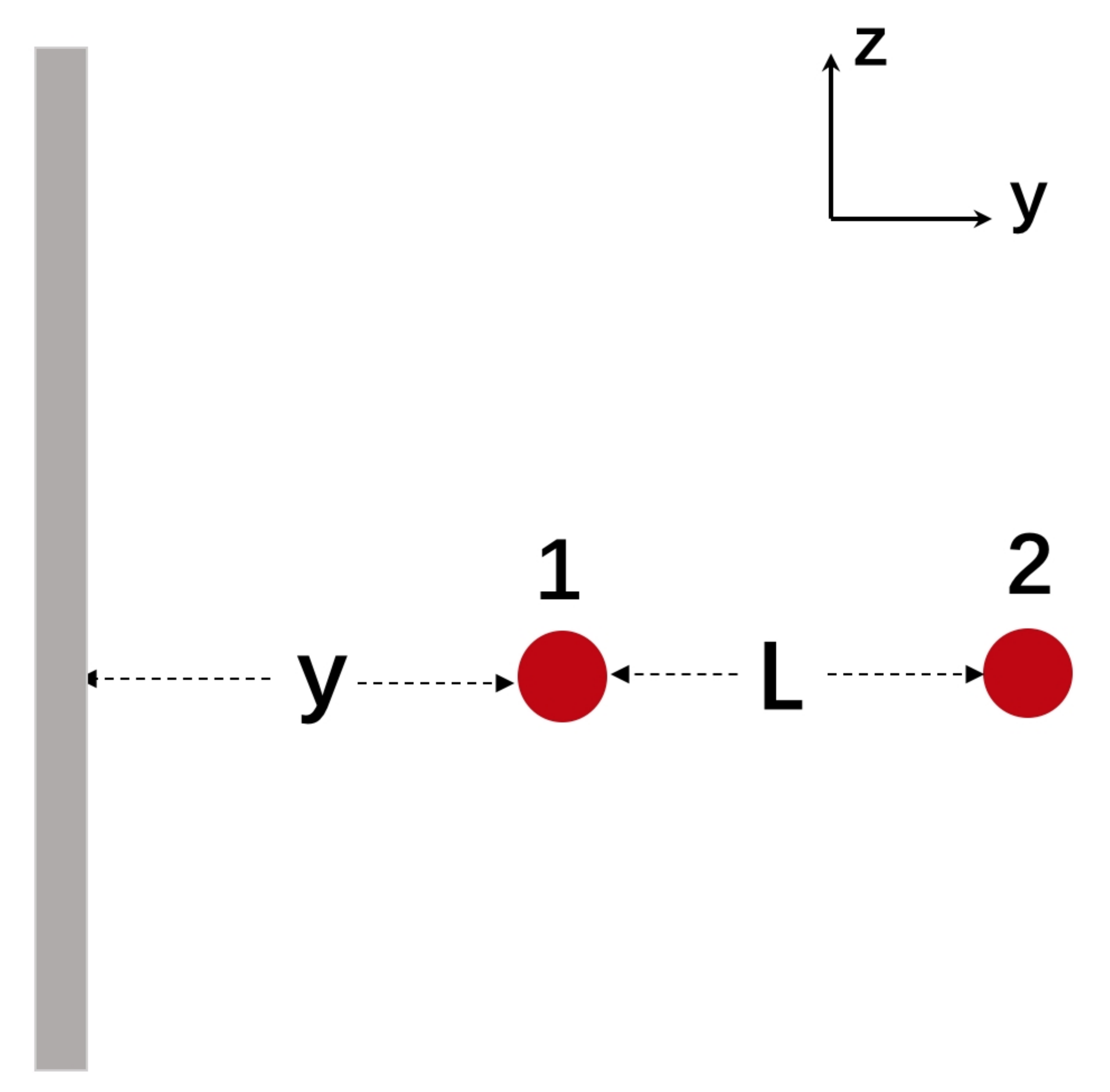}
\caption{\label{palver}
Two atoms are separated from each other at a distance of $L$, which are aligned parallel to (left) or vertically to (right) a reflecting boundary.}
\end{figure}

We consider  two atoms which  are separated from each other by a distance $L$  and are moving with a constant proper acceleration $a$ along the $x$ axis, the trajectories of which are
\begin{eqnarray}
t(\tau)=\frac{1}{a}\sinh{a\tau},\ \  x(\tau)=\frac{1}{a}\cosh{a\tau},
\end{eqnarray}
with $\tau$ being the proper time.
A reflecting boundary is located at $y=0$. We consider two configurations, see Fig. {\ref{palver}}.
In one case, the two-atom system is aligned parallel to the boundary at a distance $y$,  and in the other case, the two-atom system is placed vertically to the boundary, with the distance between the boundary and the nearer atom being $y$.

According to the discussions above, it is necessary to obtain the coefficients $A_{i}$ and $B_{i}$ in Eqs. (\ref{evolution}) in the two situations respectively in order to study how entanglement evolves, which are explicitly calculated in Appendix \ref{app}.

\section{The time evolution of concurrence}

Now we explore the entanglement dynamics of a uniformly accelerated two-atom system in the presence of a reflecting boundary. Two different configurations are considered, i.e. both parallel and vertical alignments with respect to the boundary. 
In particular, we study the cases when the two-atom system is located at different distances from the boundary, paying special attention to the effects of the boundary,  acceleration and polarization  on the entanglement dynamics.

\subsection{Two-atom system placed far from the boundary}

When the two-atom system is placed infinitely far away from  the boundary, i.e., when $y/L\rightarrow \infty$, the bounded parts $h_{ij}^{(\alpha \beta)}$ in Eqs. (\ref{pf1})-(\ref{pf2}) and $s_{ij}^{(\alpha \beta)}$ in Eqs. (\ref{vs}) are all zero, suggesting that the entanglement dynamics reduces to the free space case studied in  Ref.~\cite{Yang}.

\subsection{Two-atom system placed at a distance comparable to the interatomic separation}

When the distance between the two-atom system and the boundary is comparable to the separation of the two atoms, i.e., $y\sim L$, 
we are particularly interested in the degradation of entanglement for atoms initially prepared in a maximally entangled state and the creation of entanglement for atoms in a separable initial state.

\subsubsection{Entanglement degradation}

First, we discuss the entanglement degradation when the two-atom system is initially prepared in the symmetric state $|S\rangle$, which is maximally entangled.

\paragraph{Boundary effects}

\begin{figure}[!htbp]
\centering
\includegraphics[width=0.49\textwidth]{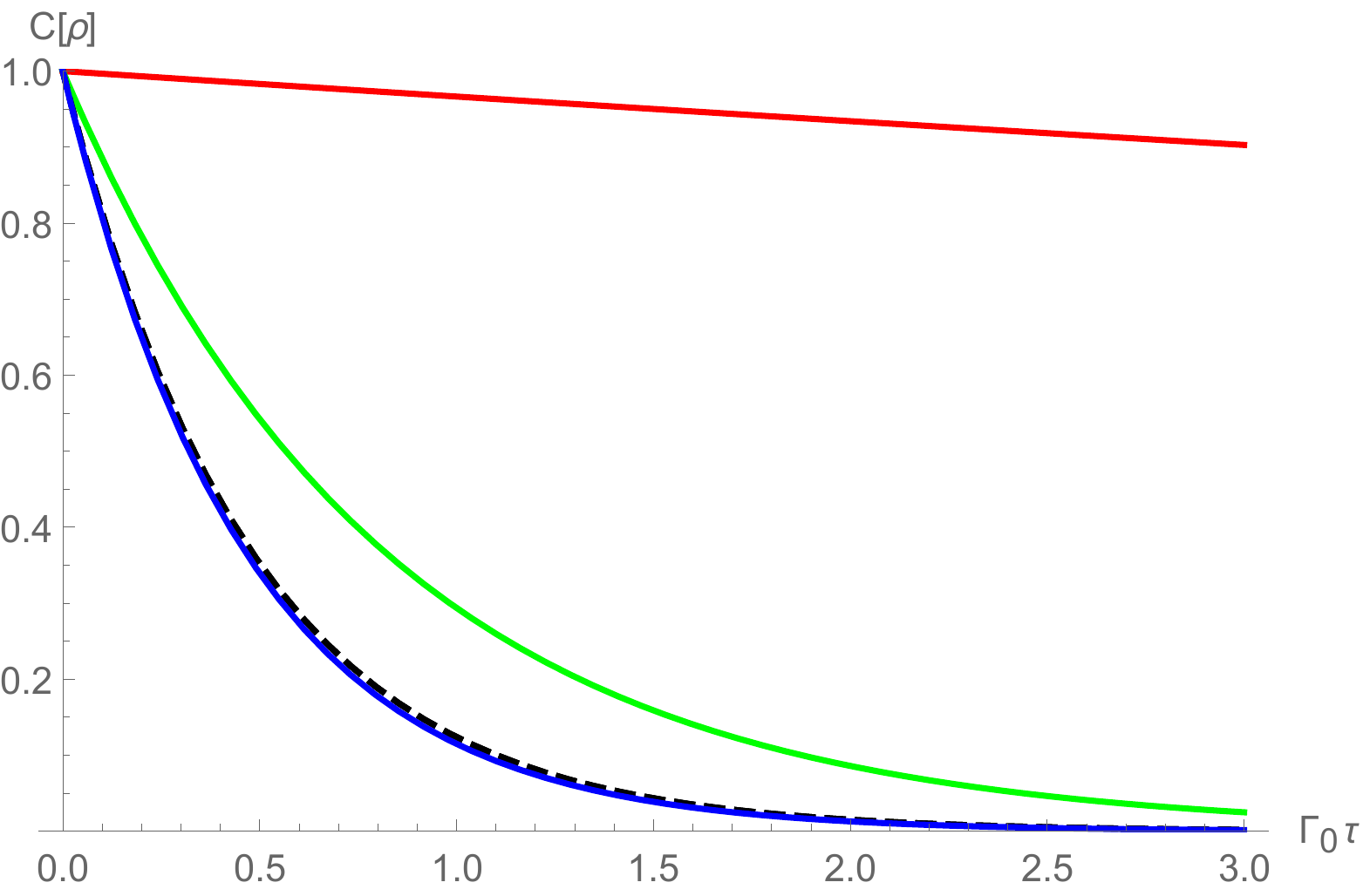}
\includegraphics[width=0.5\textwidth]{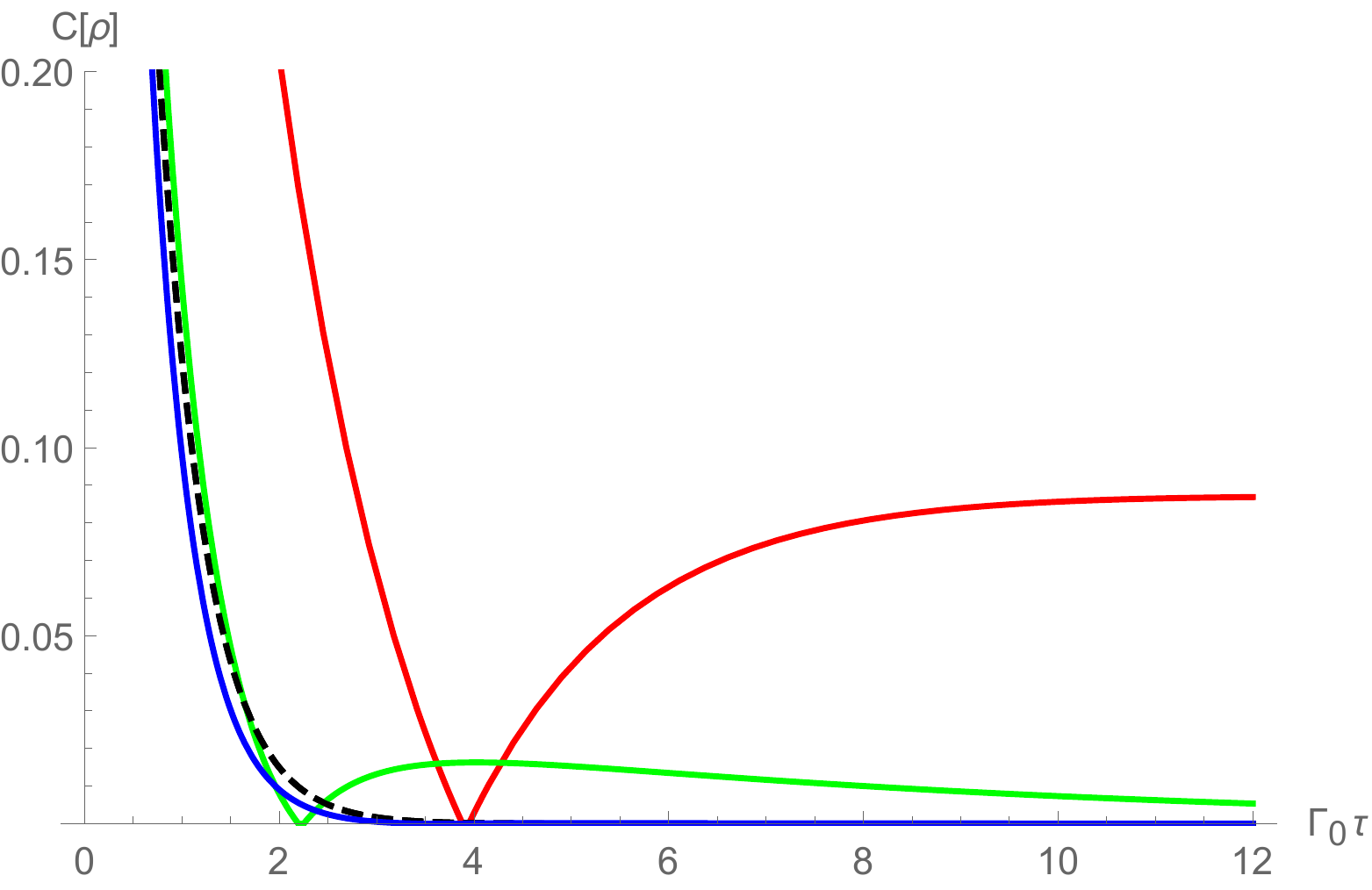}
\caption{\label{psxvsx}
Comparison between the dynamics of concurrence as a function of the dimensionless proper time $\Gamma_0\tau$ for uniformly accelerated atoms initially prepared in $|S\rangle$ aligned parallel to (left) or vertically  to (right) a reflecting boundary. Both of the two atoms are polarizable along the direction of  acceleration  (the $x$-axis). Here $\omega L=1$, $a/\omega=1/2$, $\Gamma_{0}=\omega^{3}|\mathbf{d}|^{2}/3\pi$ is the spontaneous emission rate, the real red, green and blue lines  correspond to $y/L$=1/10, 7/10, 6/5 respectively, and the dashed lines describe the corresponding ones in the free space.}
\end{figure}

First, we concentrate on the effects of the boundary on the entanglement dynamics, so we fix the interatomic separation at $\omega L=1$, the acceleration $a/\omega =1/2$, and assume that both the atoms are polarizable  along the direction of acceleration, which is parallel to the boundary. 
In Fig. \ref{psxvsx}, we compare the entanglement dynamics for  two atoms polarizable along the direction of acceleration and aligned parallel or vertically to the boundary respectively, with the initial state of the two-atom system being $|S\rangle$. Here and after the time evolution of concurrence is plotted as a function of the dimensionless proper time $\Gamma_0\tau$, where $\Gamma_{0}=\omega^{3}|\mathbf{d}|^{2}/3\pi$ is the spontaneous emission rate, see the Appendix \ref{app} for more details. 
Apparently, the decay rate of concurrence decreases as the two-atom system gets  closer  to the boundary, and the lifetime of entanglement is prolonged for the parallel two-atom system when the atoms are transversely polarizable compared with the free space case, see Fig. \ref{psxvsx} (left). However, this is not the case for atoms polarizable  vertically, see discussions in the {\it polarization effects} part for details.
In addition, in the free space case,  the concurrence of the two-atom system initially prepared in $|S\rangle$ always decays monotonically \cite{Yang}. However, in the presence of a boundary, when the atoms are vertically aligned  to the boundary, entanglement revival can be achieved for the atoms with identical polarization, see Fig. \ref{psxvsx} (right).
However, this does not happen in the case when the atoms are aligned parallel to the boundary.

\paragraph{Acceleration effects}

\begin{figure}[!htbp]
\centering
\includegraphics[width=0.49\textwidth]{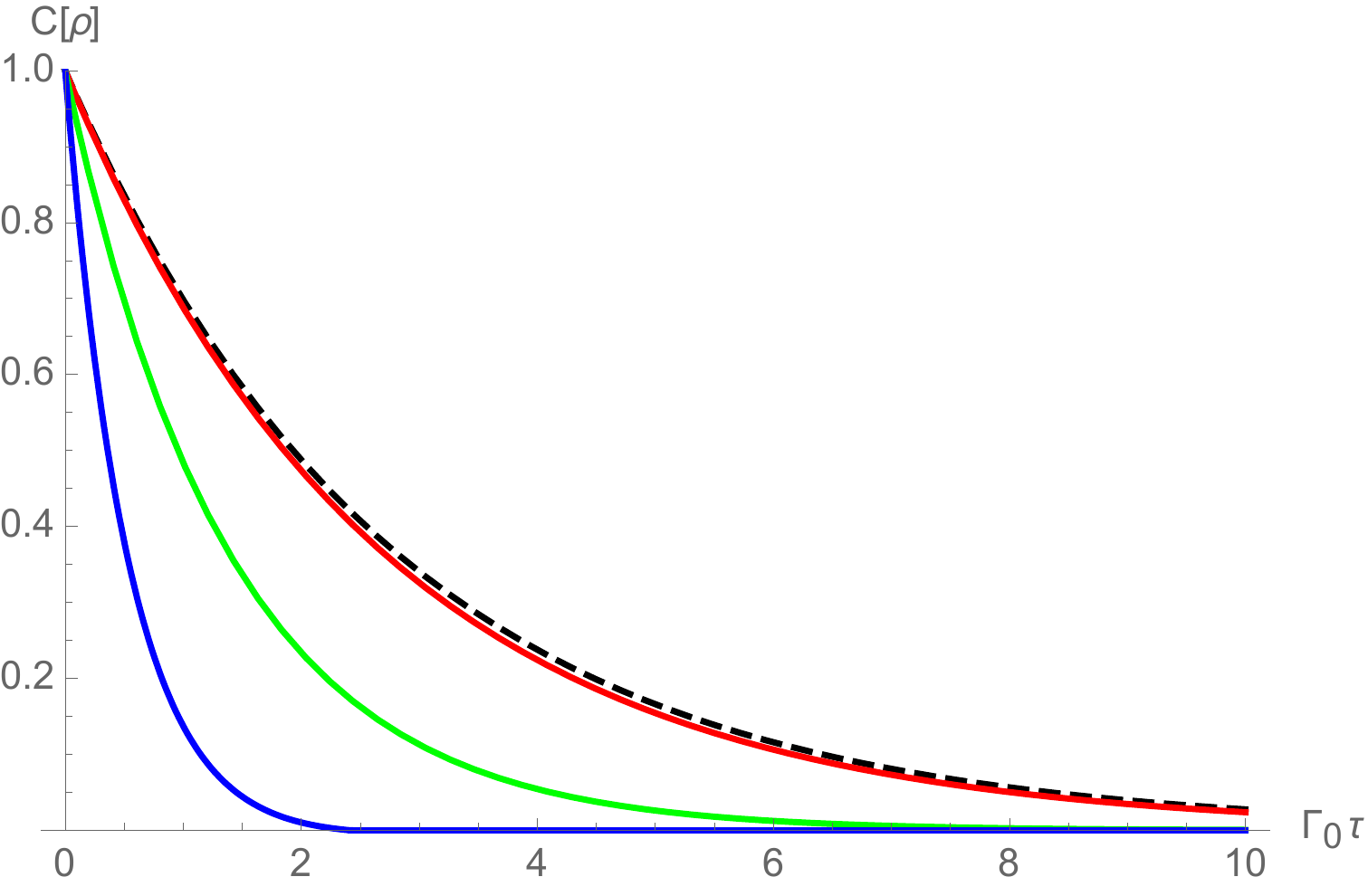}
\includegraphics[width=0.49\textwidth]{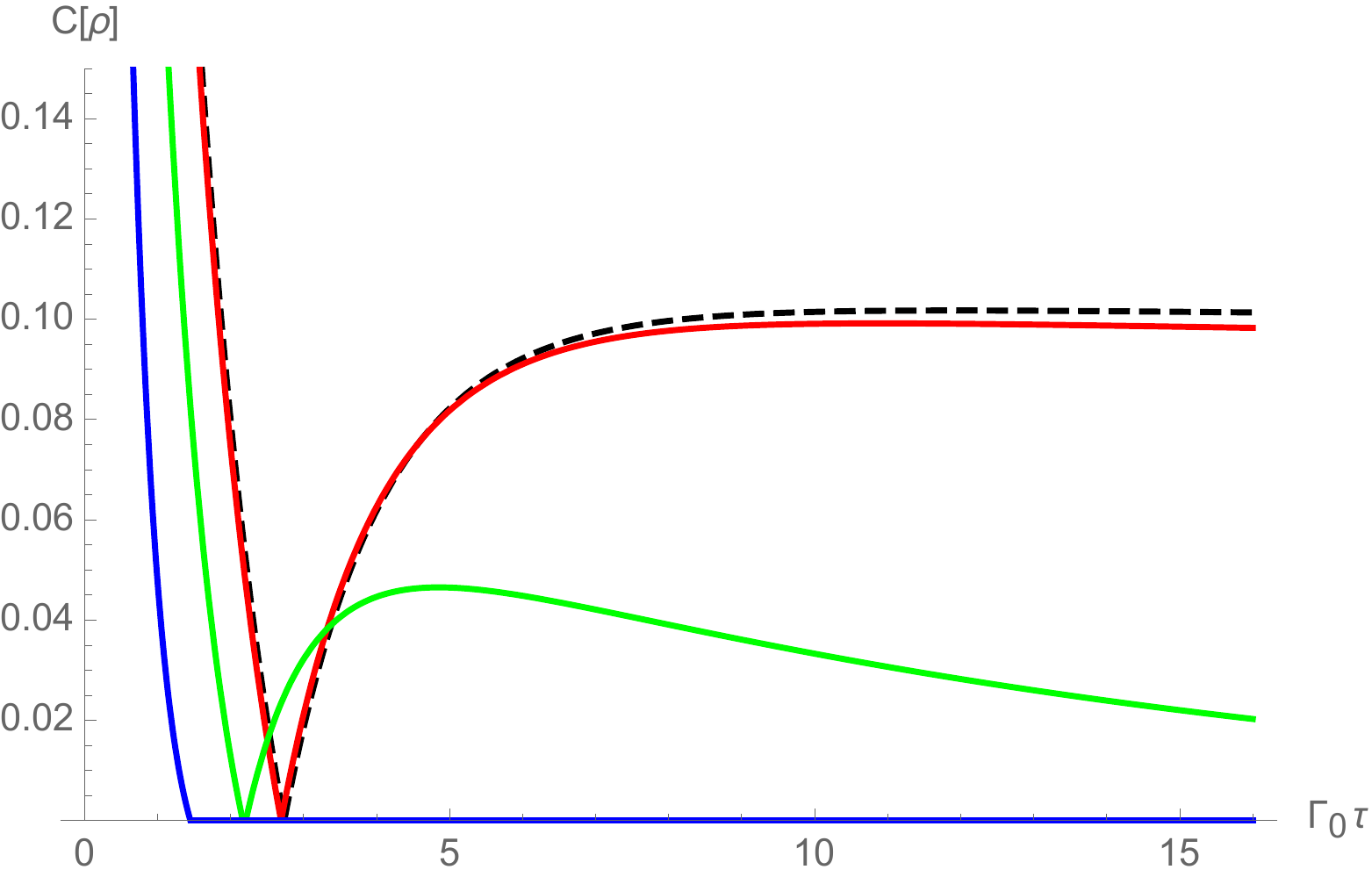}
\caption{\label{ps}
Comparison between the dynamics of concurrence for uniformly accelerated atoms initially prepared in $|S\rangle$ aligned parallel to (left) or vertically  to (right) a reflecting boundary. Both of the two atoms are polarizable along the direction of  acceleration  (the $x$-axis).
Here $\omega L=1$, $y/L=1/2$, the real red, green and blue lines correspond to $a/\omega$=1/10, 1/2, 1 respectively and the dashed lines describe the cases for inertial atoms.}
\end{figure}

In this part, we focus on the acceleration effects on the entanglement dynamics, i.e. the effects due to acceleration. 
As before, we assume that both the two atoms are polarizable along the direction of acceleration, and we set $\omega L=1$, $y/L=1/2$.
Fig.~\ref{ps} (left) shows that when the parallel-aligned two-atom system is located at a distance from the boundary comparable to the separation of the atoms, the lifetime of entanglement is shortened as the acceleration gets larger.
When the atoms are aligned vertically to the boundary, we observe from Fig.~\ref{ps} (right) that  the entanglement revival for two inertial atoms can be achieved, but it does not happen when the acceleration is large enough.
Here we note that entanglement revival can not occur for two inertial atoms with different orientations of polarization.
For both alignments, entanglement sudden death is universal for large acceleration.
As before, some of the results here are also polarization dependent, more explicit discussions are shown in the following. 

\paragraph{Polarization effects}\label{Polarization-effects}

As the entanglement dynamics is expected to be crucially dependent on the polarization directions of the atoms, in the following we  consider  the cases  with different polarizations.

\begin{figure}[!htbp]
\centering
\includegraphics[width=0.49\textwidth]{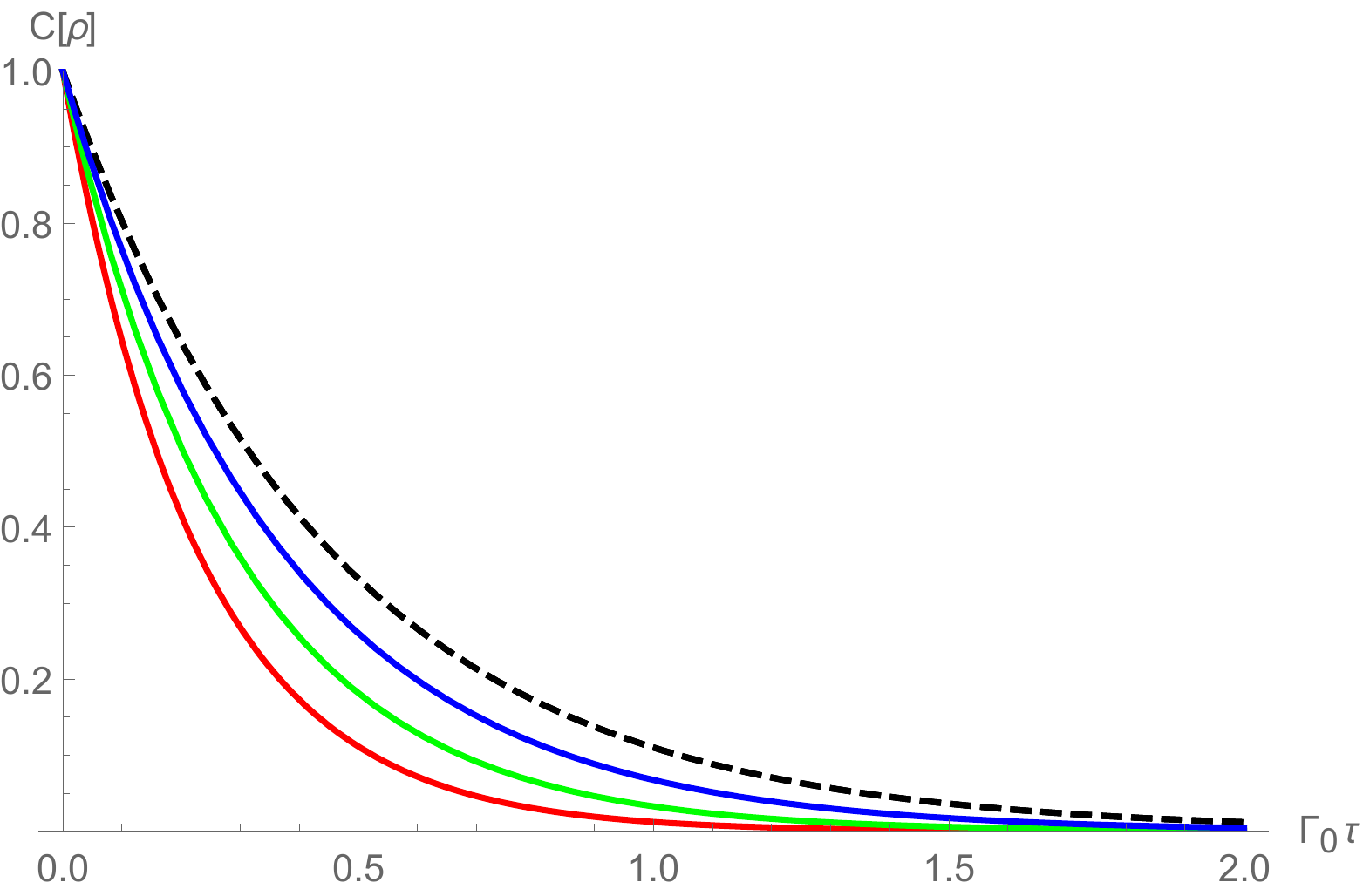}
\includegraphics[width=0.5\textwidth]{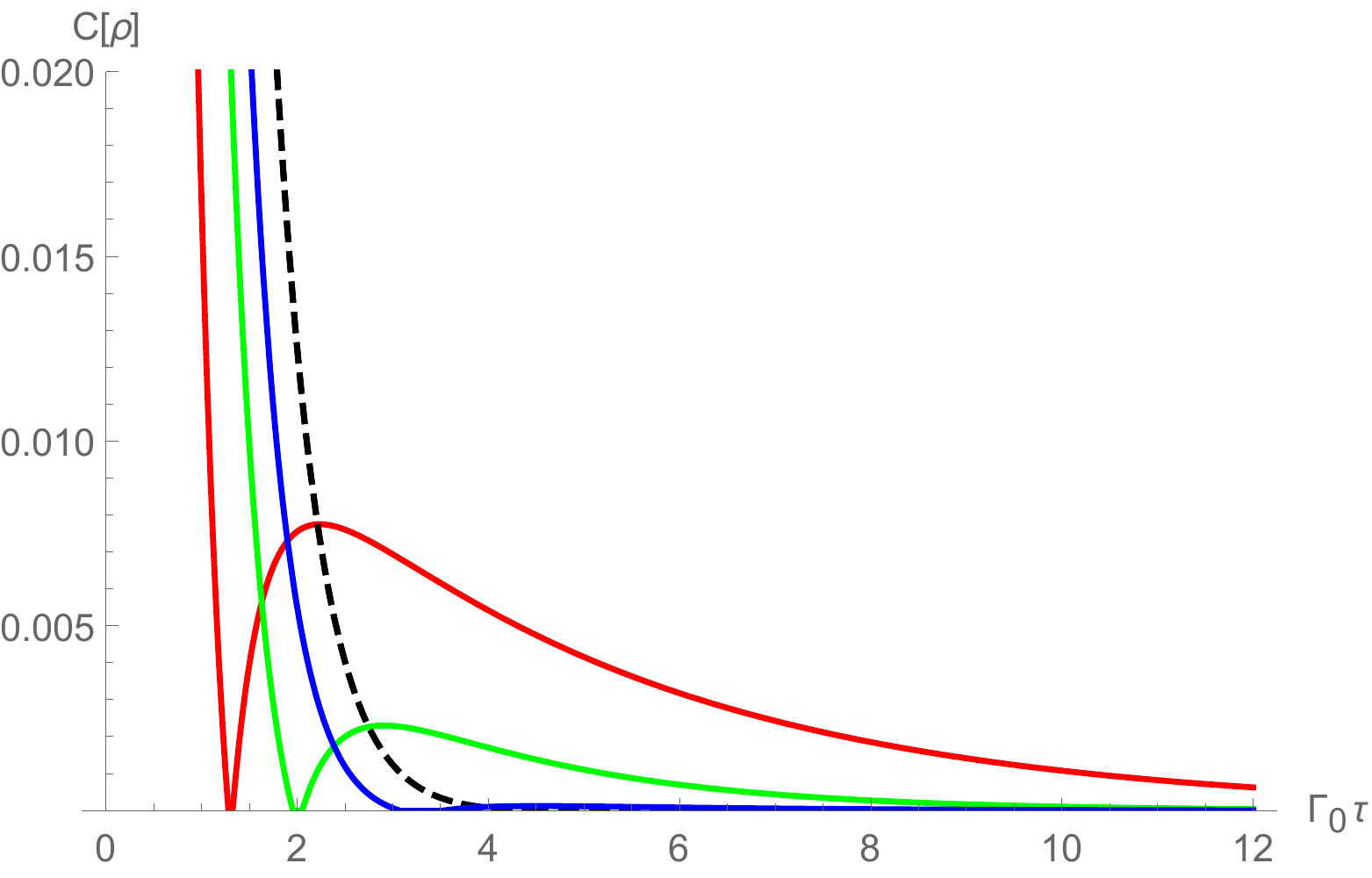}
\caption{\label{psyvsz}
Comparison between the dynamics of concurrence for uniformly accelerated atoms initially prepared in $|S\rangle$ aligned parallel to (left) or vertically  to (right) a reflecting boundary. Both of the two atoms are polarizable vertically to the boundary  (the $y$-axis). Here $\omega L=1$, $a/\omega=1/2$, the real red, green and blue lines  correspond to $y/L$=1/10, 7/10, 6/5 respectively, and the dashed lines describe the corresponding ones in the free space.}
\end{figure}

In contrast to the case when the two atoms are polarizable parallel to the boundary (see Fig. \ref{psxvsx}), for  vertically polarizable atoms, the decay rate of concurrence increases and the lifetime of entanglement becomes shorter as the parallel-aligned atoms gets closer to the   boundary, see Fig. \ref{psyvsz} (left).
That is, the lifetime of entanglement can be either prolonged or shortened compared with that in the free space case, which is a combined effect of the boundary and the atomic polarizations.

\begin{figure}[!htbp]
\centering
\includegraphics[width=0.49\textwidth]{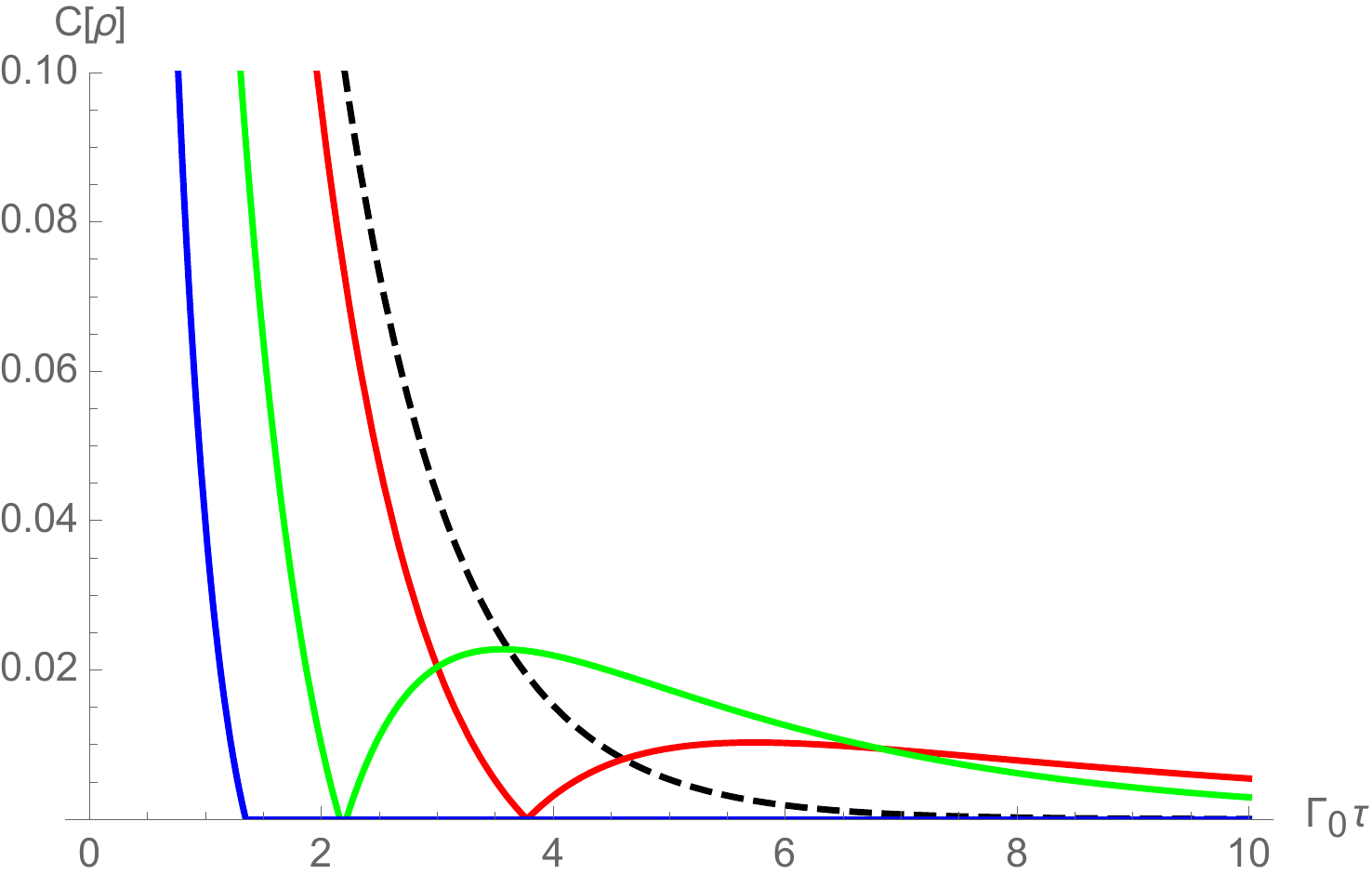}
\includegraphics[width=0.5\textwidth]{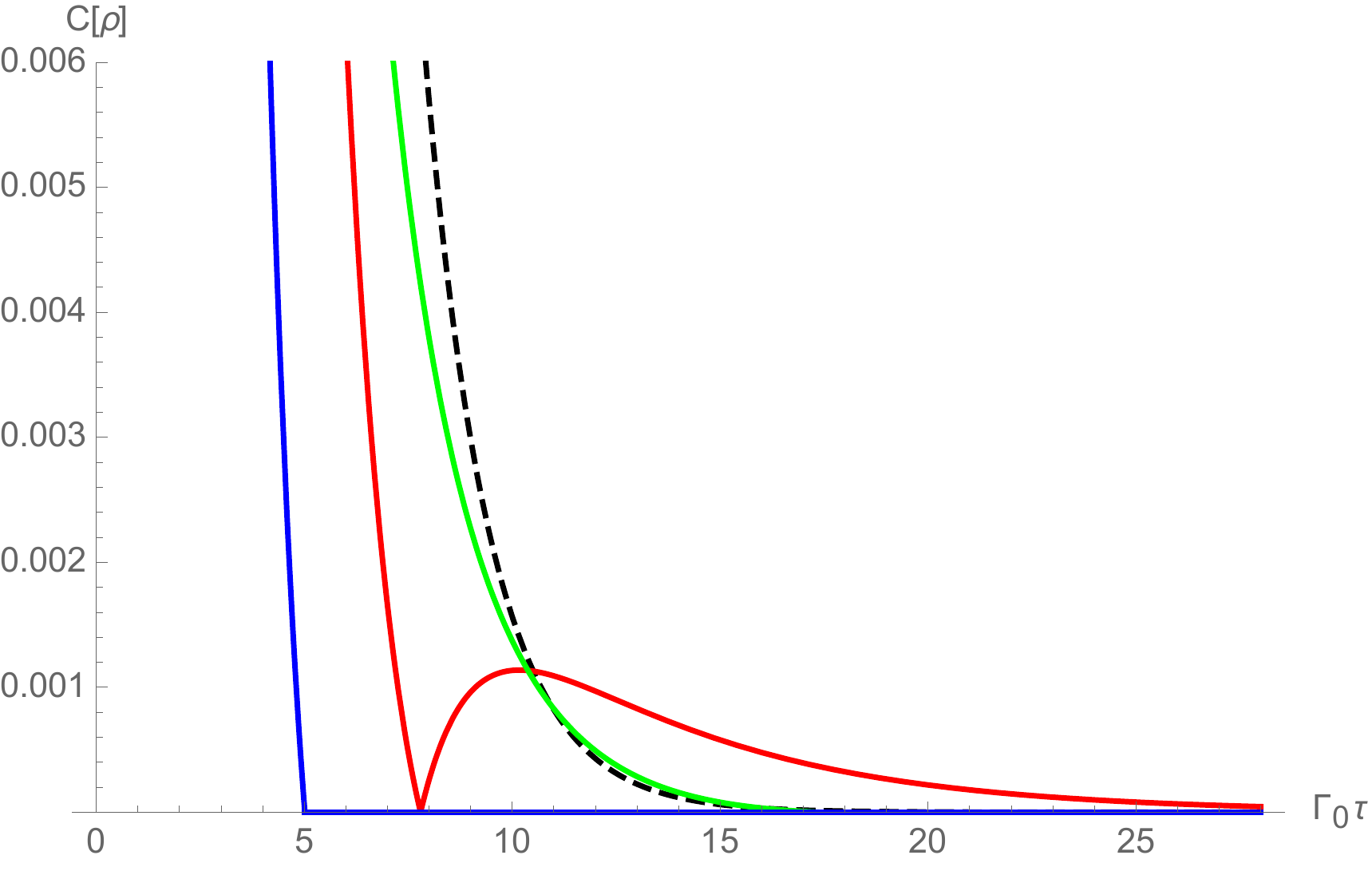}
\caption{\label{pvs1}
Comparison between the dynamics of concurrence for uniformly accelerated atoms initially prepared in $|S\rangle$ aligned parallel to (left) or vertically  to (right) a reflecting boundary.
One of the atoms (the nearer one in the vertically aligned case) is polarizable along the direction of  acceleration (the $x$-axis) and the other vertically to the boundary (the $y$-axis).
Here $\omega L=1$, $y/L=1/2$, the real red, green and blue lines correspond to $a/\omega$=1/10, 1/2, 1 respectively, and the dashed lines describe the cases for inertial atoms.}
\end{figure}

In Fig.~\ref{pvs1} we show the entanglement dynamics for atoms with different polarizations, i.e. one of the atoms is polarizable along the direction of acceleration (the $x$-axis), while the other is polarizable vertically (the $y$-axis). 
When $a \rightarrow 0$, i.e., for inertial atoms, the concurrence decays monotonically. While as the acceleration gets larger, entanglement revival appears, which is in sharp contrast to the case described in Fig. \ref{ps} (right). 
Therefore, acceleration can be either harmful [see Fig. \ref{ps} (right)] or beneficial  (see Fig.~\ref{pvs1}) to the entanglement revival depending on the atomic polarizations.
However, entanglement revival cannot happen as the acceleration gets large enough.

\begin{figure}[!htbp]
\centering
\includegraphics[width=0.49\textwidth]{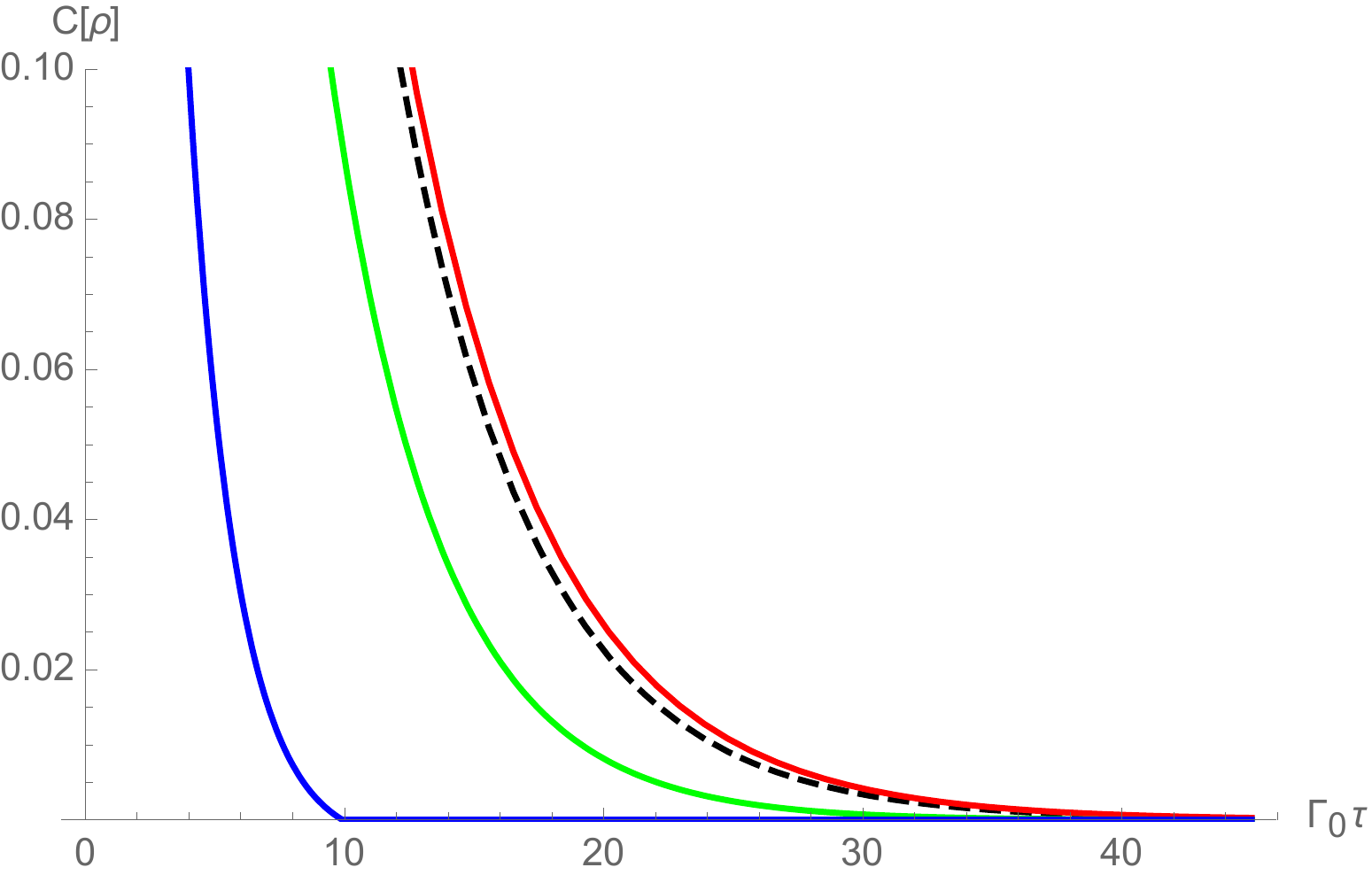}
\includegraphics[width=0.49\textwidth]{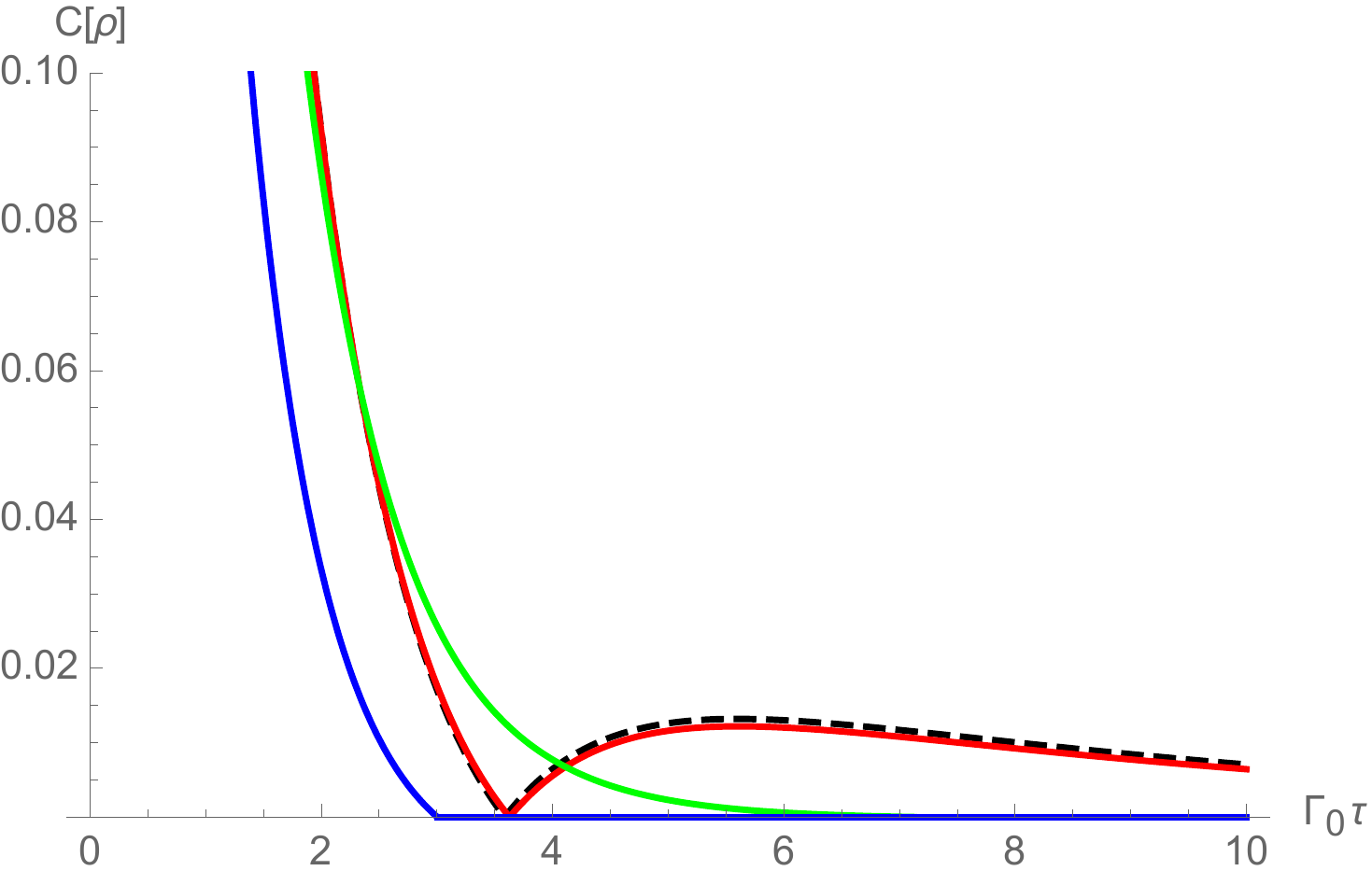}
\caption{\label{ps2}
Comparison between the dynamics of concurrence for uniformly accelerated atoms polarizable differently with the alignment parallel to a reflecting boundary.
The system is initially prepared in $|S\rangle$.
One of the two atoms is polarizable along the direction of separation (the $z$-axis) and the other along the direction of acceleration (the $x$-axis) (left) and vertically to the boundary (the $y$-axis) (right).
Here $\omega L=1$, $y/L=1/2$, the real red, green and blue lines correspond to $a/\omega$=1/10, 1/2, 1 respectively, and the dashed lines describe the cases for inertial atoms.}
\end{figure}

In addition, for parallel-aligned atoms, entanglement revival can never happen  when both atoms are polarizable along the direction of acceleration [see Fig. \ref{ps} (left)] or when the atoms are respectively polarizable along the direction of acceleration and separation [see Fig. \ref{ps2} (left)].
However, it can be achieved for atoms polarizable respectively along the direction of acceleration and vertically to the boundary [Fig. \ref{pvs1} (left)], or for atoms polarizable respectively along the direction of separation and vertically to the boundary [Fig. \ref{ps2} (right)]. 
To conclude, we find that when the two-atom system is aligned parallel to the boundary,  entanglement revival can occur only when the atoms are  polarizable differently, with one of them polarizable vertically to the boundary.

According to Ref. \cite{Anastopoulos07}, the behaviors of entanglement derived from the Markovian and  non-Markovian dynamics are obvious only when the  distance between the two atoms $L$ is much smaller than the transition wavelength $1/\omega$, i.e., $\omega L \ll 1$. Throughout the paper we assume $\omega L \sim1$, so the non-Markovian effects are expected to be negligible. 
In paticular, our discussions show that it is possible to achieve entanglement sudden death but with no revival in the free space (see  Fig. \ref{psxvsx} and Fig. \ref{psyvsz}), which is in agreement with the result in  Ref. \cite{Anastopoulos07}. In contrast, the revival of entanglement can  be achieved in the presence of a boundary.

\subsubsection{Entanglement creation}

When the  initial state is a separable state $|E\rangle$, we can see from Eqs. (\ref{K}) that entanglement can be generated when the factor $\sqrt{[\rho_{AA}(\tau)-\rho_{SS}(\tau)]^{2}-[\rho_{AS}(\tau)-\rho_{SA}(\tau)]^{2}}$ outweighs  $2\sqrt{\rho_{GG}(\tau)\rho_{EE}(\tau)}$, which necessarily takes a finite time of evolution via spontaneous emission, known as the delayed sudden birth of entanglement \cite{Tanas08}.

\paragraph{Boundary effects}

\begin{figure}[!htbp]
\centering
\includegraphics[width=0.49\textwidth]{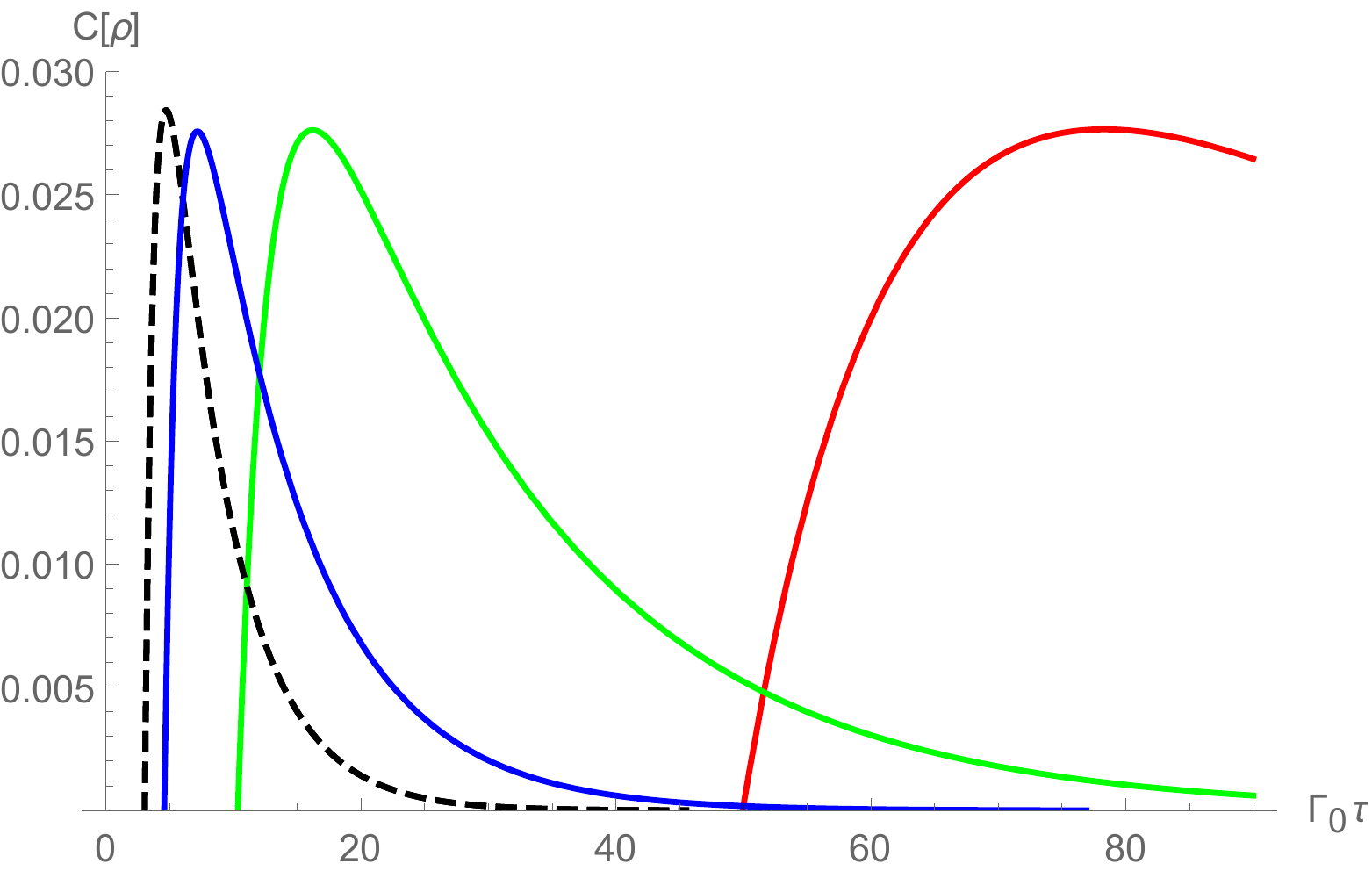}
\includegraphics[width=0.5\textwidth]{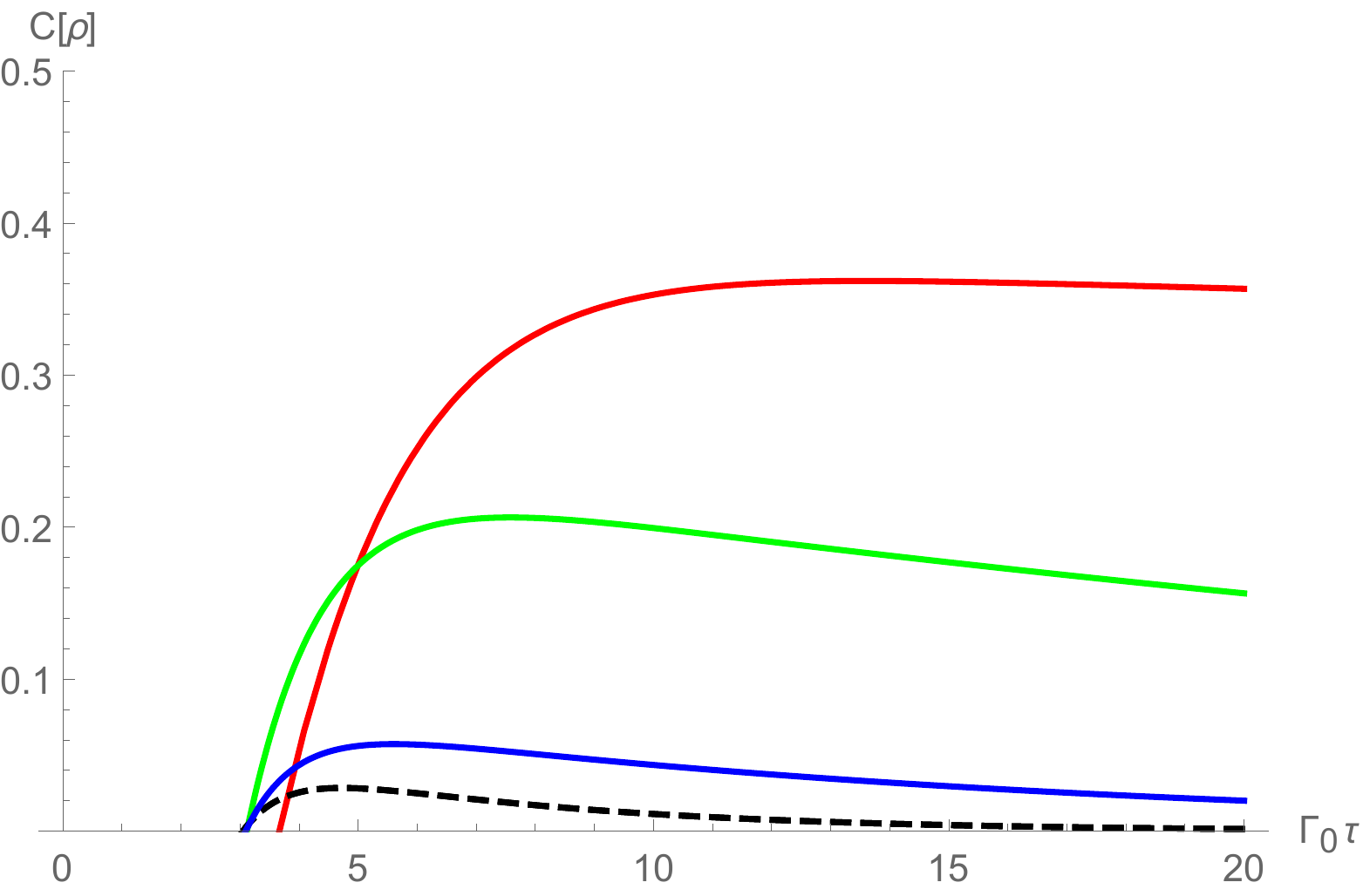}
\caption{\label{pexvex}
Comparison between the dynamics of concurrence for uniformly accelerated atoms initially prepared in $|E\rangle$ aligned parallel to (left) or vertically  to (right) a reflecting boundary. Both of the two atoms are polarizable along the direction of  acceleration  (the $x$-axis).
Here $\omega L=2/3$, $a/\omega=1/2$, the real red, green and blue lines  correspond to $y/L$=3/10, 7/10, 6/5 respectively, and the dashed lines describe the corresponding ones in the free space.}
\end{figure}

In Fig. \ref{pexvex}, we observe that, for the parallel two-atom system, the time when entanglement is generated can be apparently postponed for transversely polarizable atoms when the two atoms are aligned parallel to the boundary, while the maximal concurrence during the whole evolution is barely influenced.
In the vertical case, when the atoms are transversely polarizable, the maximal concurrence during evolution can be significantly enhanced compared with that in the free space, and the closer the two-atom system is to the boundary, the larger the maximal concurrence is. However, the birth time of entanglement is less sensitive to the distance between the two-atom system and the boundary in the vertical case.
As before, some of the above results are polarization dependent, see the following discussions for details. 

\paragraph{Acceleration effects}

\begin{figure}[!htbp]
\centering
\includegraphics[width=0.49\textwidth]{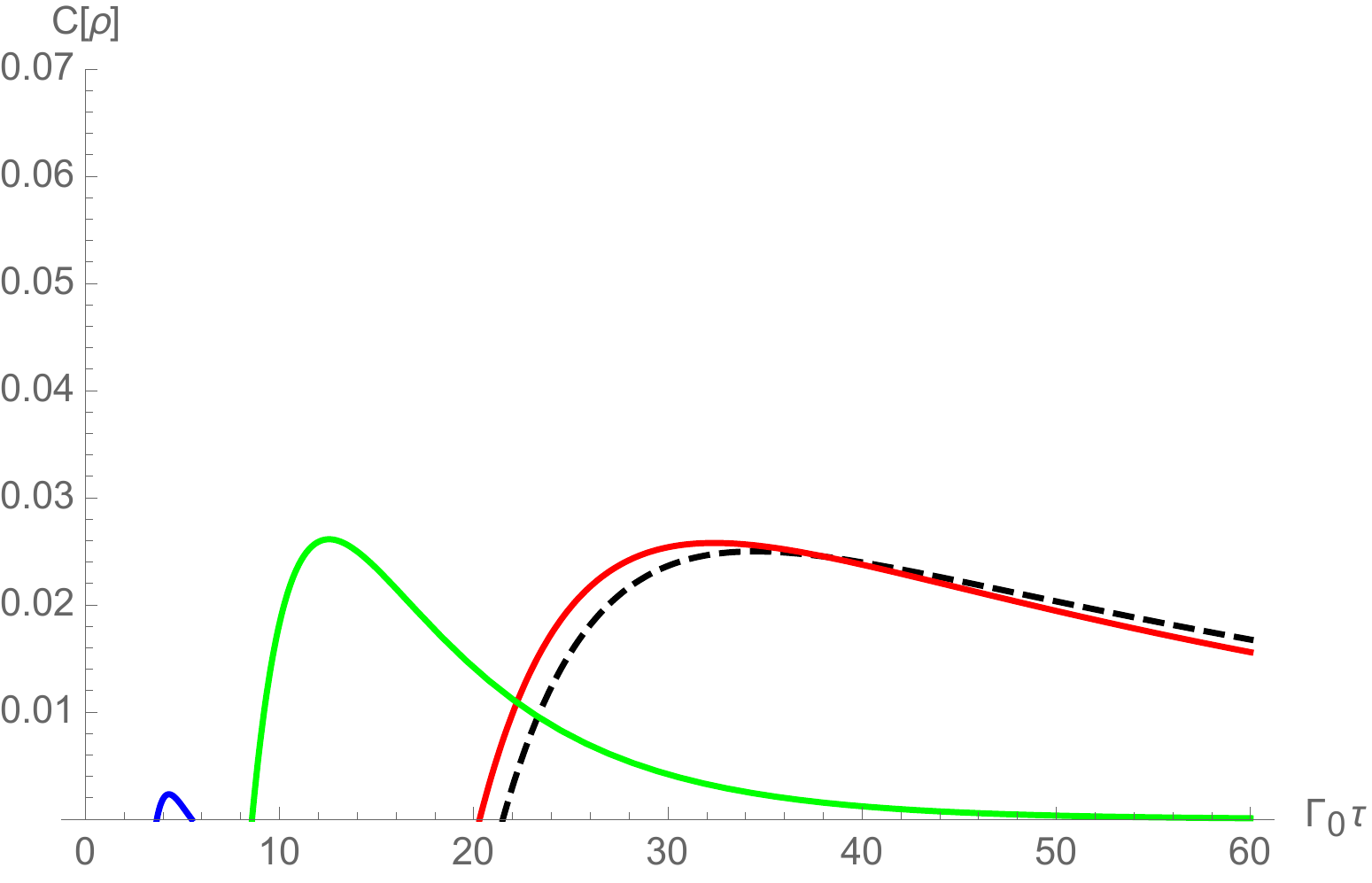}
\includegraphics[width=0.49\textwidth]{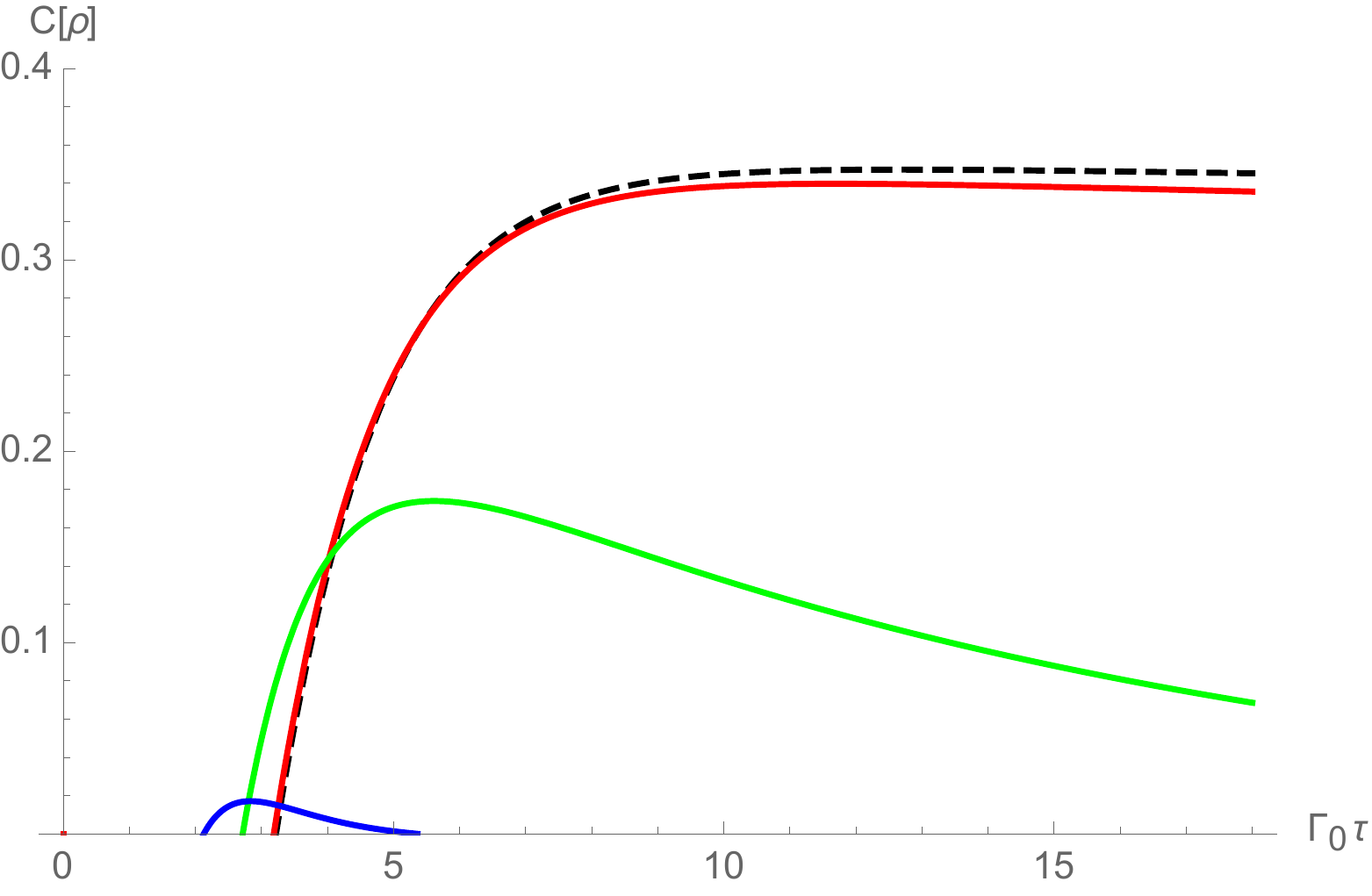}
\caption{\label{pve1}
Comparison between the dynamics of concurrence for uniformly accelerated atoms polarizable along the direction of  acceleration (the $x$-axis) with the alignment parallel to (left) or vertical to (right) a reflecting boundary.
The system is initially prepared in $|E\rangle$.
Here $\omega L=1$, $y/L=1/2$, the real red, green and blue lines correspond to $a/\omega$=1/10, 1/2, 1 respectively, and the dashed lines describe the cases for inertial atoms.}
\end{figure}

When both of the two atoms are polarizable along the direction of acceleration, the delayed sudden birth of entanglement occurs for two inertial atoms, see Fig.~\ref{pve1}. 
For parallel-aligned atoms, as the acceleration increases, the entanglement is produced earlier and preserved for a shorter time [Fig.~\ref{pve1} (left)]. 
For vertically-aligned atoms, the maximal entanglement during evolution is weakened with a larger acceleration, but the birth time of entanglement is barely influenced [Fig.~\ref{pve1} (right)].
However, this changes when the atoms are polarizable differently.
For both  two alignments, the delayed entanglement cannot be generated when the  acceleration is large enough, regardless of the polarizations.

\paragraph{Polarization effects}

\begin{figure}[!htbp]
\centering
\includegraphics[width=0.49\textwidth]{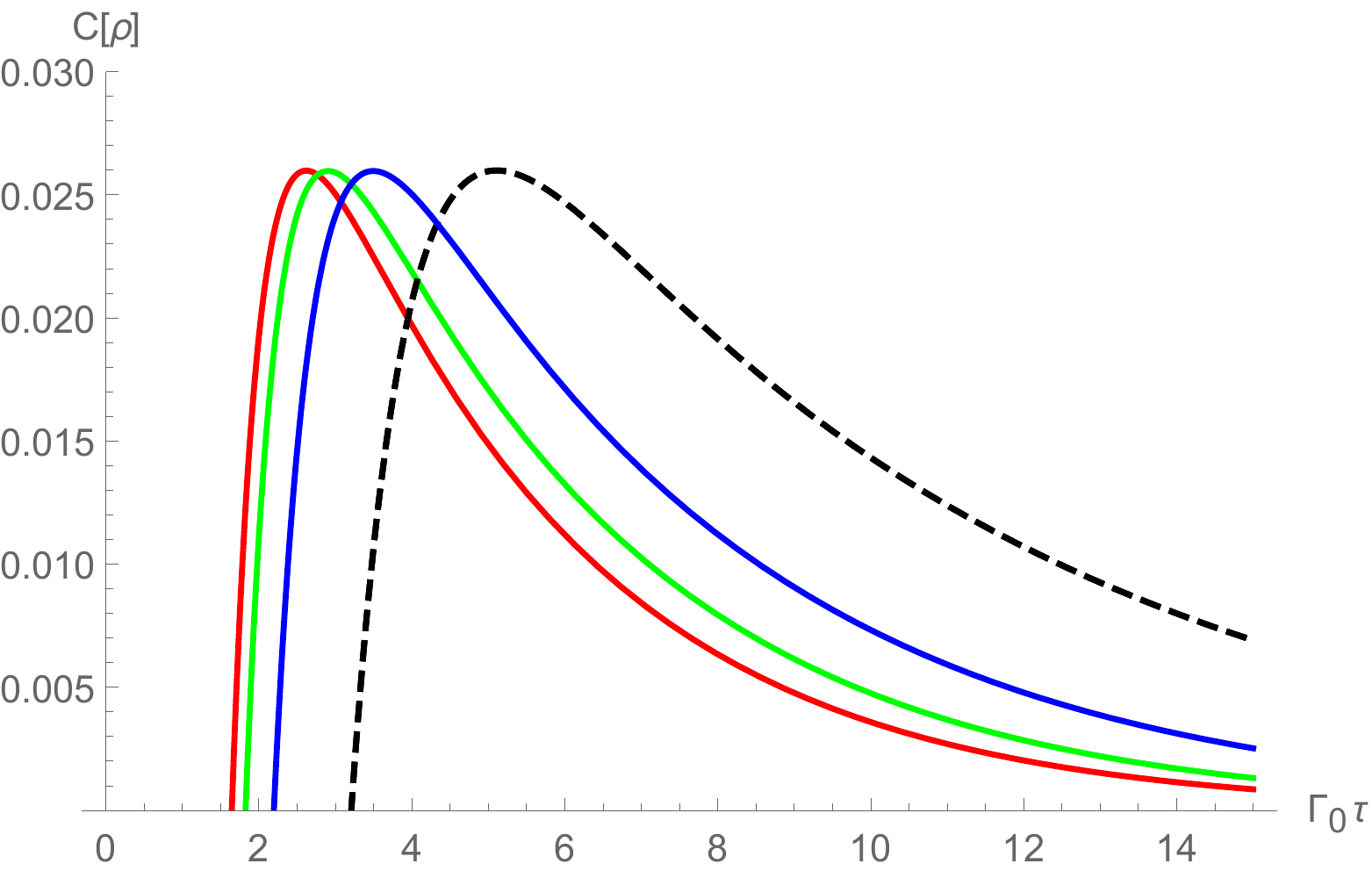}
\includegraphics[width=0.5\textwidth]{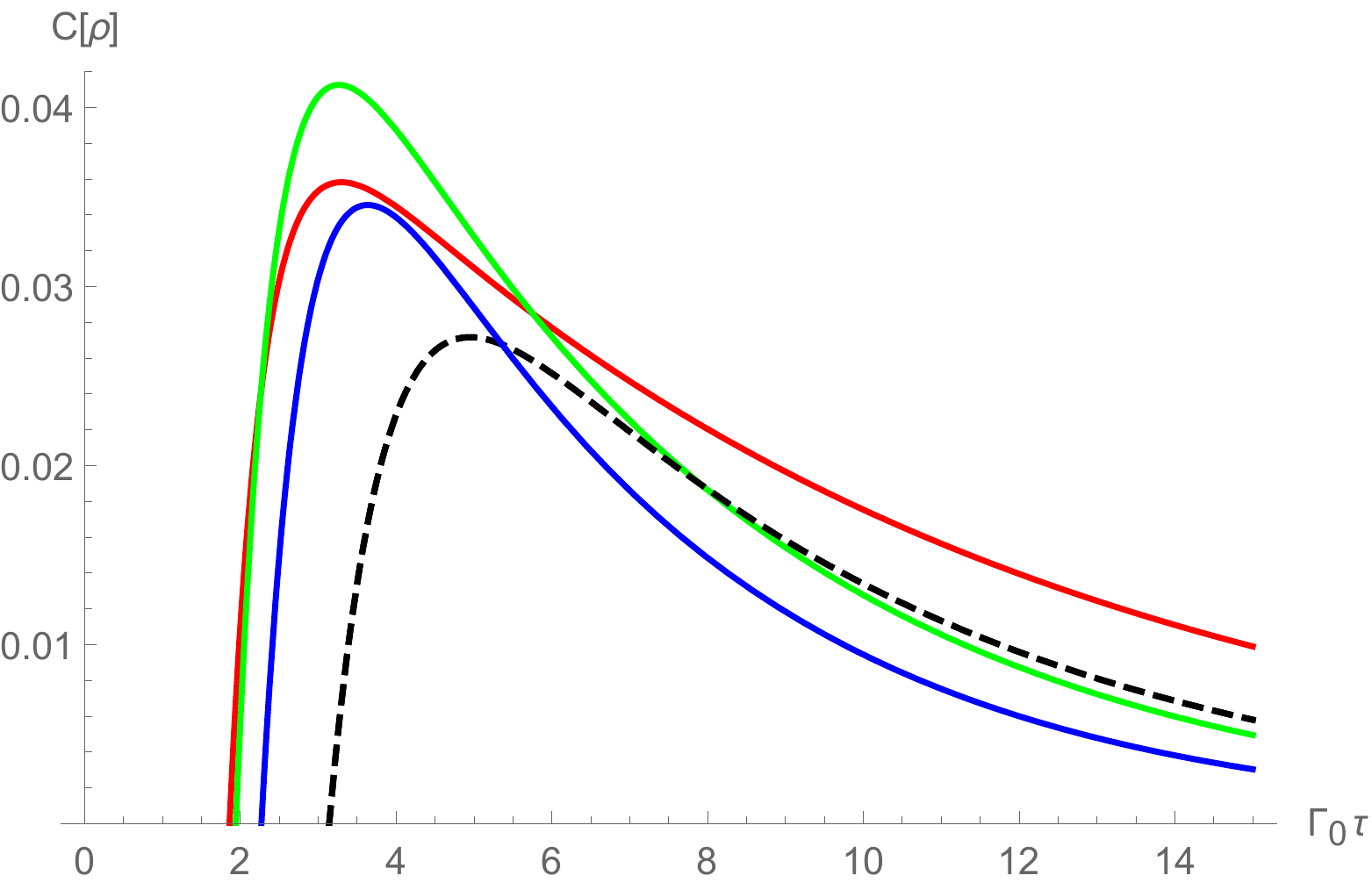}
\caption{\label{peyvey}
Comparison between the dynamics of concurrence for uniformly accelerated atoms initially prepared in $|E\rangle$ aligned parallel to (left) or vertically  to (right) a reflecting boundary. Both of the two atoms are polarizable vertically to the boundary  (the $y$-axis).
Here $\omega L=2/3$, $a/\omega=1/2$, the real red, green and blue lines  correspond to $y/L$=3/10, 7/10, 6/5 respectively, and the dashed lines describe the corresponding ones in the free space.}
\end{figure}

\begin{figure}[!htbp]
\centering
\includegraphics[width=0.5\textwidth]{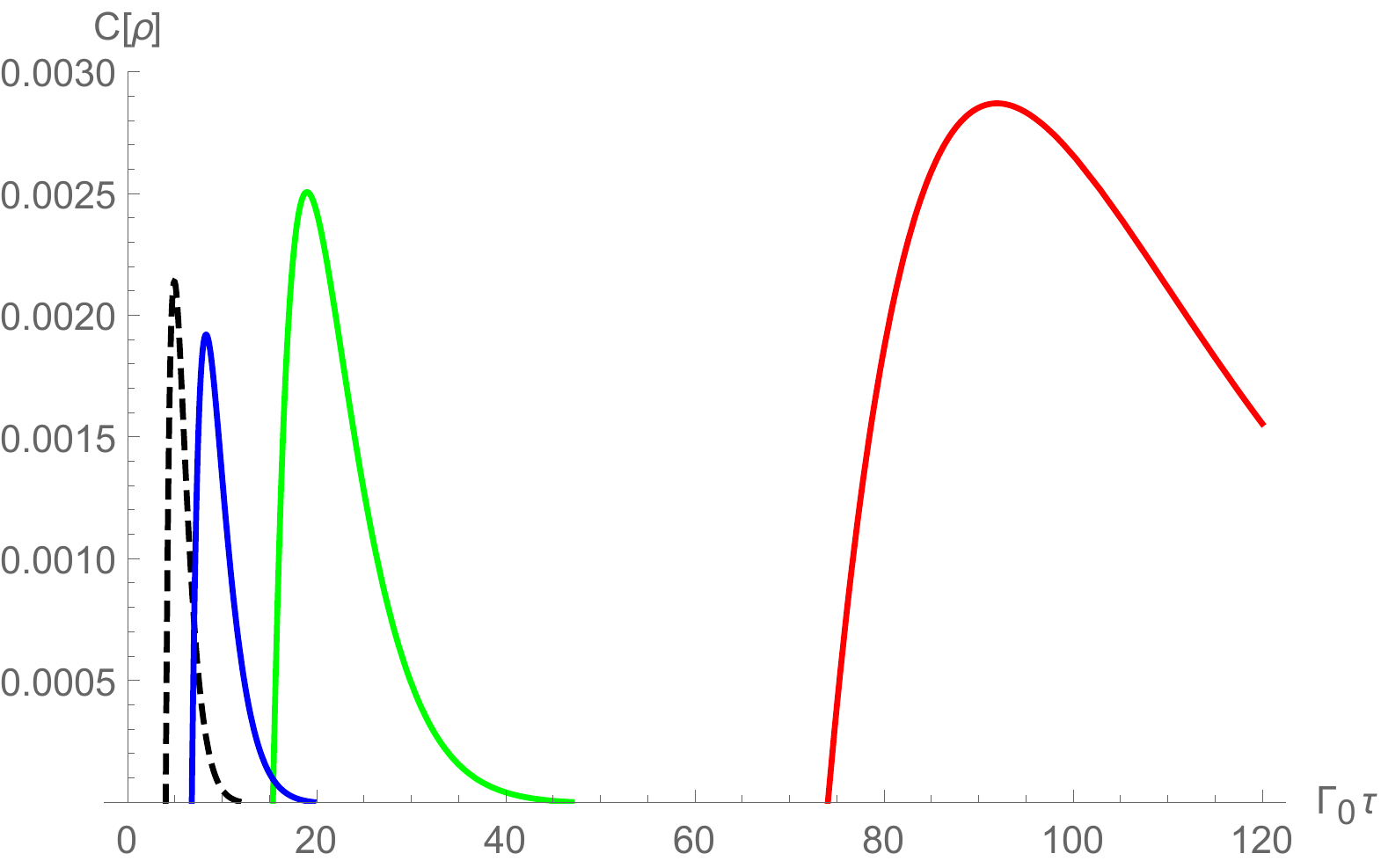}
\includegraphics[width=0.49\textwidth]{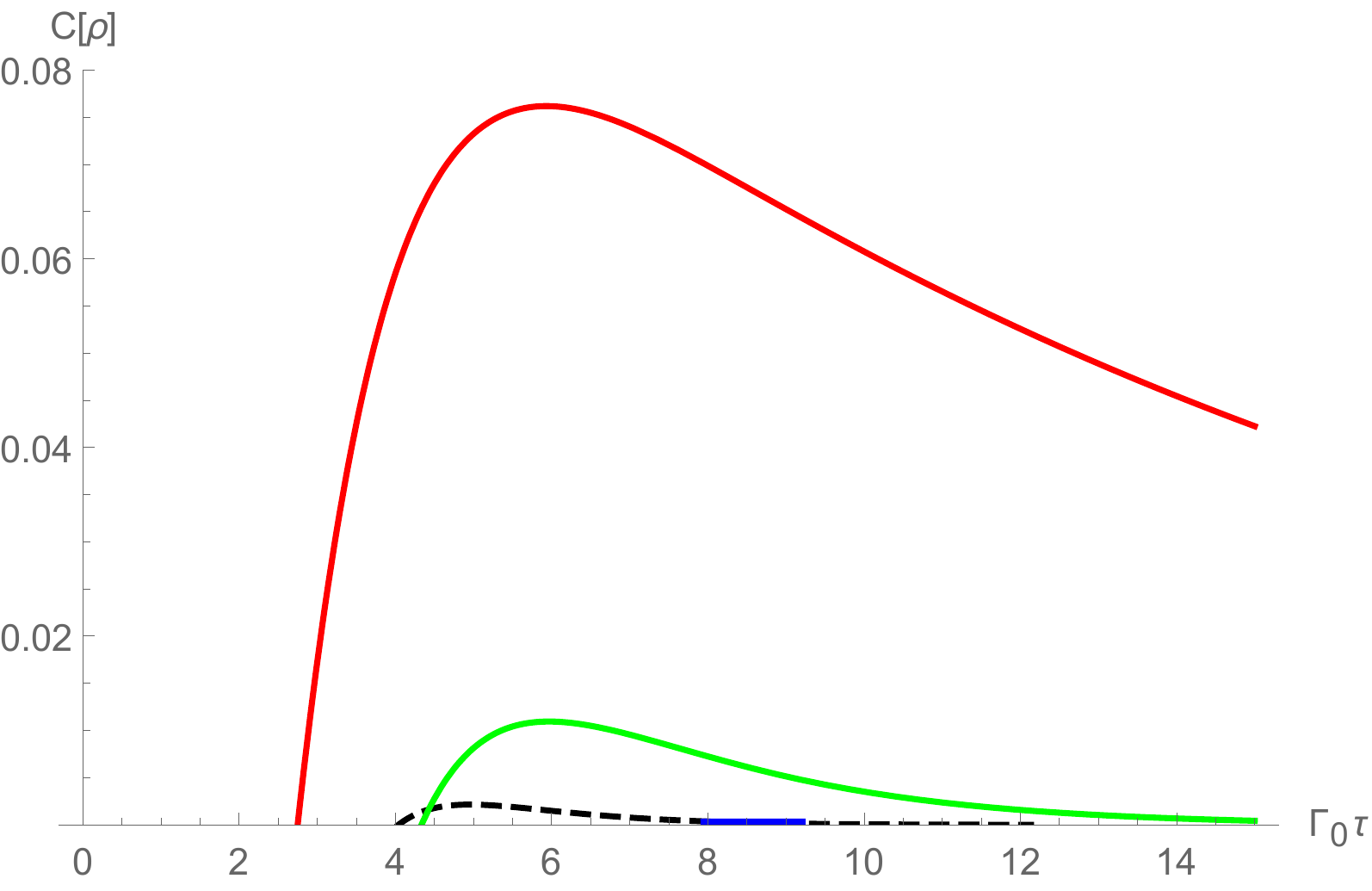}
\caption{\label{pexyvexy}
Comparison between the dynamics of concurrence for uniformly accelerated atoms initially prepared in $|E\rangle$ aligned parallel to (left) or vertically  to (right) a reflecting boundary.
One of the atoms (the nearer one in the vertically aligned case)  is polarizable along the direction of acceleration  (the $x$-axis) and the other along the direction of separation  (the $z$-axis for the parallel case and the $y$-axis for the vertical case).
Here $\omega L=2/3$, $a/\omega=1/2$, the real red, green and blue lines  correspond to $y/L$=3/10, 7/10, 6/5 respectively, and the dashed lines describe the corresponding ones in the free space.}
\end{figure}

From Fig. \ref{peyvey} (left), we observe that when the atoms are aligned parallel to the boundary, the birth time of entanglement can be advanced for vertically polarizable atoms, and the entanglement is maintained for a shorter time as the atoms gets closer to the boundary, which is in sharp contrast to the case of the atoms polarizable parallel to the boundary  [see Fig. \ref{pexvex} (left)].
Compared with the cases when the polarizations of the two atoms are the same (Fig. \ref{pexvex} and Fig. \ref{peyvey}), the maximal concurrence during  evolution of atoms aligned parallel to the boundary is more sensitive to the distance between the atoms and the boundary when the atoms are polarizable  along the direction of acceleration and separation respectively, see  Fig. \ref{pexyvexy}.

\begin{figure}[!htbp]
\centering
\includegraphics[width=0.49\textwidth]{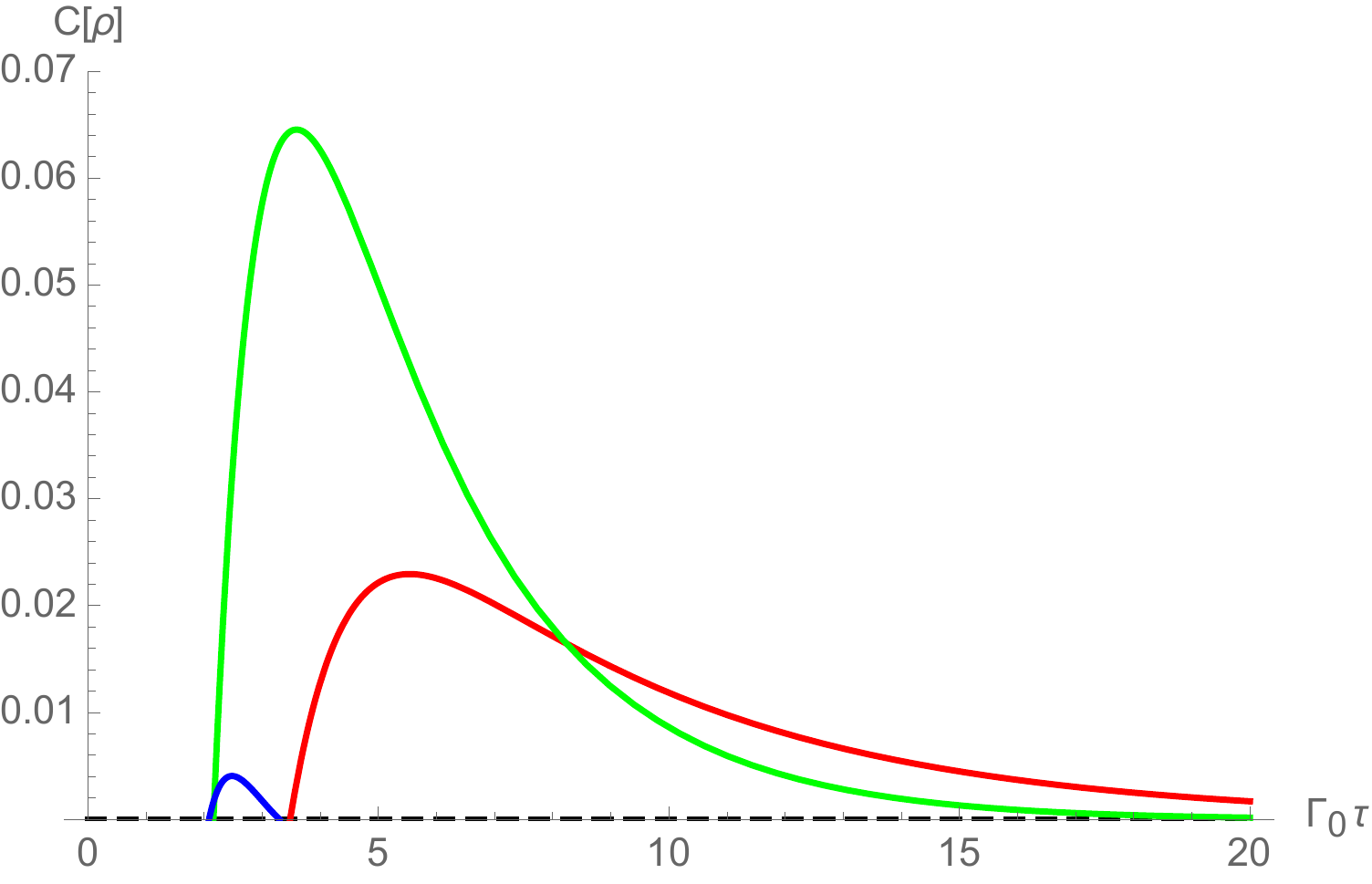}
\includegraphics[width=0.49\textwidth]{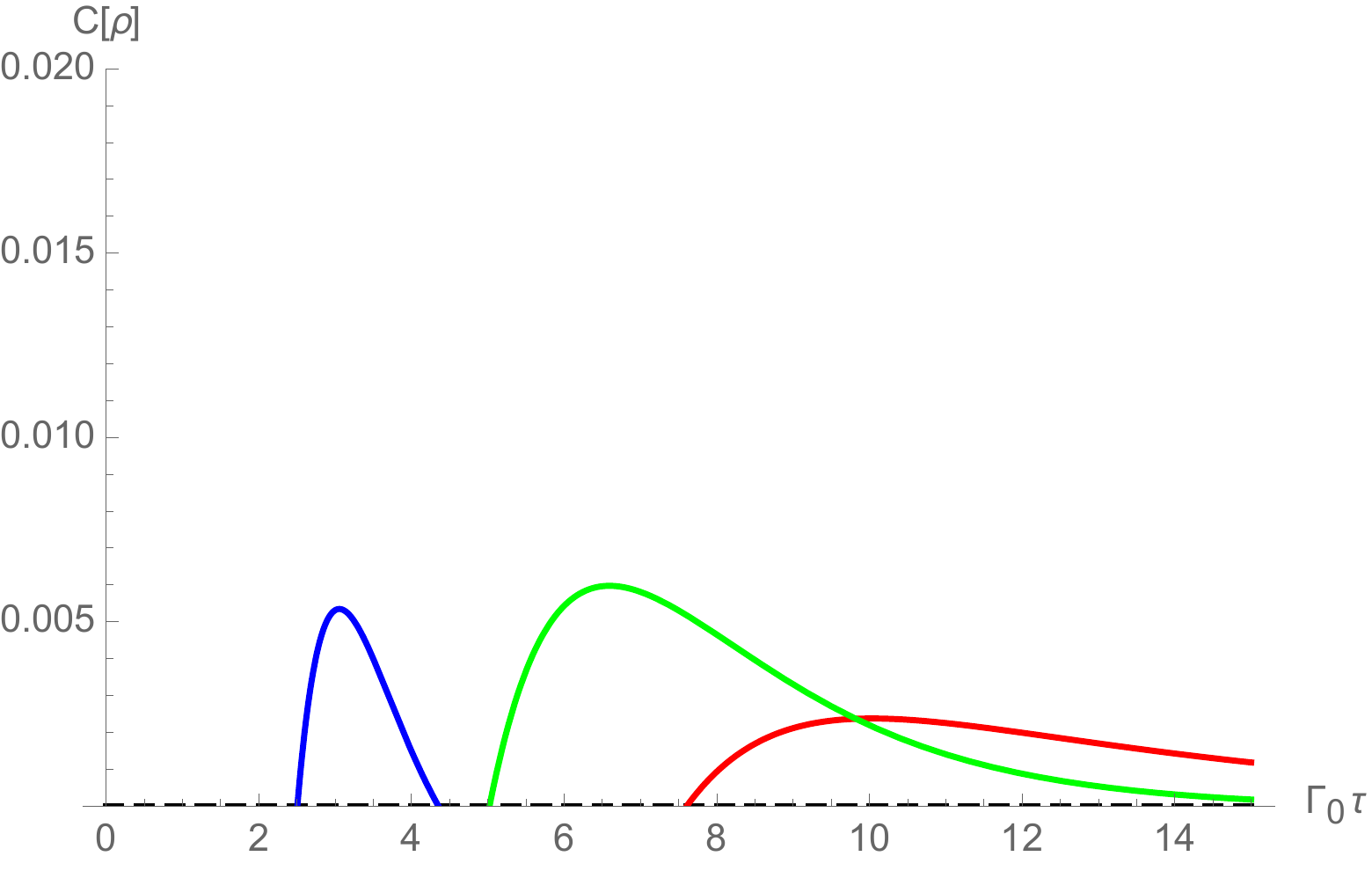}
\caption{\label{ve1}
Comparison between the dynamics of concurrence for uniformly accelerated atoms initially prepared in $|E\rangle$ aligned parallel to (left) or vertically  to (right) a reflecting boundary.
One of the atoms (the nearer one in the vertically aligned case)  is polarizable along the direction of acceleration  (the $x$-axis) and the other vertically to the boundary (the $y$-axis).
Here $\omega L=1$, $y/L=1/2$, the real red, green and blue lines correspond to $a/\omega$=1/10, 1/2, 1 respectively, and the dashed lines describe the cases for inertial atoms.}
\end{figure}

In Fig.~\ref{ve1}, we assume that one of the atoms (which is the nearer one in the vertically aligned case) is polarizable along the direction of acceleration and the other vertically to the boundary, and find that when $a\rightarrow 0$, the concurrence is zero all the time. That is, two inertial atoms remain separable and no entanglement is generated.
As the acceleration increases, the delayed birth of entanglement happens, and the nonzero concurrence can be enhanced.
The comparison between the cases in Fig. \ref{pve1} and Fig. \ref{ve1} reveals that  acceleration does not always play the role of destroying entanglement, it may  generate and enhance entanglement as well. 

\subsection{Two-atom system placed close to the boundary}

Finally, we turn to the case in which the two-atom system is placed close to the boundary.

\subsubsection{Parallel alignment with respect to the boundary}

When the two-atom system is placed  extremely close to the boundary, i.e., when $y/L \rightarrow 0$, for the case when the atoms are aligned parallel to the boundary, the leading terms of the coefficients Eqs. (\ref{pab}) when expanding in power series of $y/L$ are
\begin{eqnarray}\label{eq20}
&A_{1(p)}&\approx\frac{\Gamma_{0}\coth{\frac{\pi \omega}{a}}}{2\omega^{2}}\hat{d}_{2}^{(1)}\hat{d}_{2}^{(1)}(a^{2}+\omega^{2}),\ \ \ \ \ \ \ \ \ \
A_{2(p)}\approx\frac{\Gamma_{0}\coth{\frac{\pi \omega}{a}}}{2\omega^{2}}\hat{d}_{2}^{(2)}\hat{d}_{2}^{(2)}(a^{2}+\omega^{2}), \nonumber
\end{eqnarray}
\begin{eqnarray}
&A_{3(p)}&\approx\frac{3\Gamma_{0}\coth{\frac{\pi \omega}{a}}}{2\omega^{3}L^{3}(4+a^{2}L^{2})^{3/2}}\hat{d}_{2}^{(1)}\hat{d}_{2}^{(2)}\biggl\{\omega L \sqrt{4+a^{2}L^{2}} (2+a^{2}L^{2})\cos{\left(\frac{2\omega}{a}\sinh^{-1}\frac{aL}{2}\right)} \nonumber \\
&&\ \ +\left[-4+\omega^{2}L^{2}(4+a^{2}L^{2})\right]\sin{\left(\frac{2\omega}{a}\sinh^{-1}\frac{aL}{2}\right)}\biggl\}, \nonumber \\
&B_{i(p)}&=A_{i(p)}\tanh{\frac{\pi \omega}{a}}\ \ (i=1,2,3),
\end{eqnarray}
where the subscript $p$ denotes parallel.

\paragraph{Acceleration effects}

\begin{figure}[!htbp]
\centering
\includegraphics[width=0.49\textwidth]{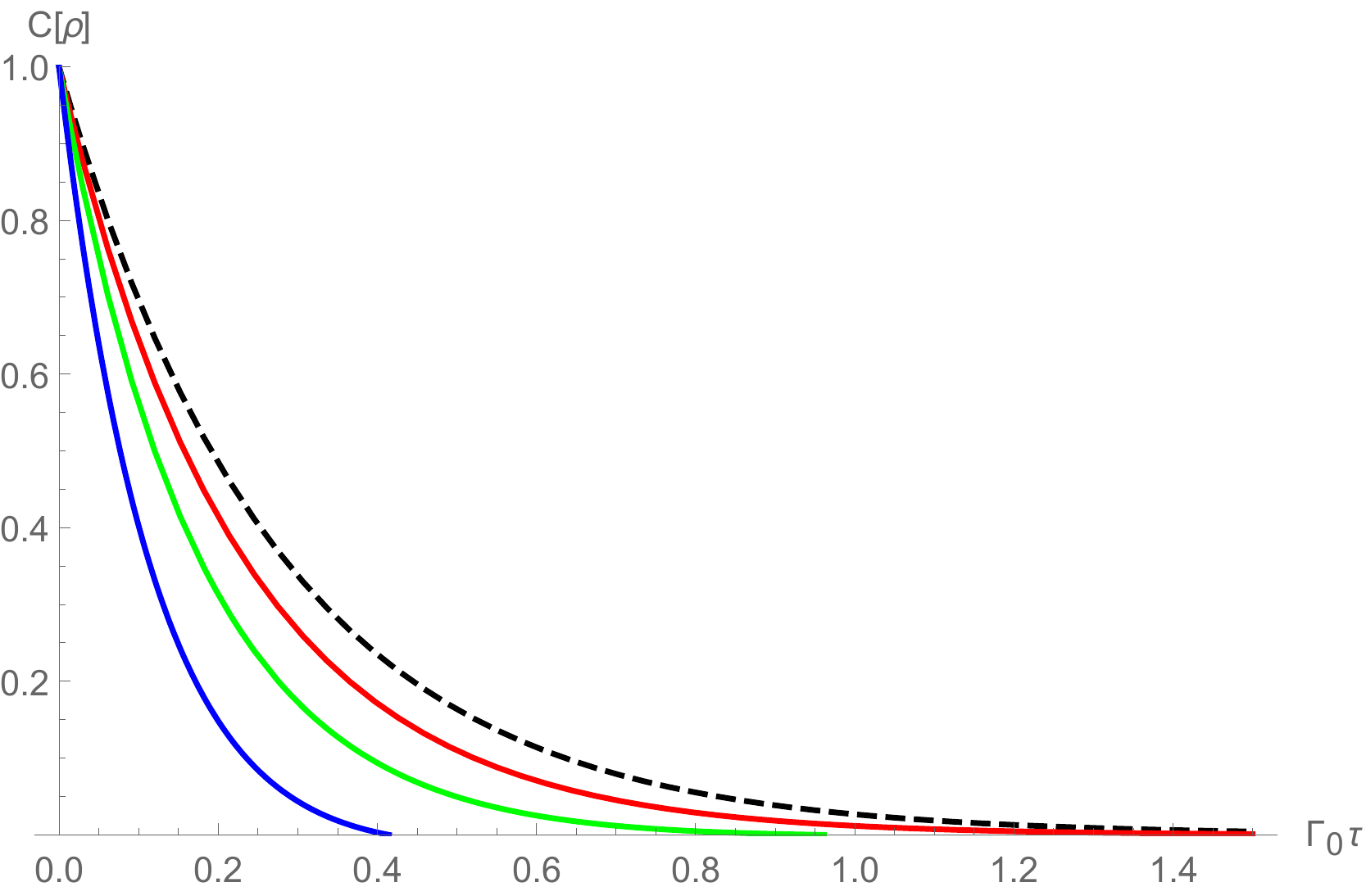}
\includegraphics[width=0.5\textwidth]{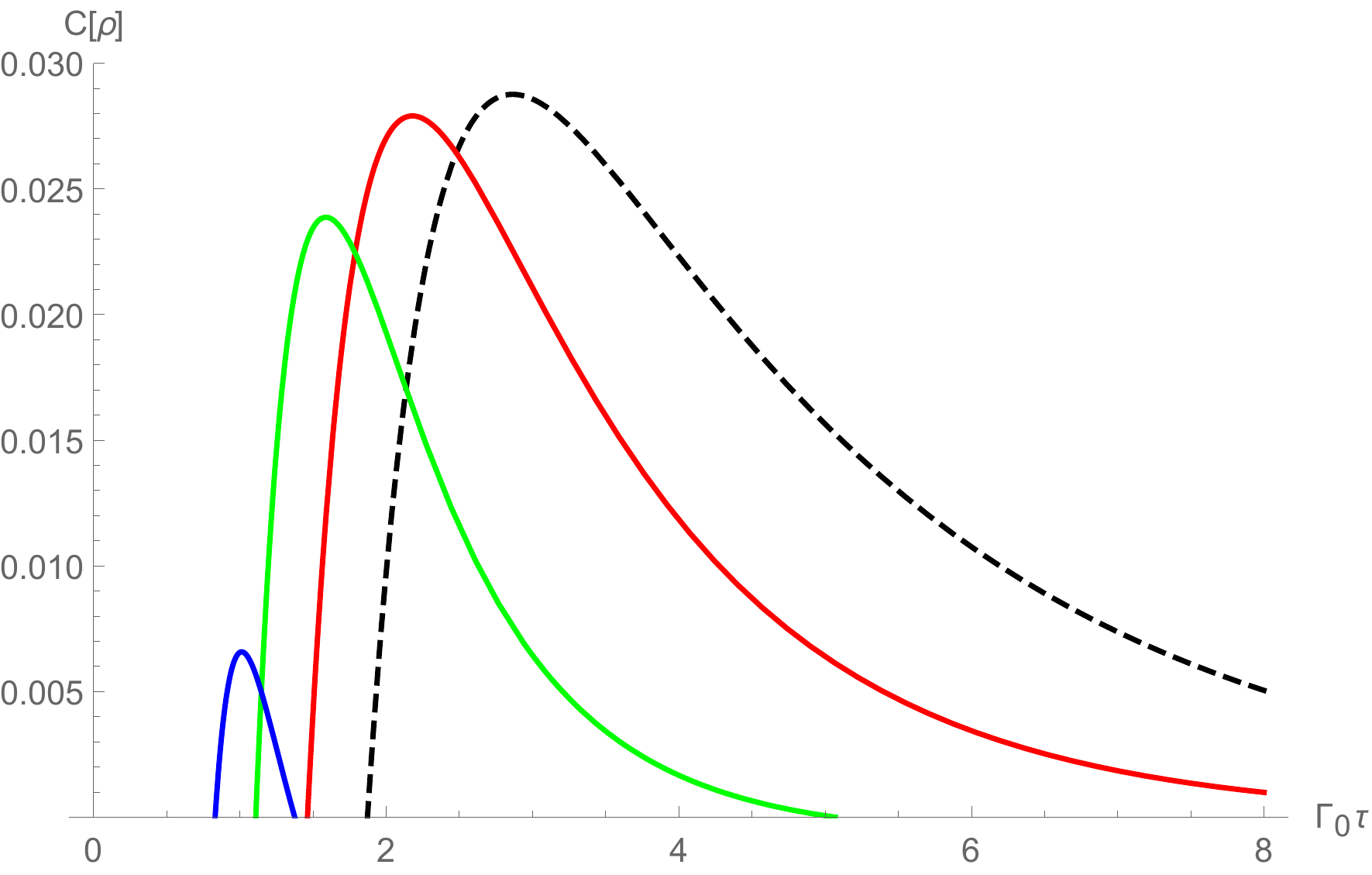}
\caption{\label{CSE1}
The figures show the dynamics of concurrence for uniformly accelerated atoms to which the parallel-aligned boundary is very close.
The atoms are initially prepared in $|S\rangle$ (left) or $|E\rangle$ (right).
Both of the two atoms are polarizable along the $y$-axis. Here $\omega L=1$, the real red, green and blue lines  correspond to $a/\omega=1/2,4/5,6/5$ respectively, and the dashed lines describe the case for two inertial atoms.}
\end{figure}

When the boundary is extremely close to the system, we  
 assume that the  polarizations of the two atoms are vertical to the boundary, i.e., $\hat{d}^{(1)}=\hat{d}^{(2)}=(0,1,0)$. As shown in Fig. \ref{CSE1}, for parallel-aligned atoms placed extremely close to the boundary ($y/L\to 0$), as the acceleration gets larger, the decay rate of entanglement increases for initially entangled atoms, while for initially  separable atoms,  entanglement is generated earlier and  maintained for a shorter time, meanwhile the maximal entanglement during evolution is weakened. 
This is consistent with the proceeding  discussions when $y\sim L$.

\paragraph{Polarization effects}

When both of the two atoms are polarizable vertically with respect to the boundary, e.g., $\hat{d}^{(1)}=\hat{d}^{(2)}=(0,1,0)$, the coefficients $A_{i(p)}$, $B_{i(p)}$ in Eqs. (\ref{eq20}) are exactly two times  those in the free space case \cite{Yang}. That is, for the two-atom system prepared in an entangled state initially, the concurrence for the parallel two-atom system decays two times as fast as that in the free space.
Similarly, for two atoms initially prepared in a separable state, the birth time of  entanglement is  earlier but maintains for a shorter time compared with that in the free space case.
When both of the atoms are polarizable parallel to the boundary, e.g., $\hat{d}^{(1)}$=(1,0,0), $\hat{d}^{(2)}$=(0,0,1), all the coefficients $A_{i(p)}$, $B_{i(p)}$  vanish, so the initially entangled state will be preserved as if it were a closed system.

\subsubsection{Vertical alignment with respect to the boundary}

As for the case of  vertically-aligned atoms, by expanding Eqs. (\ref{vab}) in power series of $y/L$, we obtain
\begin{eqnarray}\label{avi}
&A_{1(v)}&\approx\frac{\Gamma_{0}\coth{\frac{\pi \omega}{a}}}{2\omega^{2}}\hat{d}_{2}^{(1)}\hat{d}_{2}^{(1)}(a^{2}+\omega^{2}), \nonumber\\
&A_{2(v)}&\approx\frac{-3\Gamma_{0}\coth{\frac{\pi \omega}{a}}}{64\omega^{3}L^{3}(1+a^{2}L^{2})^{5/2}}\biggl\{2\omega L \sqrt{1+a^{2}L^{2}}\biggl[\hat{d}_{1}^{(2)}\hat{d}_{1}^{(2)}(1+4a^{2}L^{2})+\hat{d}_{3}^{(2)}\hat{d}_{3}^{(2)}(1+2a^{2}L^{2})\times \nonumber \\
&&(1+a^{2}L^{2})+\hat{d}_{2}^{(2)}\hat{d}_{2}^{(2)}(2+a^{2}L^{2}+2a^{4}L^{4})-2\hat{d}_{1}^{(2)}\hat{d}_{2}^{(2)}aL(2a^{2}L^{2}-1)\biggl]\cos{\left(\frac{2\omega}{a}\sinh^{-1}aL\right)} \nonumber \\
&&-\biggl[\hat{d}_{1}^{(2)}\hat{d}_{1}^{(2)}(1+2a^{2}L^{2}+4a^{4}L^{4}-4\omega^{2}L^{2}-4\omega^{2}a^{2}L^{4})+\hat{d}_{3}^{(2)}\hat{d}_{3}^{(2)}(1-4\omega^{2}L^{2}-4\omega^{2}a^{2}L^{4})  \nonumber \\
&&\times(1+a^{2}L^{2})+\hat{d}_{2}^{(2)}\hat{d}_{2}^{(2)}(2+5a^{2}L^{2}-4\omega^{2}a^{2}L^{4}-4\omega^{2}a^{4}L^{6})+2aL  \hat{d}_{1}^{(2)}\hat{d}_{2}^{(2)}(1+4a^{2}L^{2} \nonumber \\
&&+4\omega^{2}L^{2}+4\omega^{2}a^{2}L^{4})\biggl]\sin{\left(\frac{2\omega}{a}\sinh^{-1}aL\right)}\biggl\}+\frac{\Gamma_{0}\coth{\frac{\pi \omega}{a}}}{4\omega^{2}}(a^{2}+\omega^{2}), \nonumber
\end{eqnarray}
\begin{eqnarray}\label{avi}
&A_{3(v)}&\approx\frac{-3\Gamma_{0}\coth{\frac{\pi \omega}{a}}}{2\omega^{3}L^{3}(4+a^{2}L^{2})^{5/2}}\times \nonumber\\
&&\biggl\{\omega L \sqrt{4+a^{2}L^{2}}[\hat{d}_{2}^{(1)}\hat{d}_{2}^{(2)}(16+2a^{2}L^{2}+a^{4}L^{4})\nonumber
-2aL \hat{d}_{2}^{(1)}\hat{d}_{1}^{(2)}(a^{2}L^{2}-2)]\cos{\left(\frac{2\omega}{a}\sinh^{-1}\frac{aL}{2}\right)}
\nonumber \\
&&-[2aL\hat{d}_{2}^{(1)}\hat{d}_{1}^{(2)}(4+4\omega^{2}L^{2}+4a^{2}L^{2}+\omega^{2}a^{2}L^{4}) +\hat{d}_{2}^{(1)}\hat{d}_{2}^{(2)}(32+20a^{2}L^{2}-4\omega^{2}a^{2}L^{4}-\omega^{2}a^{4}L^{6})] \nonumber \\
&&\times \sin{\left(\frac{2\omega}{a}\sinh^{-1}\frac{aL}{2}\right)}\biggl\},  \nonumber \\
&B_{i(v)}&=A_{i(v)}\tanh{\frac{\pi \omega}{a}}\ \ (i=1,2,3),
\end{eqnarray}
where the subscript $v$ denotes vertical.

\paragraph{Acceleration effects}

\begin{figure}[!htbp]
\centering
\includegraphics[width=0.49\textwidth]{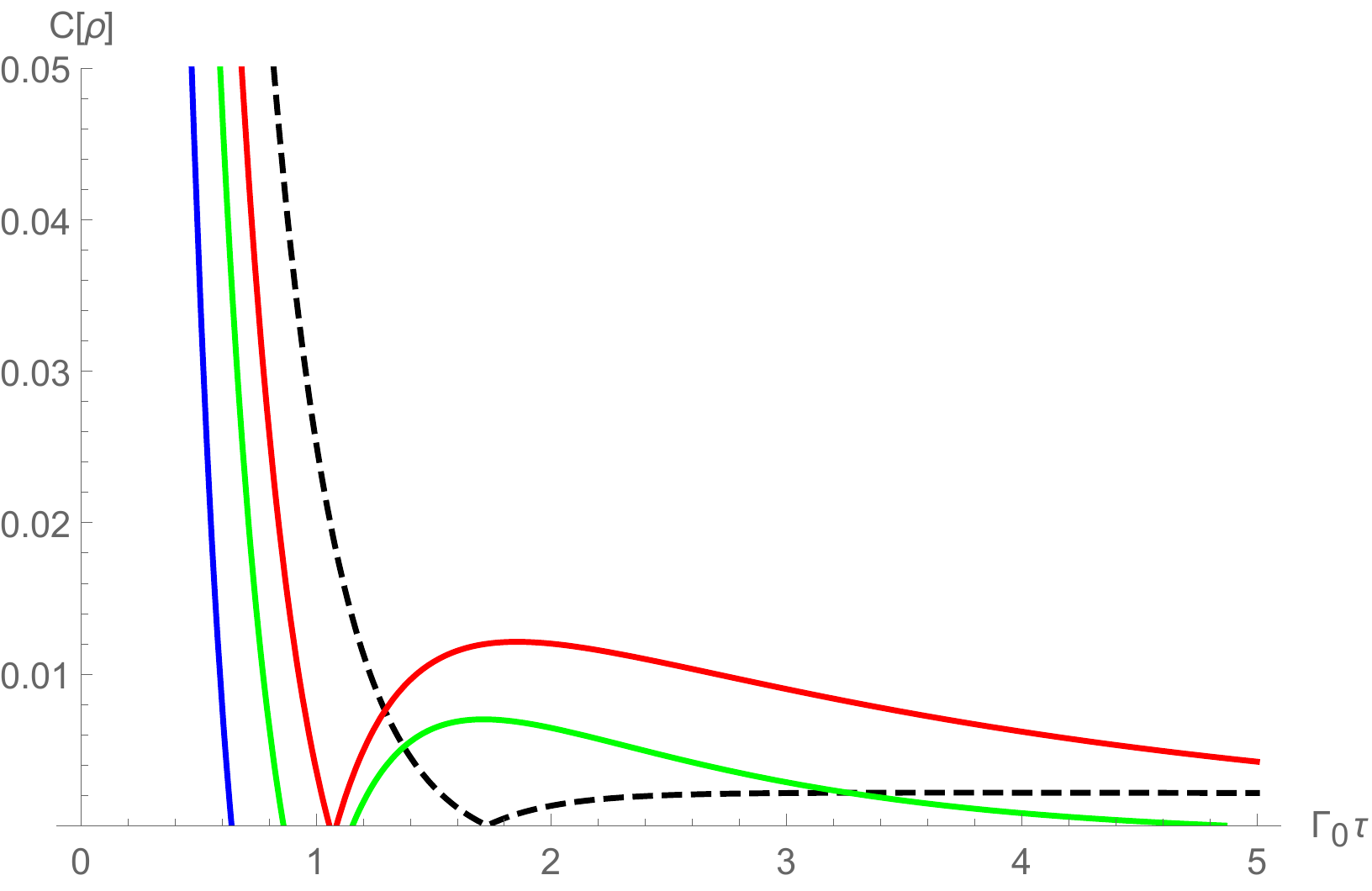}
\includegraphics[width=0.5\textwidth]{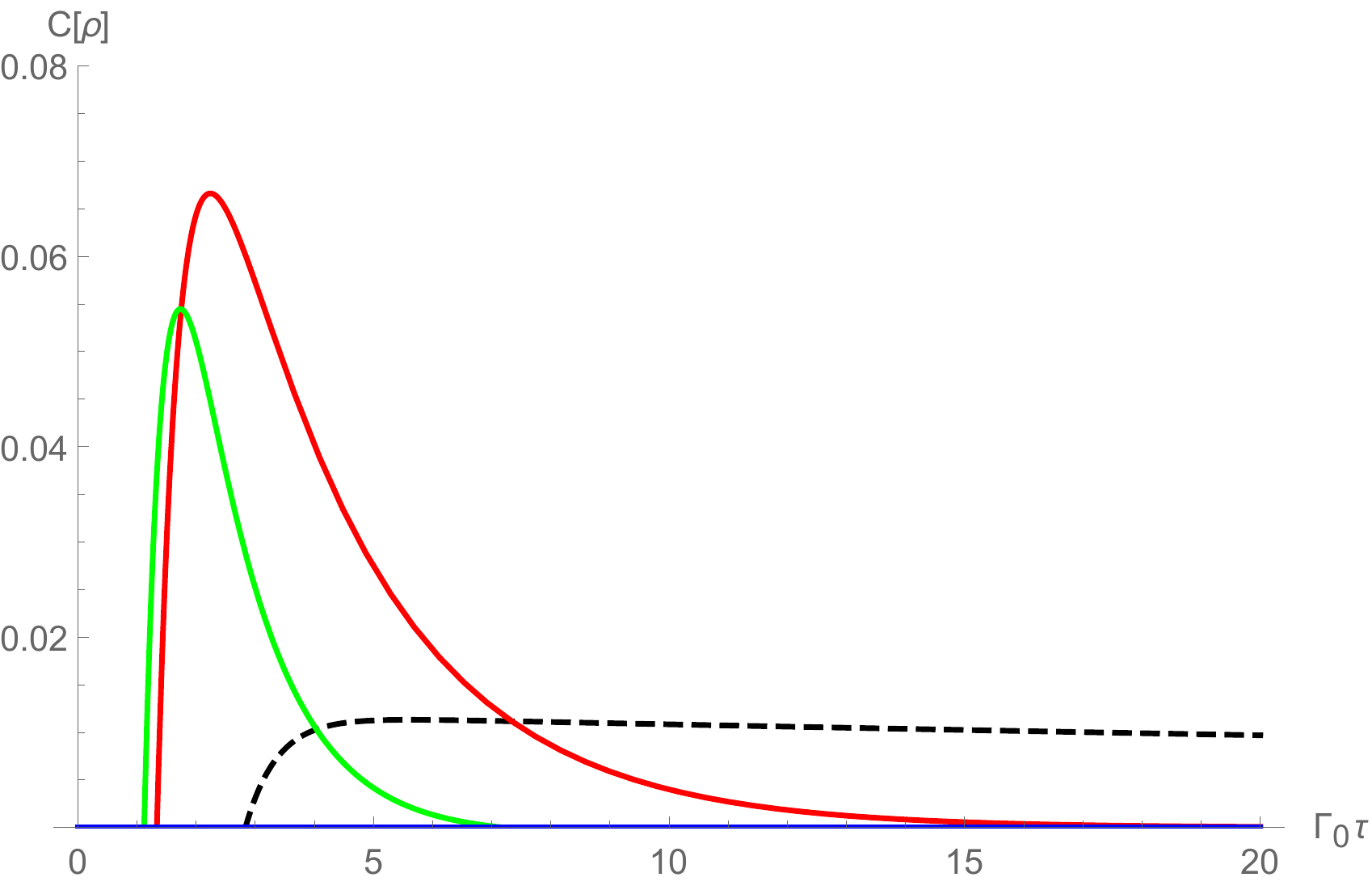}
\caption{\label{CSE2}
The figures show the dynamics of concurrence for uniformly accelerated atoms to which the vertical-aligned boundary is very close.
The atoms are initially prepared in $|S\rangle$ (left) or $|E\rangle$ (right).
Both of the two atoms are polarizable along the $y$-axis. Here $\omega L=1$, the real red, green and blue lines  correspond to $a/\omega=1/2,4/5,6/5$ respectively, and the dashed lines describe the case for two inertial atoms.}
\end{figure}

Figure \ref{CSE2} shows that when the vertically-aligned two-atom system is extremely close to the boundary, entanglement revival occurs for two inertial atoms initially in $|S\rangle$. When the acceleration is small, the revived entanglement can be larger than that in the inertial case, while as the acceleration gets large enough, entanglement revival never happens.
For two inertial atoms initially in $|E\rangle$, the lifetime of entanglement is extremely long. 
From Fig. \ref{CSE2} we also conclude that in certain cases acceleration can enhance the  entanglement revival and the  maximal entanglement during evolution.

\paragraph{Polarization effects}

The above calculations [see Eqs. (\ref{avi})] indicate that when a vertically-aligned two-atom system is placed very close to the boundary, only the nearer atom with its polarization parallel to the boundary is protected from the influence of vacuum fluctuations. Similar conclusions have been drawn in the context of quantum Fisher information \cite {Jin}.

\section{The maximal concurrence during evolution}

In the following, we study the effects of the boundary, acceleration and polarization on the maximal entanglement during  evolution when the two-atom system is initially prepared in $|E\rangle$.




\subsection{Boundary effects}

When discussing the boundary effects, to be specific, we assume that the atoms are placed close to the boundary ($y/L=1/100$), and compare the results with those in the free space.

\subsubsection{Atoms with identical orientation of polarization}

\begin{figure}[!htbp]
\centering
\includegraphics[width=0.49\textwidth]{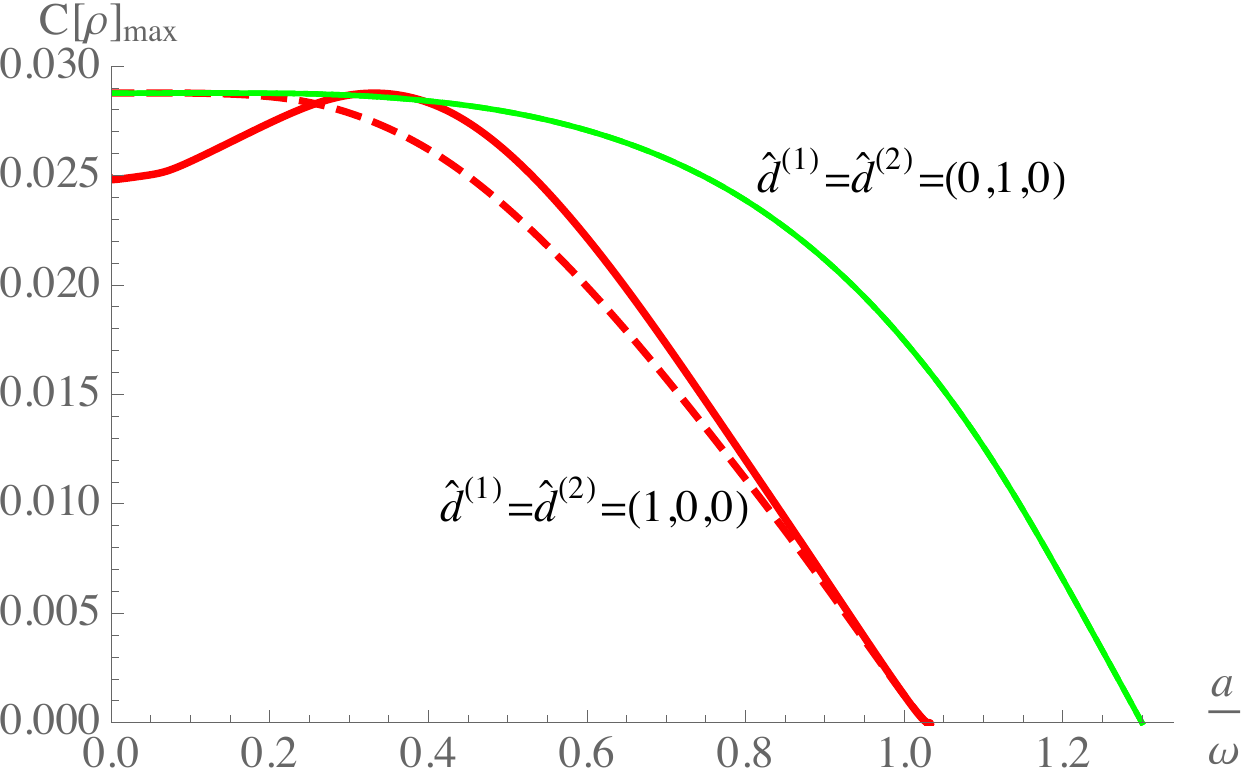}
\includegraphics[width=0.5\textwidth]{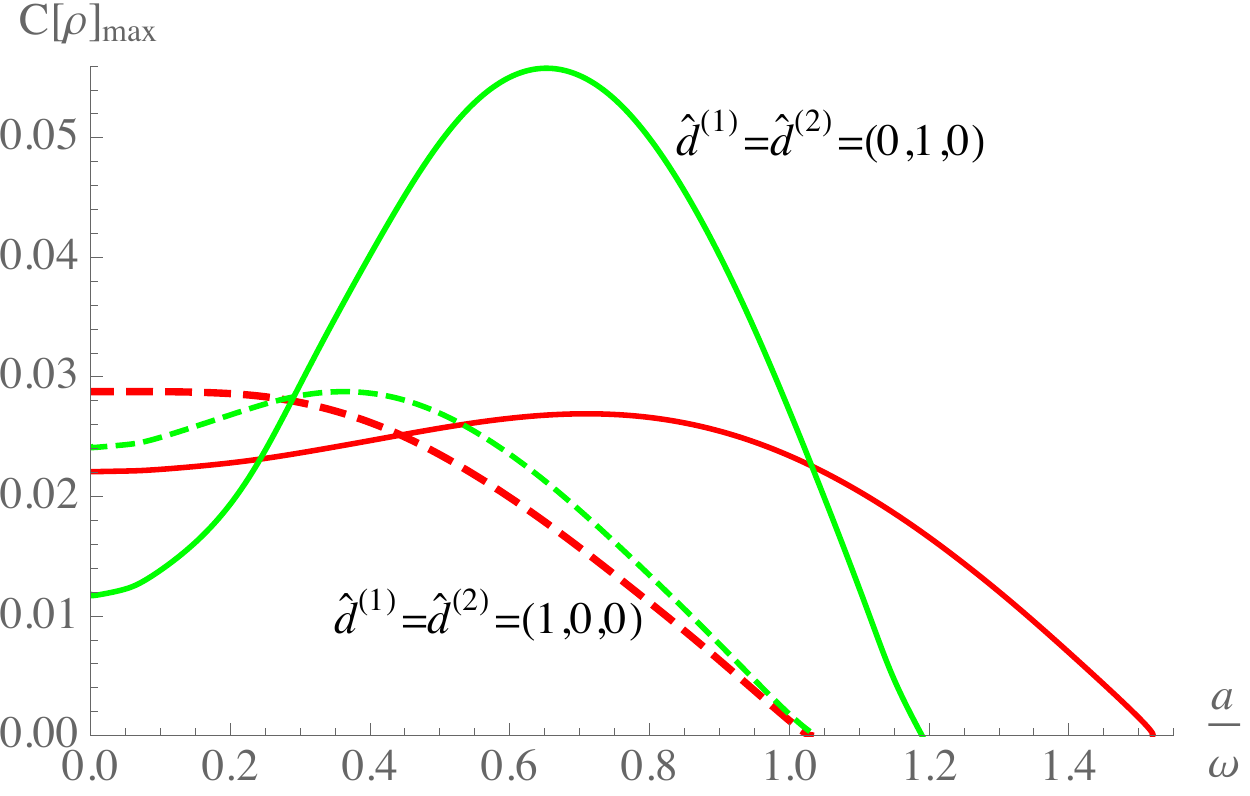}
\caption{\label{a/w1}
Comparison between the maximal concurrence during evolution in the free space (dashed lines) and that in the presence of a boundary (solid lines). The two-atom system initially prepared in $|E\rangle$ is aligned  parallel to (left) or vertically to (right) the boundary.
Both of the two atoms are polarizable  along the direction of acceleration  (the $x$-axis)  (red lines) or vertically to the boundary  (the $y$-axis)  (green lines).
The green dashed and solid lines in the left figure coincide with each other.
Here $\omega L=1$,  and $y/L=1/100$.}
\end{figure}

\begin{figure}[!htbp]
\centering
\includegraphics[width=0.48\textwidth]{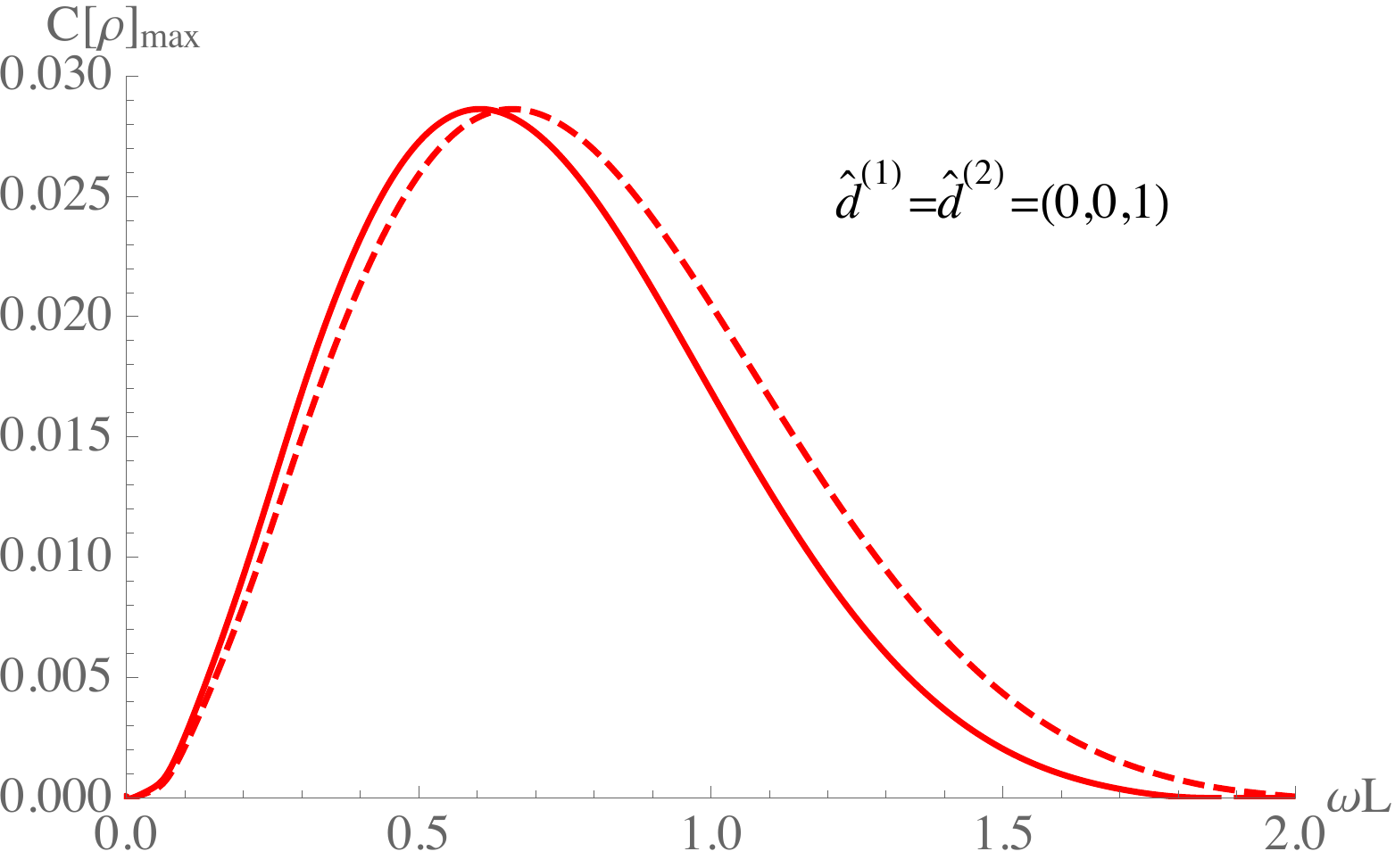}
\includegraphics[width=0.48\textwidth]{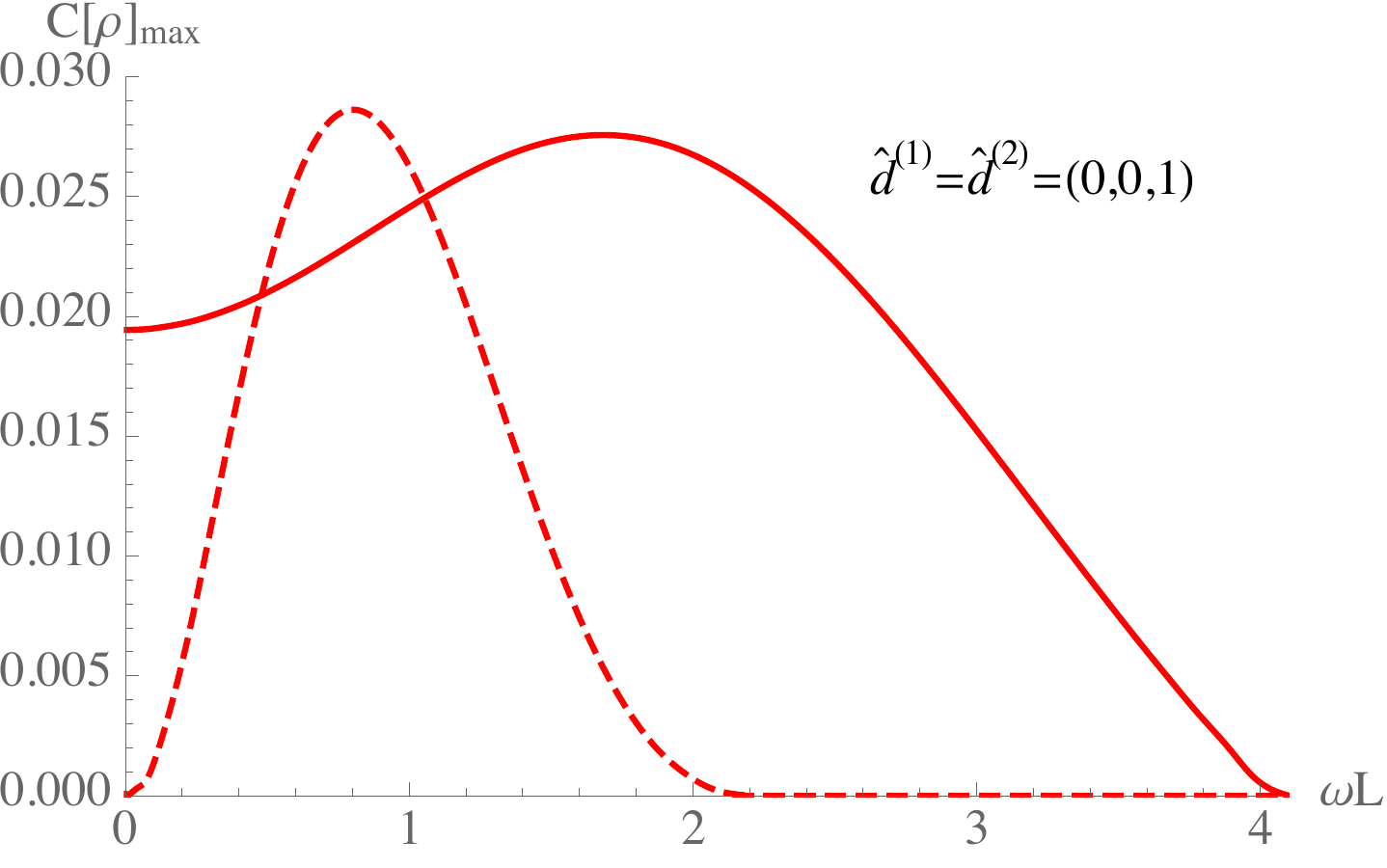}
\caption{\label{wl1}
Comparison between the maximal concurrence during evolution for uniformly accelerated atoms initially prepared in $|E\rangle$ aligned parallel to (left) or vertically  to (right) a reflecting boundary.
Here $a/\omega=2/3$, $y/L=1/100$, the solid and dashed lines  describe the cases with and without the presence of the boundary respectively.
}
\end{figure}

First we assume that the polarizations of the two atoms are the same.
From Figs. \ref{a/w1}-\ref{wl1}, we see that the  maximal entanglement during evolution can be either enhanced or weakened by the presence of boundary, depending on the polarizations of the atoms, the acceleration, the atomic separation.
Especially, we observe that the green dashed line coincides with the solid line in Fig. \ref{a/w1}  (left), i.e., when the parallel two-atom system is placed  extremely close to the boundary, the maximal entanglement during evolution is not altered in the presence of the boundary if both of the atoms are polarizable vertically to the boundary  (the $y$-axis).
Compared with the results in the free space case, the range of  atomic separation within which entanglement generation happens can be effectively broadened in the presence of a boundary if the atoms are vertically aligned, while entanglement generation does not happen for the parallel case, as shown in Fig. \ref{wl1}.

\subsubsection{Atoms with different orientations of polarization}

\begin{figure}[!htbp]
\centering
\includegraphics[width=0.49\textwidth]{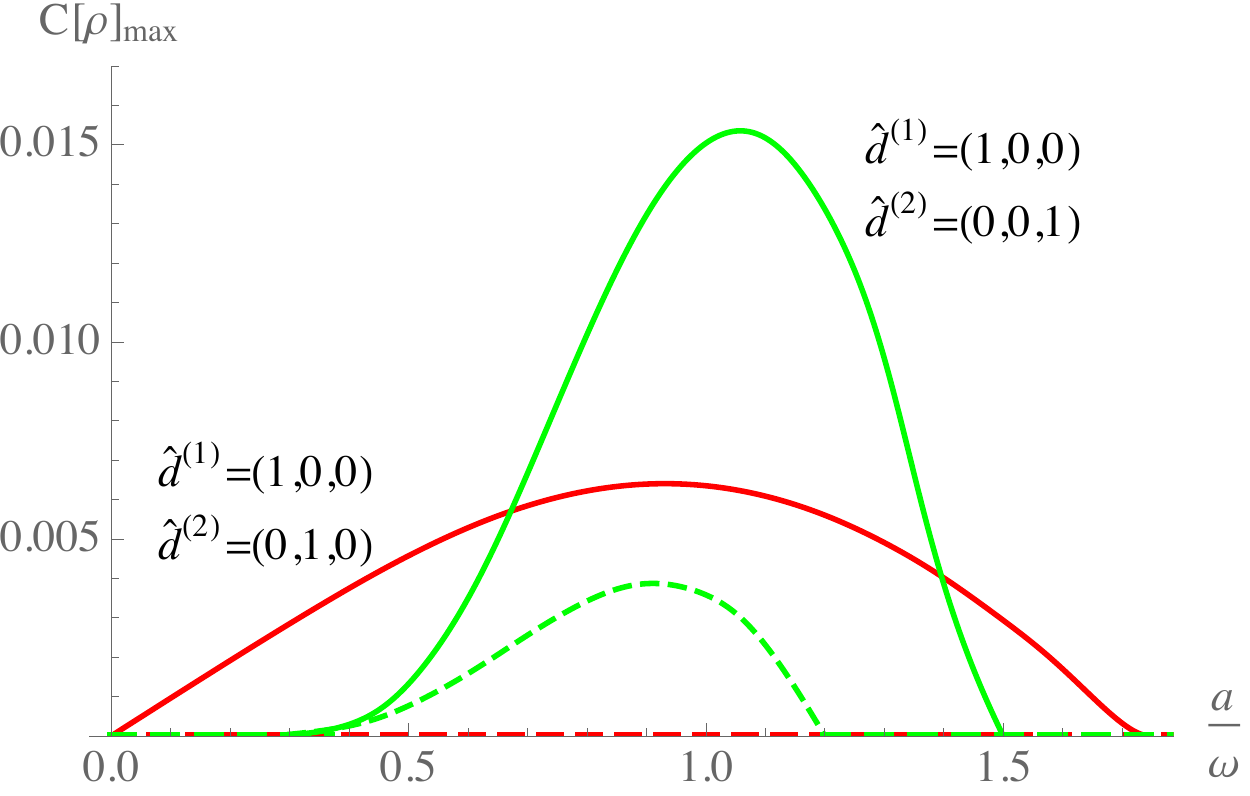}
\includegraphics[width=0.5\textwidth]{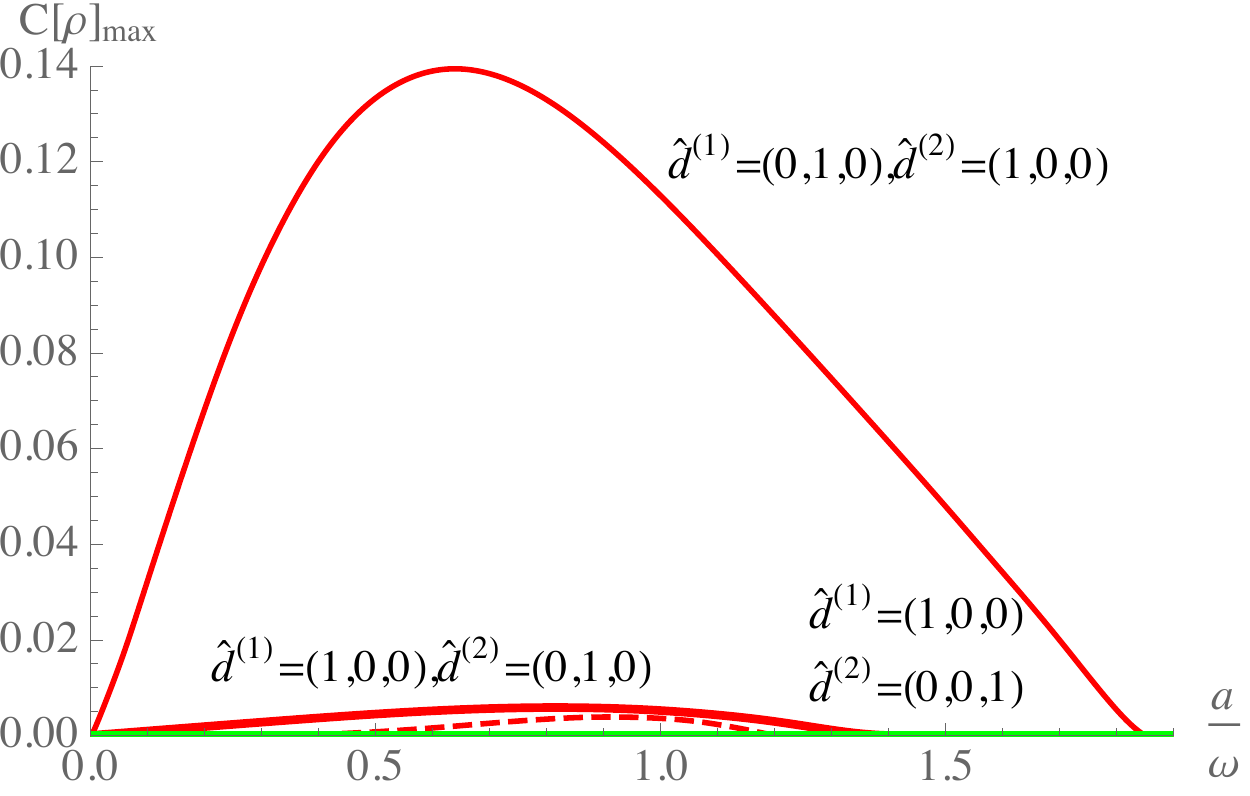}
\caption{\label{a/w2}
Comparison between the maximal concurrence during evolution in the free space (dashed lines) and that in the presence of a boundary (solid lines). The two-atom system initially prepared in $|E\rangle$ is aligned  parallel to (left) or vertically to (right) the boundary.
One atom is polarizable along the direction of acceleration  (the $x$-axis), while the other is polarizable vertically to  (the $y$-axis) (red lines) or parallel to  (the $z$-axis) (green lines) the boundary.
The green dashed and solid lines in the right figure coincide with each other.
Here $\omega L=1/2$,  and $y/L=1/100$.}
\end{figure}

\begin{figure}[!htbp]
\centering
\includegraphics[width=0.48\textwidth]{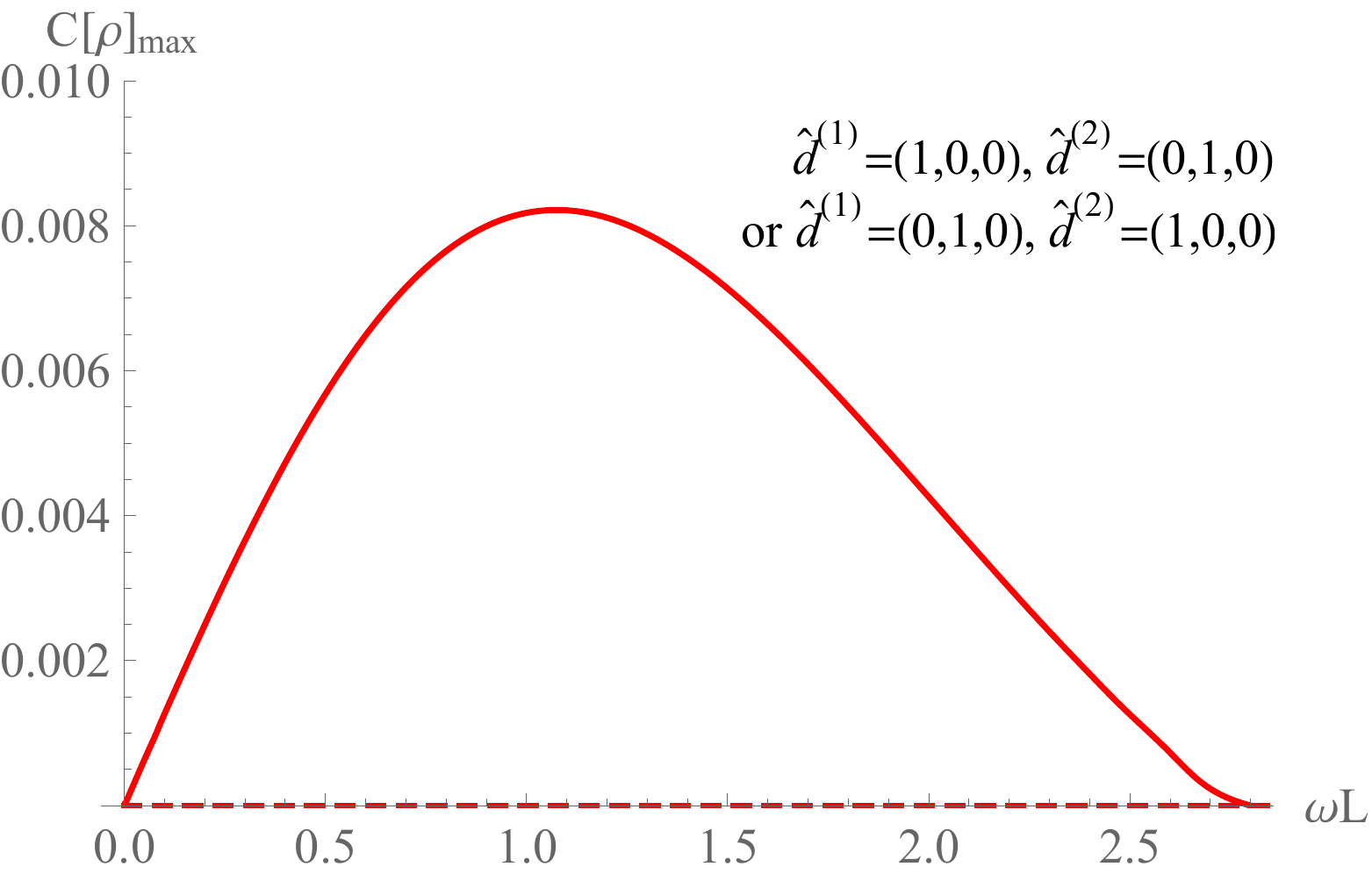}
\includegraphics[width=0.48\textwidth]{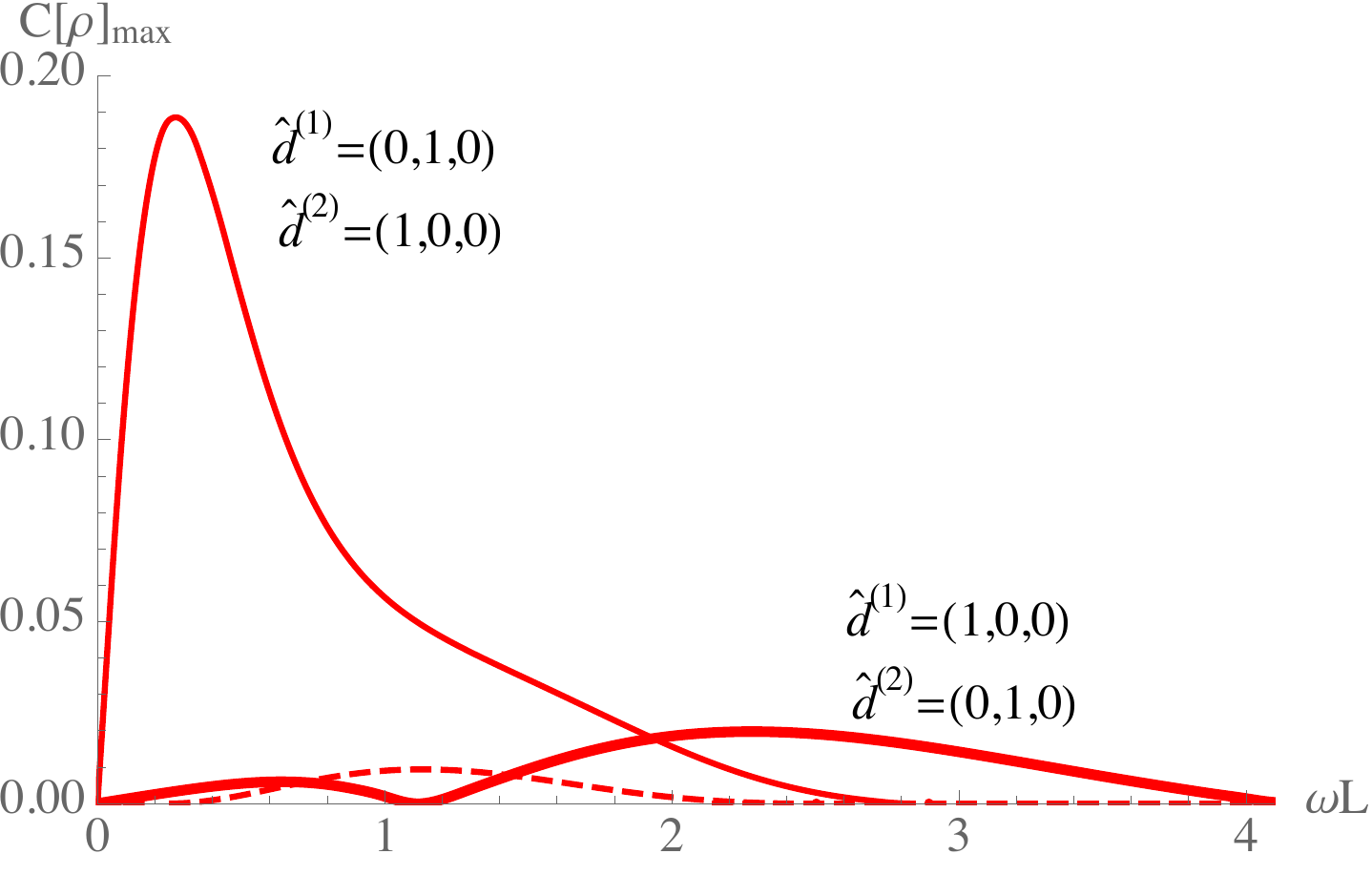}
\caption{\label{wl2}
Comparison between the maximal concurrence during evolution for uniformly accelerated atoms initially prepared in $|E\rangle$ aligned parallel to (left) or vertically  to (right) a reflecting boundary.
One of the atoms is polarizable along the direction of acceleration  (the $x$-axis), while the other is polarizable vertically to the boundary  (the $y$-axis).
Here $a/\omega=2/3$, $y/L=1/100$, the solid and dashed lines  describe the cases with and without the presence of the boundary respectively.
}
\end{figure}

To illustrate the case when the polarizations of the two atoms are different, we assume that one of the atoms is polarizable along the direction of acceleration  (the $x$-axis), while the other is polarizable vertically to the direction of acceleration  (the $yoz$ plane).
In  free space, two accelerated atoms can never get entangled for any given acceleration $a$ and separation $L$ when one of the atoms is polarizable  along  the direction of acceleration  (the $x$-axis) and the other vertically to the plane determined by the direction of acceleration and the atomic separation (the $y$-axis) \cite{Yang}.
However, they can get entangled in the presence of a boundary when they are aligned parallel to the boundary, as is indicated by Figs. \ref{a/w2}-\ref{wl2} (left).
When the atoms are vertically aligned,  the concurrence which is nonzero in circumstances in the free space can be enhanced by the presence of a boundary, see Figs. \ref{a/w2}-\ref{wl2} (right).


\subsection{Acceleration effects}

Now we discuss how the maximal concurrence during evolution is affected by the acceleration of the atoms.

\subsubsection{Atoms with identical orientation of polarization}

\begin{figure}[!htbp]
\centering
\includegraphics[width=0.48\textwidth]{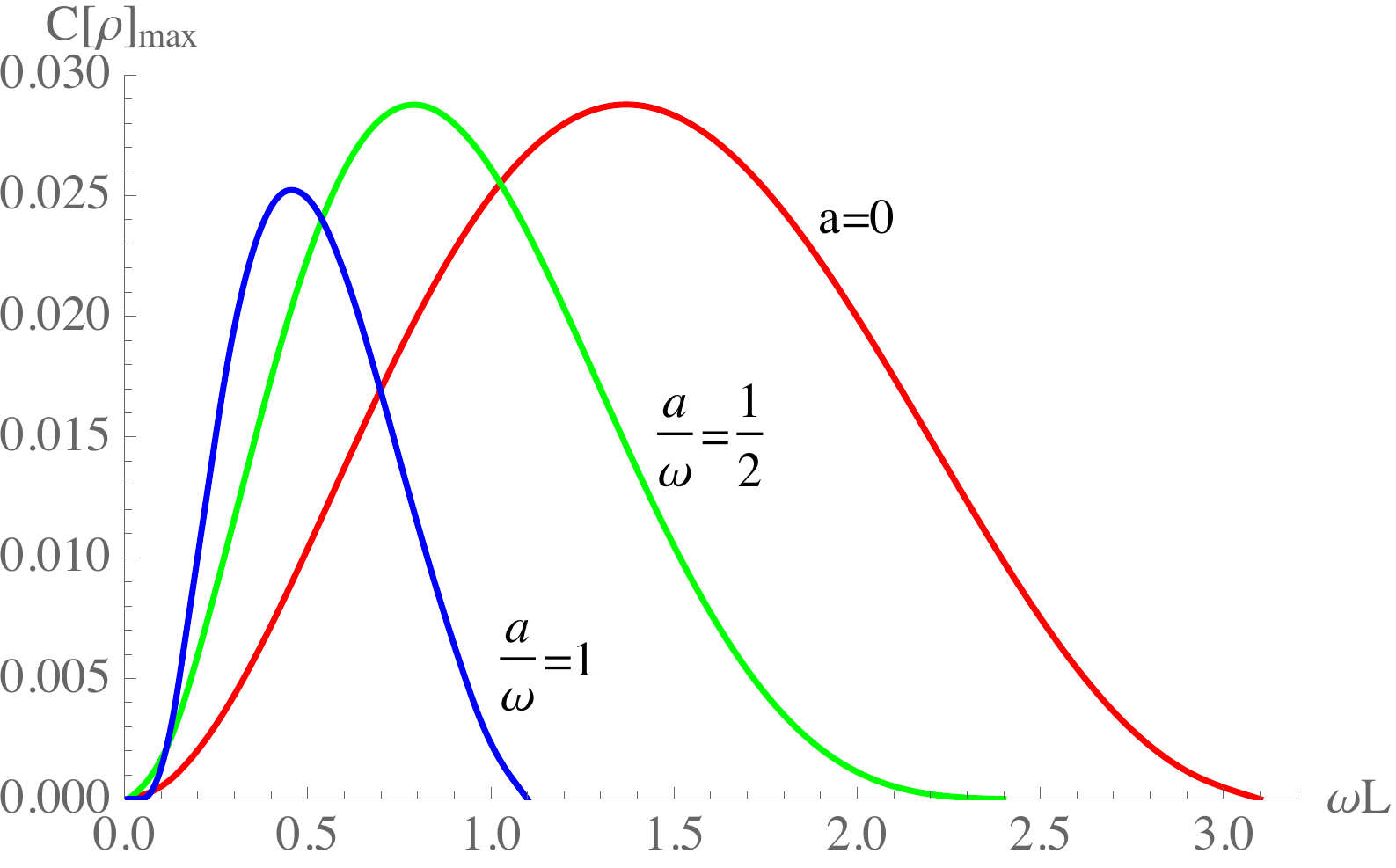}
\includegraphics[width=0.48\textwidth]{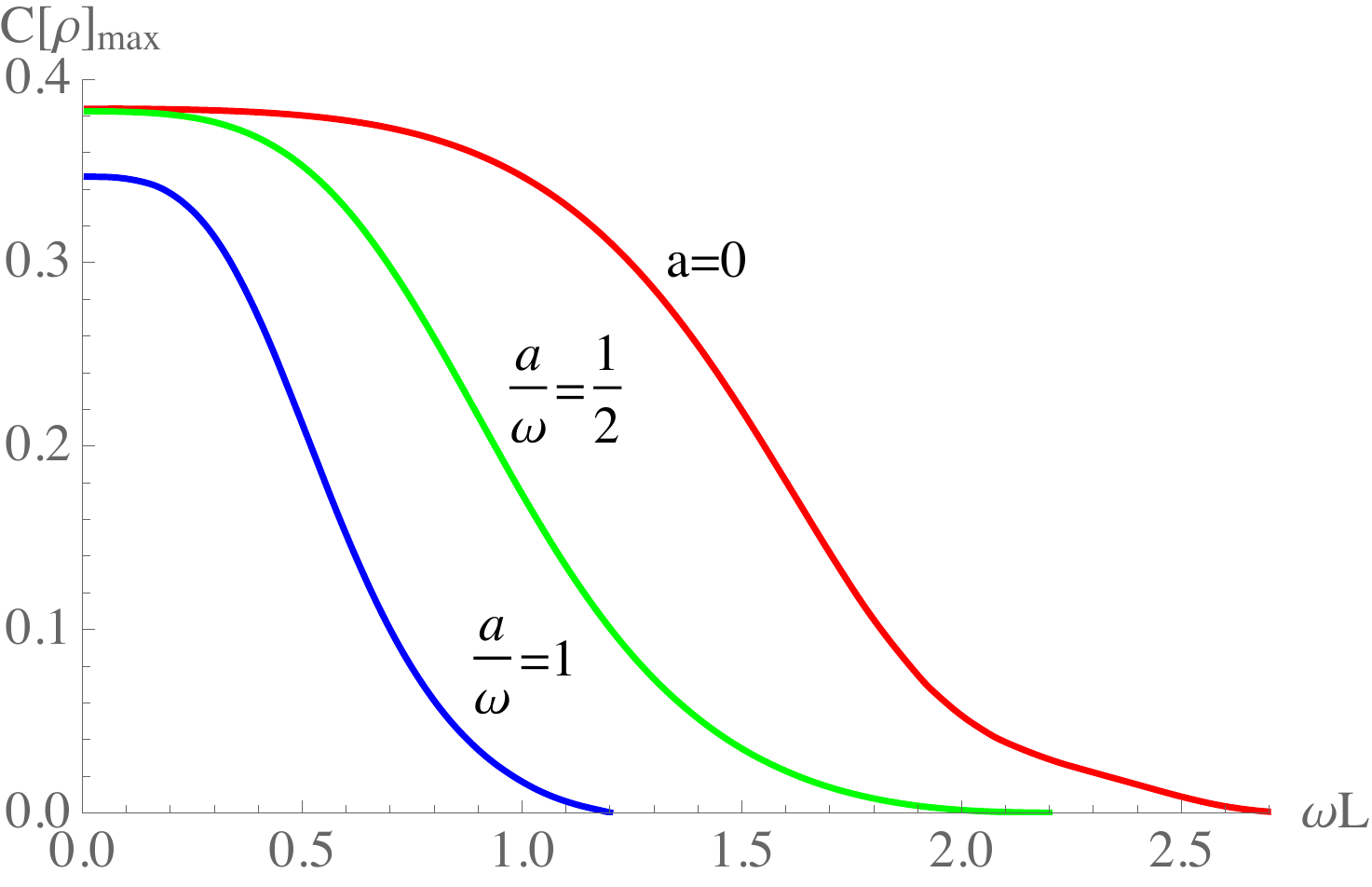}
\caption{\label{wl3}
Comparison between the maximal concurrence during evolution for uniformly accelerated atoms initially prepared in $|E\rangle$ aligned parallel to (left) or vertically  to (right) a reflecting boundary.
Both atoms are polarizable along the direction of acceleration  (the $x$-axis).
Here $y/L=1/2$, the red, green and blue lines correspond to $a
/\omega$=0, 1/2, 1 respectively.
}
\end{figure}

\begin{figure}[!htbp]
\centering
\includegraphics[width=0.48\textwidth]{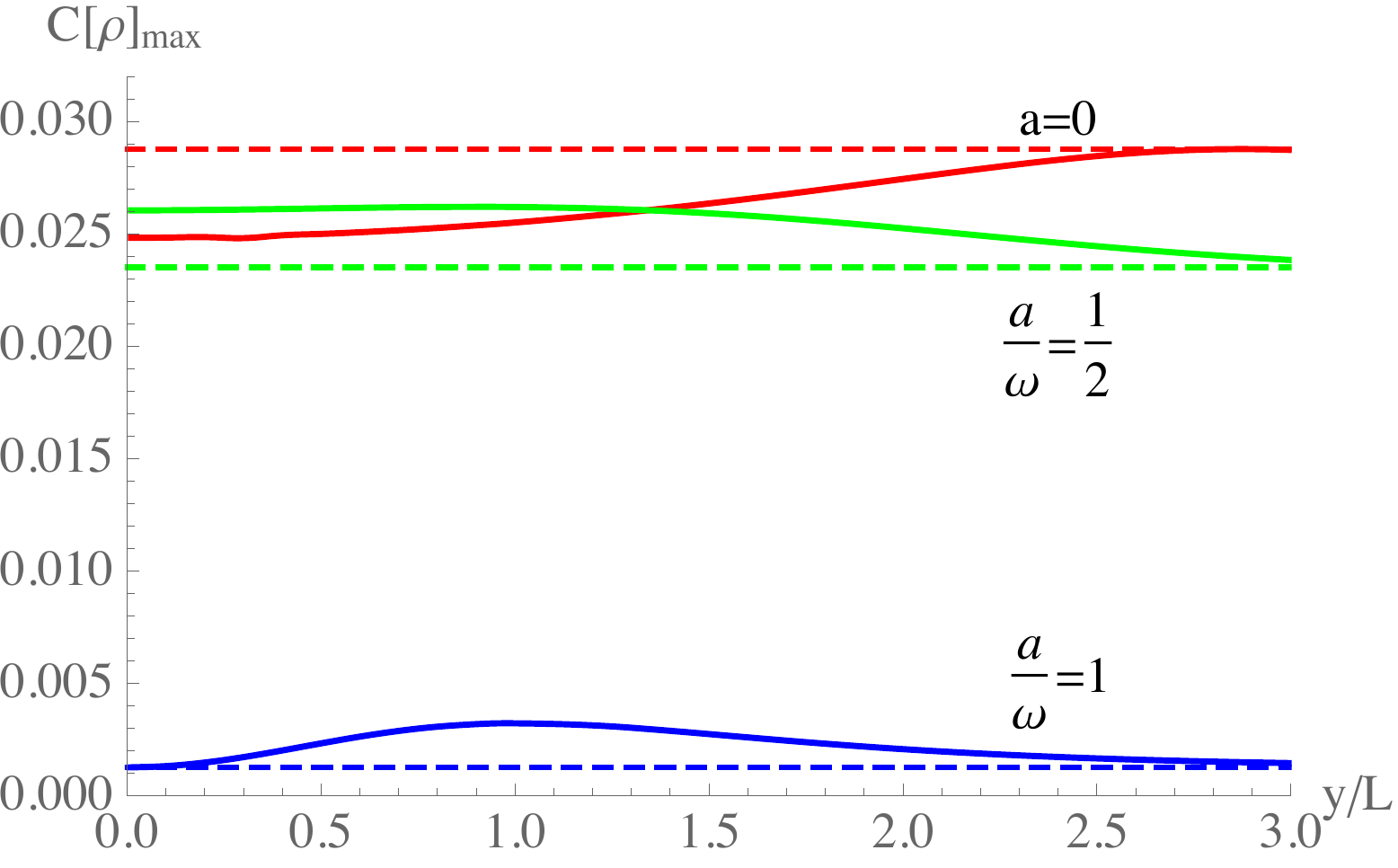}
\includegraphics[width=0.48\textwidth]{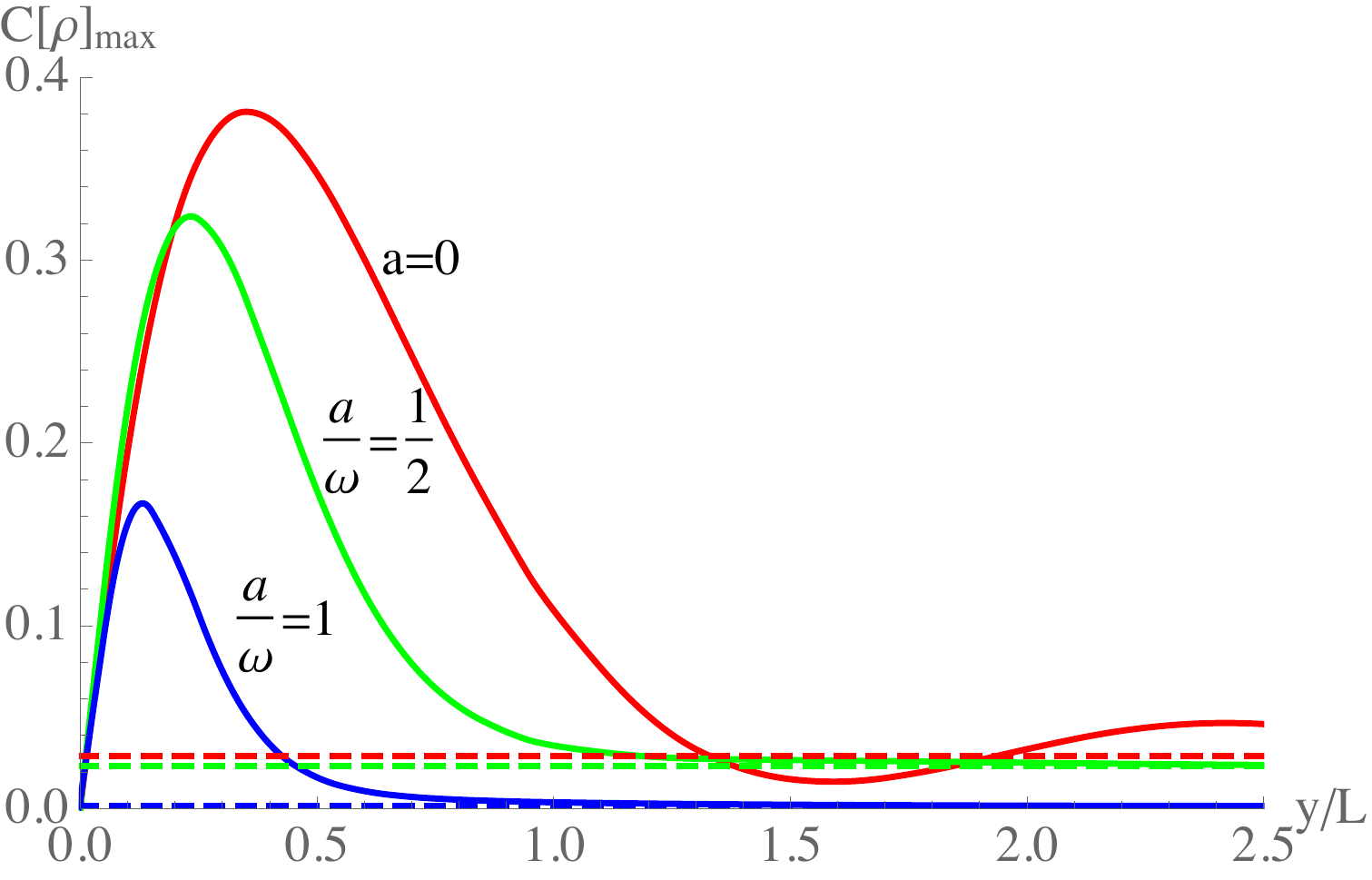}
\caption{\label{yl1}
Comparison between the maximal concurrence during evolution for uniformly accelerated atoms initially prepared in $|E\rangle$ aligned parallel to (left) or vertically  to (right) a reflecting boundary.
Both atoms are polarizable along the direction of acceleration  (the $x$-axis).
Here $\omega L=1$, the red, green and blue lines correspond to $a
/\omega$=0, 1/2, 1 respectively.
}
\end{figure}

In Figs.  \ref{wl3}-\ref{yl1}, we assume that both the atoms are polarizable along the direction of acceleration.
When the atoms are aligned parallel to the boundary, the maximal entanglement during evolution can be either enhanced or weakened by the acceleration, depending on the atomic separation and the distance between the boundary and the system, as shown in Figs.  \ref{wl3}-\ref{yl1} (left). However, in the vertically-aligned case, as acceleration increases, the maximal entanglement is weakened, see Fig.  \ref{wl3} (right).
As acceleration gets larger, the range of atomic separation within which entanglement generation happens with a boundary is apparently narrowed for both alignments.
Figure \ref{yl1} shows that, once entanglement can be created in the free space, the maximal  concurrence during  evolution can be significantly enhanced in the presence of a  boundary if they are vertically aligned.

\subsubsection{Atoms with different orientations of polarization}

\begin{figure}[!htbp]
\centering
\includegraphics[width=0.48\textwidth]{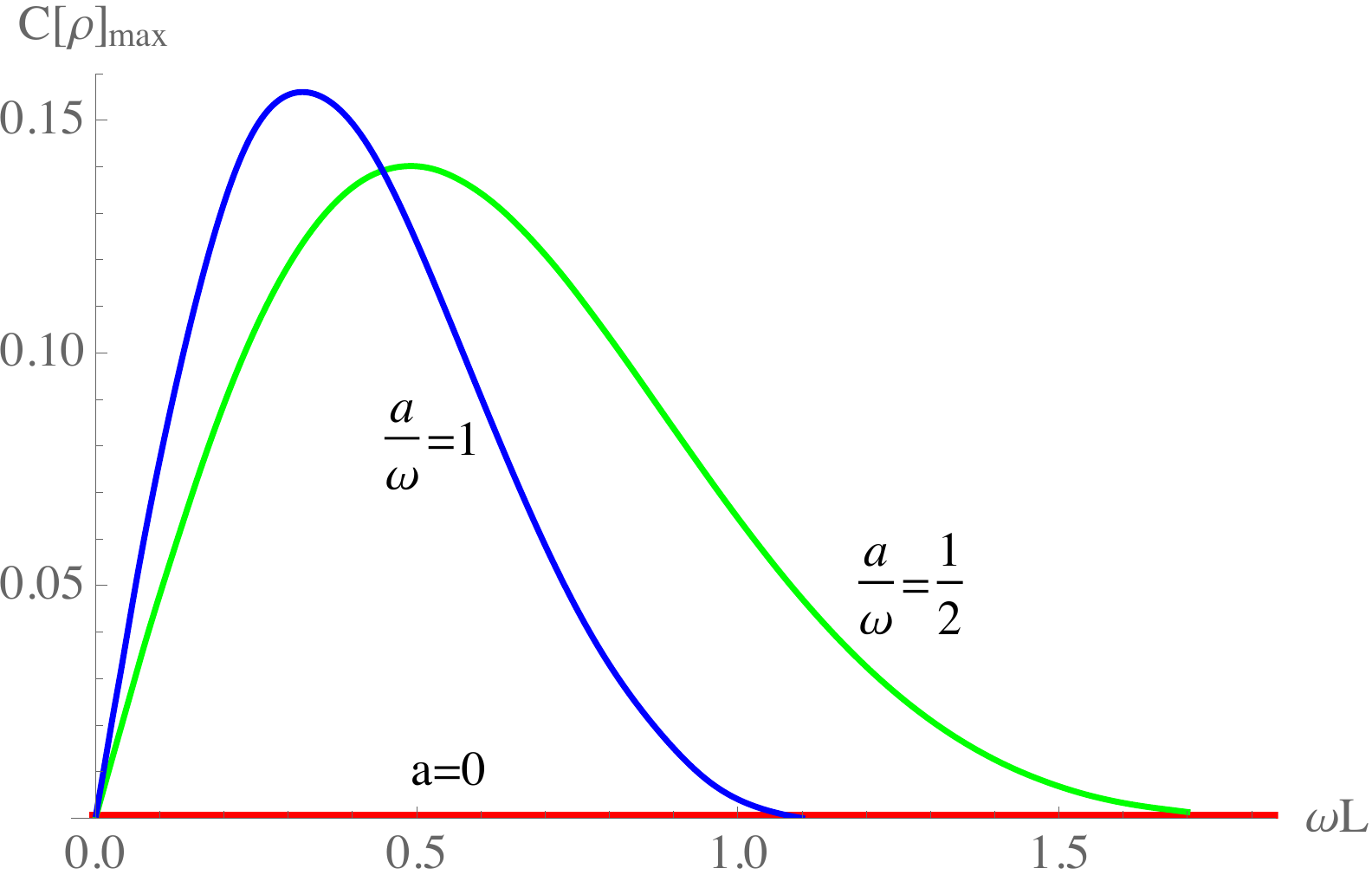}
\includegraphics[width=0.48\textwidth]{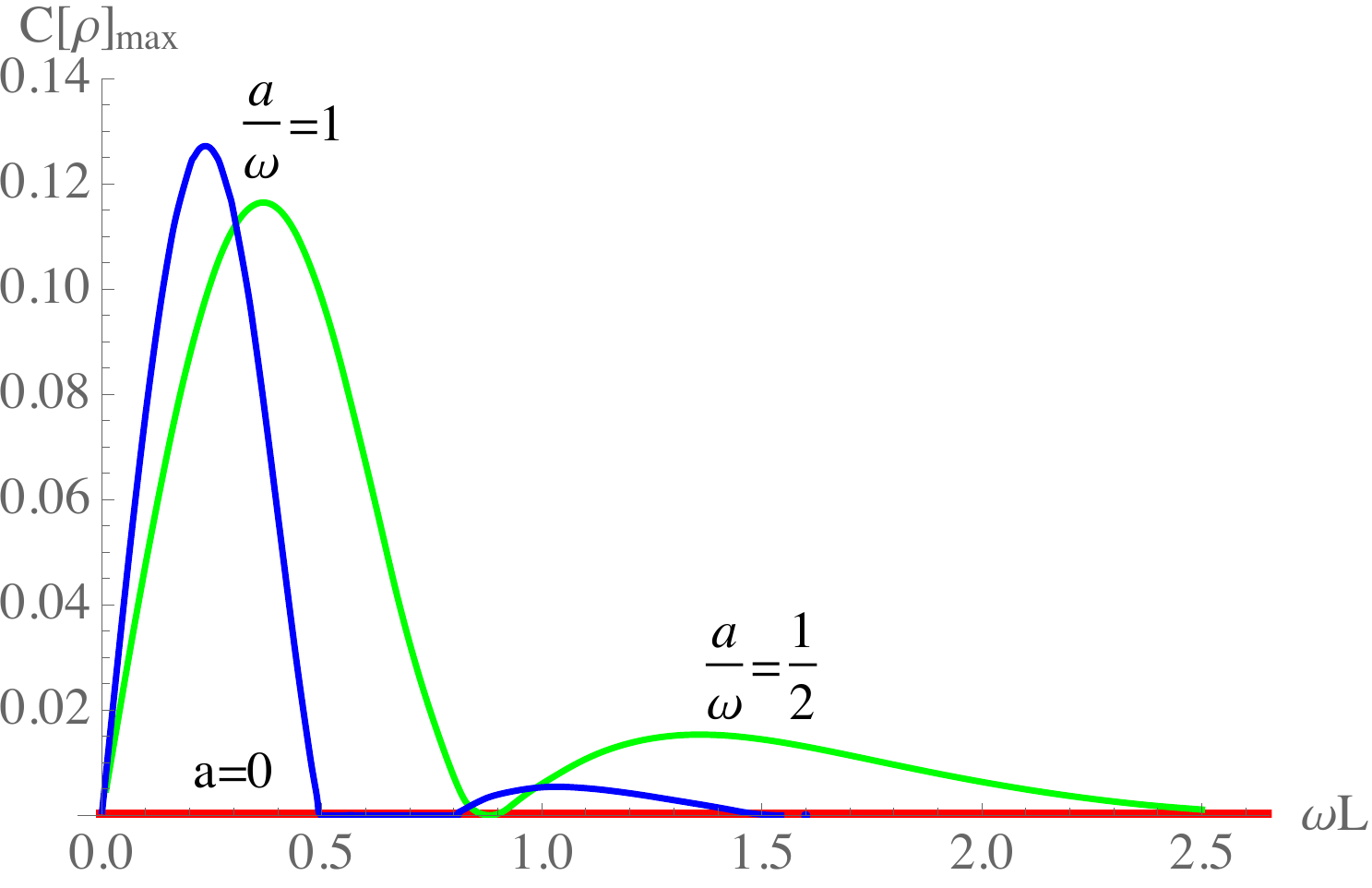}
\caption{\label{wl4}
Comparison between the maximal concurrence during evolution for uniformly accelerated atoms initially prepared in $|E\rangle$ aligned parallel to (left) or vertically  to (right) a reflecting boundary.
One of the atoms (the nearer one in the vertically aligned case) is polarizable along the direction of acceleration (the $x$-axis) and the other vertically to the boundary (the $y$-axis).
Here $y/L=1/2$, the red, green and blue lines correspond to $a
/\omega$=0, 1/2, 1 respectively.}
\end{figure}

\begin{figure}[!htbp]
\centering
\includegraphics[width=0.48\textwidth]{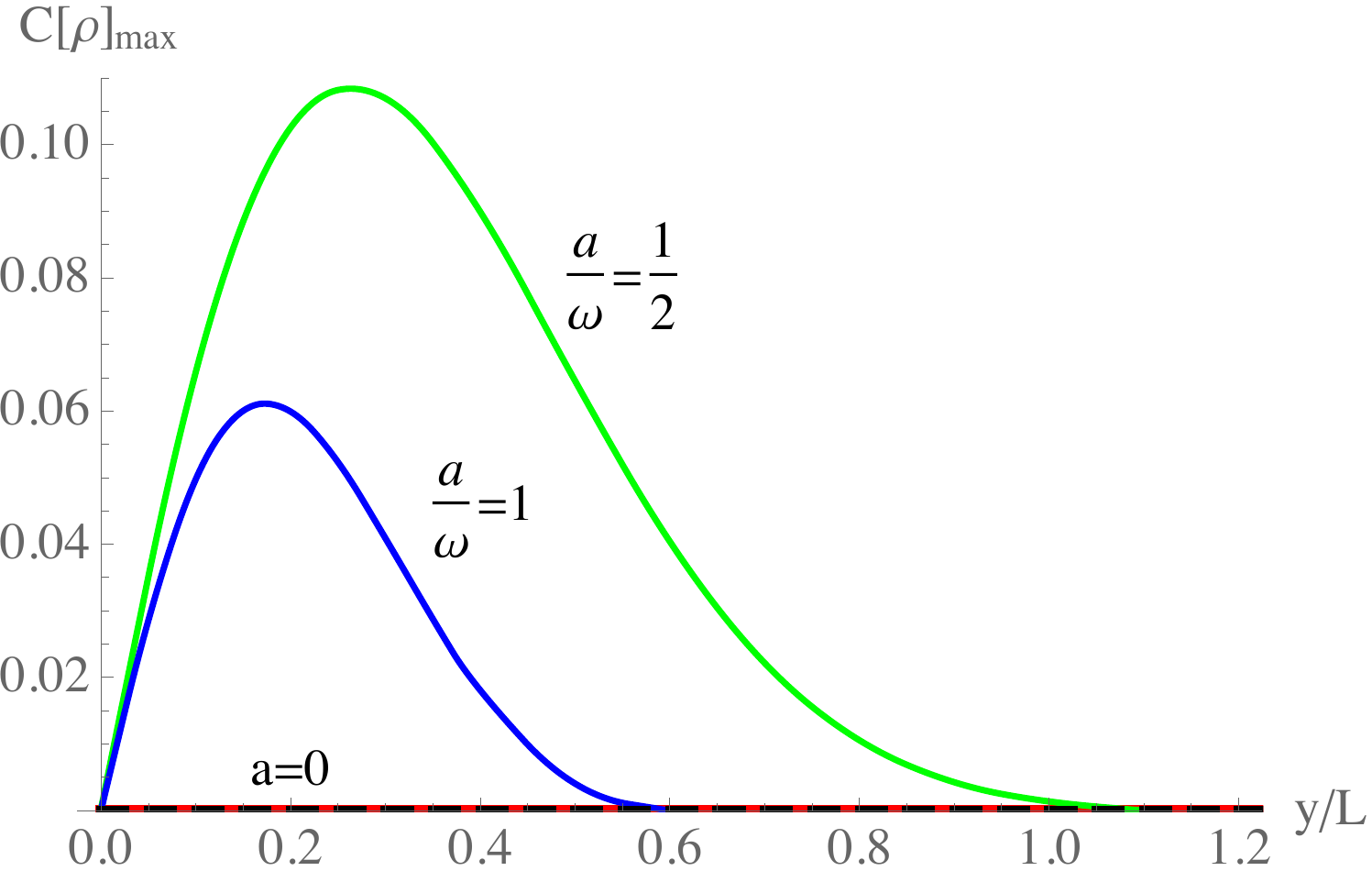}
\includegraphics[width=0.48\textwidth]{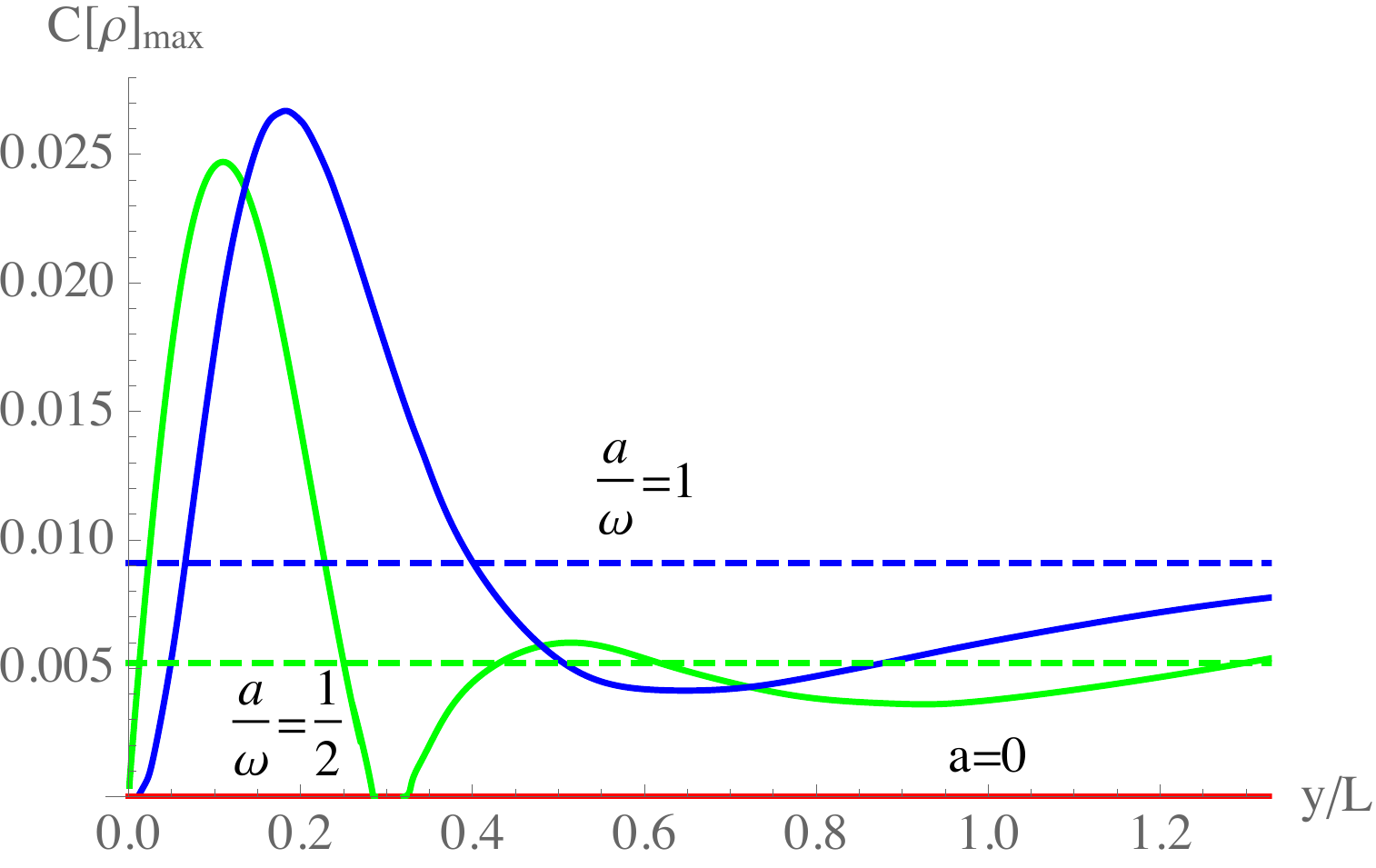}
\caption{\label{yl2}
Comparison between the maximal concurrence during evolution for uniformly accelerated atoms initially prepared in $|E\rangle$ aligned parallel to (left) or vertically  to (right) a reflecting boundary.
One of the atoms (the nearer one in the vertically aligned case) is polarizable along the direction of acceleration (the $x$-axis) and the other vertically to the boundary (the $y$-axis).
Here $\omega L=1$, the red, green and blue lines correspond to $a
/\omega$=0, 1/2, 1 respectively.
}
\end{figure}

As for the case of the two atoms with different orientations of polarization, we assume that one of the atoms (the nearer one in the vertically aligned case) is polarizable along the direction of acceleration (the $x$-axis), and the other is polarizable vertically to the boundary (the $y$-axis).
Figures \ref{wl4}-\ref{yl2} show that for the given polarizations,  entanglement generation between two inertial atoms can never happen, irrespective of the  separation $L$, distance $y$ and the alignment of the boundary. However,  the two atoms can get entangled with an appropriate  acceleration. 
Furthermore, for certain accelerations, there exists an interval of separation or distance within which entanglement cannot be generated when the atoms are vertically aligned, see Figs.~\ref{wl4}-\ref{yl2} (right).

\subsection{Polarization effects}

In the presence of a boundary, the maximal entanglement during evolution can be either enhanced or weakened, depending on the polarizations of the atoms, the acceleration, and the atomic separation, see Figs.  \ref{a/w1}-\ref{wl2}. 
When the two atoms are polarizable differently, entanglement generation can never happen in the free space with any given acceleration and separation, but it does happen when the two atoms are aligned parallel to the boundary. Meanwhile, the maximal entanglement during  evolution can be greatly enhanced for vertically-aligned atoms, compared with the case for atoms with the same polarizations.
In particular, in  Fig. \ref{wl2} (right), we show that a much larger concurrence can be reached when the nearer atom is polarizable vertically to the boundary  (the $y$-axis) and the farther one along the direction of acceleration  (the $x$-axis).

\section{Summary}

In this paper, we have investigated, in the framework of open quantum systems, the entanglement dynamics for two uniformly accelerated two-level atoms in weak interaction with a bath of fluctuating  electromagnetic fields in vacuum in the presence of a reflecting boundary.
In particular, two different alignments of atoms are considered, i.e. parallel and vertical alignments with respect to the boundary.
We focus on the effects of the boundary, and acceleration on the entanglement dynamics,  which are closely related to the atomic polarization.

The presence of a boundary greatly enriches dynamics of entanglement. 
When the atoms are placed far away from the boundary, the results reduce to those in the free space case as expected. 
When the atoms are placed extremely close to the boundary, for the parallel case, the initial entanglement of two transversely polarizable atoms can be preserved as if it were a closed system, while the concurrence of two vertically polarizable atoms evolves two times as fast as that in the free space, with its maximum during evolution remains the same.
In the presence of a parallel-aligned boundary, the revival of entanglement between two accelerated atoms initially in the symmetric state can  occur only if the atoms are polarizable differently, with one of them polarizable vertically to the boundary.
However, the entanglement revival can happen for the vertical two-atom system when the atomic polarizations are the same.
This is in sharp contrast to the fact that the concurrence of the two-atom system initially prepared in the symmetric state always decays monotonically in absence of a boundary.
Remarkably, two separable atoms both of which are initially in the excited state, for which entanglement generation can never happen in the free space with any given acceleration and separation, can get entangled in the presence of a boundary if they are aligned parallel to the boundary.
Moreover, the birth time of entanglement can be noticeably advanced or postponed for the parallel two-atom system placed close to the boundary, while the maximal concurrence during evolution can be significantly enhanced when the atoms are vertically aligned.

For the effect of acceleration, we find that it does not always play the role of destroying entanglement, but can generate and enhance entanglement as well. 
When the two atoms are polarizable differently, e.g., one of them is polarizable along the direction of acceleration and the other vertically to the boundary, entanglement revival appears with an appropriate acceleration, which cannot happen for inertial atoms. 
Similar situations occur when the system is initially in a separable state. That is, two inertial atoms with different polarizations remain separable all the time,  while as the acceleration increases, the delayed birth of entanglement happens, and the nonzero concurrence can be enhanced, in contrast to the fact that the delayed birth of entanglement can only be achieved for two inertial atoms with identical polarizations.
However, entanglement generation and revival cannot happen as the acceleration gets large enough.


\begin{acknowledgments}

This work was supported in part by the NSFC under Grants No. 11435006 and No. 11690034.

\end{acknowledgments}

\appendix

\section{CALCULATIONS OF THE COEFFICIENTS $A_i$ AND $B_i$ }\label{app}

The appendix is devoted to the calculations of the coefficients $A_{i}$ and $B_{i}$ in Eqs.~(\ref{evolution}) for two different alignments of the two-atom system.

We assume that a conducting boundary is placed at $y=0$. The two point function of the vector potential $A^{\mu}(x)$ can be obtained with the help of the method of images as
\begin{equation}
D^{\mu \nu}(x,x')=\langle 0|A^{\mu}(x)A^{\nu}(x')|0\rangle=D^{\mu \nu}_{free}(x-x')+D^{\mu \nu}_{bnd}(x,x'),
\end{equation}
where
\begin{equation}
D^{\mu \nu}_{free}(x-x')=\frac{\eta^{\mu\nu}}{4\pi^{2}\left[-(x-x')^{2}-(y-y')^{2}-(z-z')^{2}+(t-t'-i\varepsilon)^{2}\right]},
\end{equation}
and
\begin{equation}
D^{\mu \nu}_{bnd}(x,x')=-\frac{\eta^{\mu\nu}+2n^{\mu}n^{\nu}}{4\pi^{2}\left[-(x-x')^{2}-(y+y')^{2}-(z-z')^{2}+(t-t'-i\varepsilon)^{2}\right]},
\end{equation}
with $\varepsilon \rightarrow +0$, $\eta^{\mu\nu}=\mathrm{diag}(1,-1,-1,-1)$ and the unit normal vector $n^{\mu}=(0,0,1,0)$.
Here $D^{\mu \nu}_{free}(x-x')$ and $D^{\mu \nu}_{bnd}(x,x')$ are  the two point function in the free space and the correction due to the reflecting boundary respectively. The two point function of the electric-field strength then takes the following  form
\begin{equation}\label{EE}
\langle0|E_{m}(x(\tau))E_{n}(x(\tau'))|0\rangle=\langle0|E_{m}(x(\tau))E_{n}(x(\tau'))|0\rangle_{free}+\langle0|E_{m}(x(\tau))E_{n}(x(\tau'))|0\rangle_{bnd},
\end{equation}
where
\begin{eqnarray}
\langle0|E_{m}(x(\tau))E_{n}(x(\tau'))|0\rangle_{free}&=&\frac{1}{4\pi^{2}}\left[\delta_{mn}\partial _{0}\partial _{0}'-\partial _{m}\partial _{n}'\right]\nonumber\\
&&\times\frac{1}{(x-x')^{2}+(y-y')^{2}+(z-z')^{2}-(t-t'-i\varepsilon)^{2}},
\end{eqnarray}
and
\begin{eqnarray}
\langle0|E_{m}(x(\tau))E_{n}(x(\tau'))|0\rangle_{bnd}&=&\frac{1}{4\pi ^{2}}\left[(\delta_{mn}-2n_{m}n_{n})\partial _{0}\partial _{0}'-\partial _{m}\partial _{n}'\right] \nonumber \\
&&\times\frac{-1}{(x-x')^{2}+(y+y')^{2}+(z-z')^{2}-(t-t'-i\varepsilon)^{2}}.
\end{eqnarray}

\subsection{ The parallel case}

First, we consider a parallel two-atom system with a separation $L$,  whose distance to the boundary is $y$, as shown in Fig.~\ref{palver}.
Through a Lorentz transformation from the laboratory frame to the proper frame of the  atoms, and the Fourier transforms of the two point functions Eq. (\ref{EE}), we have, for $\alpha=\beta$,
\begin{equation}\label{pf1}
\mathcal{G}_{mn}^{(11)}(\omega)=\mathcal{G}_{mn}^{(22)}(\omega)=\frac{\omega^{3}}{3\pi\left(1-e^{-2\pi\omega/a}\right)}\left[f_{mn}^{(11)}(\omega,a)-h_{mn}^{(11)}(\omega,a,y)\right],
\end{equation}
and for $\alpha\neq\beta$,
\begin{equation}\label{pf2}
\mathcal{G}_{mn}^{(\alpha\beta)}(\omega)=\frac{\omega^{3}}{3\pi\left(1-e^{-2\pi\omega/a}\right)}\left[f_{mn}^{(\alpha\beta)}(\omega,a,L)-h_{mn}^{(\alpha\beta)}(\omega,a,y,L)\right],
\end{equation}
where
$f_{mn}^{(11)}(\omega,a)$ and $f_{mn}^{(\alpha\beta)}(\omega,a,L)$ correspond to the results of the free space case, see Eqs. (19)-(25) in Ref. \cite{Yang}, while $h_{mn}^{(11)}(\omega,a,y)$ and $h_{mn}^{(\alpha\beta)}(\omega,a,y,L)$ are the modifications due to the presence of the boundary.
Some straightforward calculations show that $h_{mn}^{(11)}(\omega,a,y)$ can be expressed as
\begin{eqnarray}
&&h_{11}^{(11)}(\omega,a,y)=f_{11}^{(12)}\left(\omega,a,\frac{L}{2}\right),\ \ \ \ \ \ h_{12}^{(11)}(\omega,a,y)=h_{21}^{(11)}(\omega,a,y)=-f_{13}^{(12)}\left(\omega,a,\frac{L}{2}\right),\nonumber\\
&&h_{22}^{(11)}(\omega,a,y)=-f_{33}^{(12)}\left(\omega,a,\frac{L}{2}\right),\ \ \ \ h_{13}^{(11)}(\omega,a,y)=h_{31}^{(11)}(\omega,a,y)=0,\nonumber\\
&&h_{33}^{(11)}(\omega,a,y)=f_{22}^{(12)}\left(\omega,a,\frac{L}{2}\right),\ \ \ \ \ \ h_{23}^{(11)}(\omega,a,y)=h_{32}^{(11)}(\omega,a,y)=0,
\end{eqnarray}
and all the nonzero components of $h_{mn}^{(\alpha \beta)}(\omega,a,y,L)$ ($\alpha\neq\beta$) are
\begin{eqnarray}
h_{11}^{(12)}(\omega,a,y,L)&=&\frac{12}{\omega^{3}R^{3}(4+a^{2}R^{2})^{5/2}}\nonumber\\
&&\times\bigg\{2\omega R \sqrt{4+a^{2}R^{2}} (1+a^{2}R^{2})\cos{\left(\frac{2\omega}{a}\sinh^{-1}\frac{aR}{2}\right)}\nonumber\\
&& +\left[-4-R^{2}(2a^{2}+a^{4}R^{2}-4\omega^{2}-\omega^{2}a^{2}R^{2})\right]\sin{\left(\frac{2\omega}{a}\sinh^{-1}\frac{aR}{2}\right)}\bigg\},\nonumber\\
h_{22}^{(12)}(\omega,a,y,L)&=&\frac{3}{\omega^{3}R^{5}(4+a^{2}R^{2})^{5/2}}\times\bigg\{-\omega R \sqrt{4+a^{2}R^{2}}\nonumber\\
&&\left[(2+a^{2}R^{2})(4L^{2}+a^{2}L^{4}-16a^{2}y^{4})-64y^{2}\right] \cos{\left(\frac{2\omega}{a}\sinh^{-1}\frac{aR}{2}\right)}\nonumber\\
&&-[64(2+a^{2}L^{2})y^{2}+320a^{2}y^{4}-4L^{2}(4+a^{2}L^{2})\nonumber\\
&& +\omega^{2}R^{2}(4+a^{2}R^{2})(4L^{2}+a^{2}L^{4}-16a^{2}y^{4})]\sin{\left(\frac{2\omega}{a}\sinh^{-1}\frac{aR}{2}\right)}\bigg\},\nonumber
\end{eqnarray}
\begin{eqnarray}
h_{33}^{(12)}(\omega,a,y,L)&=&\frac{3}{\omega^{3}R^{5}(4+a^{2}R^{2})^{5/2}}\times\bigg\{[20a^{2}L^{4}+32L^{2}(1+2a^{2}y^{2})-64y^{2}(1+a^{2}y^{2}) \nonumber\\
&&+(4+a^{2}R^{2})(16y^{2}-a^{2}L^{4}+16a^{2}y^{4})\omega^{2}R^{2}]\sin{\left(\frac{2\omega}{a}\sinh^{-1}\frac{aR}{2}\right)}\nonumber\\
&&+\omega R \sqrt{4+a^{2}R^{2}} [-a^{4}L^{6}-2L^{4}(a^{2}+2a^{4}y^{2})+16L^{2}(a^{2}y^{2}+a^{4}y^{4}-1)\nonumber\\
&&+32(y^{2}+3a^{2}y^{4}+2a^{4}y^{6})]\cos{\left(\frac{2\omega}{a}\sinh^{-1}\frac{aR}{2}\right)}\bigg\},\nonumber\\
h_{12}^{(12)}(\omega,a,y,L)&=&\frac{-12ay}{\omega^{3}R^{3}(4+a^{2}R^{2})^{5/2}}\nonumber\\
&&\times\bigg\{\omega R \sqrt{4+a^{2}R^{2}} (a^{2}R^{2}-2)\cos{\left(\frac{2\omega}{a}\sinh^{-1}\frac{aR}{2}\right)}\nonumber\\
&& +\left[4+R^{2}(4\omega^{2}+4a^{2}+\omega^{2}a^{2}R^{2})\right]\sin{\left(\frac{2\omega}{a}\sinh^{-1}\frac{aR}{2}\right)}\bigg\},\nonumber\\
h_{13}^{(12)}(\omega,a,y,L)&=&\frac{-6aL}{\omega^{3}R^{3}(4+a^{2}R^{2})^{5/2}}\nonumber\\
&&\times\bigg\{\omega R \sqrt{4+a^{2}R^{2}} (-2+a^{2}R^{2})\cos{\left(\frac{2\omega}{a}\sinh^{-1}\frac{aR}{2}\right)}\nonumber\\
&& +\left[4+R^{2}(4\omega^{2}+4a^{2}+\omega^{2}a^{2}R^{2})\right]\sin{\left(\frac{2\omega}{a}\sinh^{-1}\frac{aR}{2}\right)}\bigg\},\nonumber\\
h_{23}^{(12)}(\omega,a,y,L)&=&\frac{12Ly}{\omega^{3}R^{5}(4+a^{2}R^{2})^{5/2}}\nonumber\\
&&\times\bigg\{(2+a^{2}R^{2})\left[\omega^{2}R^{2}(4+a^{2}R^{2})-12\right]\sin{\left(\frac{2\omega}{a}\sinh^{-1}\frac{aR}{2}\right)}\nonumber\\
&&+\omega R \sqrt{4+a^{2}R^{2}} (12+4a^{2}R^{2}+a^{4}R^{4})\cos{\left(\frac{2\omega}{a}\sinh^{-1}\frac{aR}{2}\right)}\bigg\},\nonumber\\
h_{11}^{(12)}(\omega,a,y,L)&=&h_{11}^{(21)}(\omega,a,y,L),\ \
h_{22}^{(12)}(\omega,a,y,L)=h_{22}^{(21)}(\omega,a,y,L),
\nonumber\\
h_{33}^{(12)}(\omega,a,y,L)&=&h_{33}^{(21)}(\omega,a,y,L),
\nonumber\\
h_{12}^{(12)}(\omega,a,y,L)&=&h_{12}^{(21)}(\omega,a,y,L)=h_{21}^{(12)}(\omega,a,y,L)=h_{21}^{(21)}(\omega,a,y,L),
\nonumber\\
h_{13}^{(12)}(\omega,a,y,L)&=&-h_{13}^{(21)}(\omega,a,y,L)=-h_{31}^{(12)}(\omega,a,y,L)=h_{31}^{(21)}(\omega,a,y,L),\nonumber\\ h_{23}^{(12)}(\omega,a,y,L)&=&-h_{23}^{(21)}(\omega,a,y,L)=-h_{32}^{(12)}(\omega,a,y,L)=h_{32}^{(21)}(\omega,a,y,L),
\end{eqnarray}
with $R=\sqrt{L^{2}+4y^{2}}$. Then the coefficients of the master equations are
\begin{eqnarray}\label{pab}
A_{1(p)}&=&\frac{\Gamma_{0}\coth{\frac{\pi \omega}{a}}}{4}\sum_{i,j=1}^{3}\left(f^{(11)}_{ij}-h^{(11)}_{ij}
\right)\hat{d}_{i}^{(1)}\hat{d}_{j}^{(1)},\ \ \ B_{1(p)}=\frac{\Gamma_{0}}{4}\sum_{i,j=1}^{3}\left(f^{(11)}_{ij}-h^{(11)}_{ij}\right)\hat{d}_{i}^{(1)}\hat{d}_{j}^{(1)},\nonumber\\
A_{2(p)}&=&\frac{\Gamma_{0}\coth{\frac{\pi \omega}{a}}}{4}\sum_{i,j=1}^{3}\left(f^{(22)}_{ij}-h^{(22)}_{ij}\right)\hat{d}_{i}^{(2)}\hat{d}_{j}^{(2)},\ \ \ B_{2(p)}=\frac{\Gamma_{0}}{4}\sum_{i,j=1}^{3}\left(f^{(22)}_{ij}-h^{(22)}_{ij}\right)\hat{d}_{i}^{(2)}\hat{d}_{j}^{(2)},\nonumber\\
A_{3(p)}&=&\frac{\Gamma_{0}\coth{\frac{\pi \omega}{a}}}{4}\sum_{i,j=1}^{3}\left(f^{(12)}_{ij}-h^{(12)}_{ij}\right)\hat{d}_{i}^{(1)}\hat{d}_{j}^{(2)},\ \ \ B_{3(p)}=\frac{\Gamma_{0}}{4}\sum_{i,j=1}^{3}\left(f^{(12)}_{ij}-h^{(12)}_{ij}\right)\hat{d}_{i}^{(1)}\hat{d}_{j}^{(2)},\nonumber\\
&&\
\end{eqnarray}
where the subscript $p$ denotes the case of the atoms aligned parallel to the boundary, $\Gamma_{0}=\omega^{3}|\mathbf{d}|^{2}/3\pi$ is the spontaneous emission rate, and $\hat{d}_{i}^{(\alpha)}$ is a unit vector defined as $\hat{d}_{i}^{(\alpha)}=d_{i}^{(\alpha)}/|\mathbf{d}|$. Here we have assumed that the magnitudes of the electric dipoles of the atoms are assumed to be the same, i.e., $|\mathbf{d}^{(1)}|=|\mathbf{d}^{(2)}|=|\mathbf{d}|$,

\subsection{ The vertical case}

Next, we consider the situation in which the two atoms are aligned vertically to the boundary.
The distance between the boundary and the nearer atom is $y$,  see Fig.~\ref{palver}.
Similarly,  the Fourier transforms of the two point functions can be expressed as
\begin{eqnarray}\label{vs}
\mathcal{G}_{mn}^{(11)}(\omega)&=&\frac{\omega^{3}}{3\pi(1-e^{-2\pi\omega/a})}\left[g_{mn}^{(11)}(\omega,a)-s_{mn}^{(11)}(\omega,a,y)\right],\nonumber\\
\mathcal{G}_{mn}^{(22)}(\omega)&=&\frac{\omega^{3}}{3\pi(1-e^{-2\pi\omega/a})}\left[g_{mn}^{(22)}(\omega,a)-s_{mn}^{(22)}(\omega,a,y,L)\right],\nonumber\\
\mathcal{G}_{mn}^{(12)}(\omega)&=&\frac{\omega^{3}}{3\pi(1-e^{-2\pi\omega/a})}\left[g_{mn}^{(12)}(\omega,a,L)-s_{mn}^{(12)}(\omega,a,y,L)\right].
\end{eqnarray}
The unbounded parts $g_{mn}^{(\alpha \beta)}$ can be related to  $f_{mn}^{(\alpha \beta)}$ as
\begin{eqnarray}
&&g_{mn}^{(11)}(\omega,a)=f_{mn}^{(11)}(\omega,a),\ \ \ \ \ \ \ \ \ \ \ \
g_{mn}^{(22)}(\omega,a)=f_{mn}^{(11)}(\omega,a),\nonumber \\
&&g_{11}^{(12)}(\omega,a,L)=f_{11}^{(12)}(\omega,a,L),\ \ \ \ \ \
g_{22}^{(12)}(\omega,a,L)=f_{33}^{(12)}(\omega,a,L),\nonumber\\
&&g_{33}^{(12)}(\omega,a,L)=f_{22}^{(12)}(\omega,a,L),\ \ \ \ \ \
g_{12}^{(12)}(\omega,a,L)=f_{13}^{(12)}(\omega,a,L),
\end{eqnarray}
and for the bounded parts $s_{mn}^{(\alpha \beta)}$,  we find the following relations
\begin{eqnarray}
&&s_{mn}^{(11)}(\omega,a,y)=h_{mn}^{(11)}(\omega,a,y),\nonumber \\
&&s_{mn}^{(22)}(\omega,a,y,L)=h_{mn}^{(11)}(\omega,a,y+L),\nonumber \\
&&s_{mn}^{(12)}(\omega,a,y,L)=h_{mn}^{(11)}\left(\omega,a,y+\frac{L}{2}\right).
\end{eqnarray}
With Eqs. (\ref{b1})-(\ref{b2}), the corresponding coefficients can be calculated as
\begin{eqnarray}\label{vab}
A_{1(v)}&=&\frac{\Gamma_{0}\coth{\frac{\pi \omega}{a}}}{4}\sum_{i,j=1}^{3}\left(g^{(11)}_{ij}-s^{(11)}_{ij}
\right)\hat{d}_{i}^{(1)}\hat{d}_{j}^{(1)},\ \ \ B_{1(v)}=\frac{\Gamma_{0}}{4}\sum_{i,j=1}^{3}\left(g^{(11)}_{ij}-s^{(11)}_{ij}\right)\hat{d}_{i}^{(1)}\hat{d}_{j}^{(1)},\nonumber\\
A_{2(v)}&=&\frac{\Gamma_{0}\coth{\frac{\pi \omega}{a}}}{4}\sum_{i,j=1}^{3}\left(g^{(22)}_{ij}-s^{(22)}_{ij}\right)\hat{d}_{i}^{(2)}\hat{d}_{j}^{(2)},\ \ \ B_{2(v)}=\frac{\Gamma_{0}}{4}\sum_{i,j=1}^{3}\left(g^{(22)}_{ij}-s^{(22)}_{ij}\right)\hat{d}_{i}^{(2)}\hat{d}_{j}^{(2)},\nonumber\\
A_{3(v)}&=&\frac{\Gamma_{0}\coth{\frac{\pi \omega}{a}}}{4}\sum_{i,j=1}^{3}\left(g^{(12)}_{ij}-s^{(12)}_{ij}\right)\hat{d}_{i}^{(1)}\hat{d}_{j}^{(2)},\ \ \ B_{3(v)}=\frac{\Gamma_{0}}{4}\sum_{i,j=1}^{3}\left(g^{(12)}_{ij}-s^{(12)}_{ij}\right)\hat{d}_{i}^{(1)}\hat{d}_{j}^{(2)},\nonumber\\
&&\
\end{eqnarray}
where the subscript $v$ denotes the case of the atoms  aligned vertically to the boundary.

\end{document}